\documentclass[twocolumn]{aastex61}
\usepackage{morefloats}
\usepackage{color}
\usepackage{amsmath}
\usepackage{multirow}
\usepackage{graphicx}
\usepackage{xspace}

\newcommand{\cm}{cm$^{-2}$\xspace}
\newcommand{\HI}{H\,{\sc i}\xspace}
\newcommand{\kms}{km s$^{-1}$\xspace}
\newcommand{\msun}{$M_{\odot}$\xspace}

\shorttitle{HI in the M81 triplet}
\shortauthors{de Blok et al.}
\submitjournal{and accepted for publication in The Astronomical Journal}
\begin{document}
\title{A high-resolution mosaic of the neutral hydrogen in the M81 triplet.} 
\author{W.J.G. de Blok}
\affiliation{Netherlands Institute for Radio Astronomy (ASTRON), Postbus 2, 7990 AA Dwingeloo, the Netherlands}
\affiliation{Dept.\ of Astronomy, Univ.\ of Cape Town, Private Bag X3, Rondebosch 7701, South Africa}
\affiliation{Kapteyn Astronomical Institute, University of Groningen, Postbus 800, 9700 AV Groningen, The Netherlands}
\author{Fabian Walter}
\affiliation{Max-Planck Institut f\"ur Astronomie, K\"onigstuhl 17, 69117, Heidelberg, Germany}
\author{Annette M. N. Ferguson}
\affiliation{Institute for Astronomy, University of Edinburgh, Royal Observatory, Blackford Hill, Edinburgh EH9 3HJ, UK}
\author{Edouard J. Bernard}
\affiliation{Universit\'e C\^ote d'Azur, OCA, CNRS, Lagrange, France}
\author{J.M. van der Hulst}
\affiliation{Kapteyn Astronomical Institute, University of Groningen, Postbus 800, 9700 AV Groningen, The Netherlands}
\author{Marcel Neeleman}
\affiliation{Max-Planck Institut f\"ur Astronomie, K\"onigstuhl 17, 69117, Heidelberg, Germany}
\author{Adam K. Leroy}
\affiliation{Department of Astronomy, The Ohio State University, 140 W. 18th Ave., Columbus, OH 43210, USA}
\author{J\"urgen Ott}
\affiliation{National Radio Astronomy Observatory, 1003 Lopezville Rd., Socorro, NM 87801, USA}
\author{Laura K. Zschaechner}
\affiliation{Max-Planck Institut f\"ur Astronomie, K\"onigstuhl 17, 69117, Heidelberg, Germany}
\affiliation{University of Helsinki, P.O. Box 64, Gustaf H\"{a}llstr\"{o}min katu 2a, FI-00014 University of Helsinki, Finland}
\affiliation{Finnish Center for Astronomy with ESO, FI-20014 Turun yliopisto, Finland}
\author{Martin A. Zwaan}
\affiliation{European Southern Observatory, Karl-Schwarzschild-Str. 2, D-85748 Garching Near Munich, Germany}
\author{Min S. Yun}
\affiliation{Department of Astronomy, University of Massachusetts, Amherst, MA 01003, USA}
\author{Glen Langston}
\affiliation{National Science Foundation, Division of Astronomical Sciences, Arlington, VA 22230, USA}
\author{Katie M. Keating}
\affiliation{Rincon Research Corporation, 101 North Wilmot Road, Suite 101, Tucson, AZ 85711, USA}


\begin{abstract}
  We present a $3^{\circ} \times 3^{\circ}$, 105-pointing,
  high-resolution neutral hydrogen (\HI) mosaic of the M81 galaxy
  triplet, (including the main galaxies M81, M82 and NGC 3077, as well
  as dwarf galaxy NGC 2976) obtained with the Very Large Array (VLA) C
  and D arrays. This \HI synthesis mosaic uniformly covers the entire
  area and velocity range of the triplet. The observations have a
  resolution of $\sim 20''$ or $\sim 420$ pc. The data reveal many
  small-scale anomalous velocity features highlighting the complexity
  of the interacting M81 triplet.  We compare our data with Green Bank
  Telescope (GBT) observations of the same area. This comparison
  provides evidence for the presence of a substantial reservoir of
  low-column density gas in the northern part of the triplet, probably
  associated with M82. Such a reservoir is not found in the southern
  part. We report a number of newly discovered kpc-sized low-mass \HI
  clouds with \HI masses of a few times $10^6$ \msun. A detailed
  analysis of their velocity widths show that their dynamical masses
  are much larger than their baryonic masses, which could indicate the
  presence of dark matter if the clouds are rotationally
  supported. However, due to their spatial and kinematical association
  with \HI tidal features, it is more likely that the velocity widths
  indicate tidal effects or streaming motions. We do not find any
  clouds that are not associated with tidal features down to an \HI
  mass limit of a few times $10^4$ \msun. We compare the \HI column
  densities with resolved stellar density maps and find a star
  formation threshold around $3-6 \cdot 10^{20}$ cm$^{-2}$.  We
  investigate the widths of the \HI velocity profiles in the triplet
  and find that extreme velocity dispersions can be explained by a
  superposition of multiple components along the line of sight near
  M81 as well as winds or outflows around M82. The velocity
  dispersions found are high enough that these processes could explain
  the linewidths of Damped-Lyman-$\alpha$ absorbers observed at high
  redshift.
\end{abstract}


\section{Introduction}

The evolution of galaxies is affected by their environment.  Galaxy
interactions and mergers are probably the most obvious examples of
this.  These processes can severely alter or completely transform a
galaxy's properties. Signs of galaxy interactions are most easily
detected in the component of the galaxy that is the most sensitive to
them, namely the extended reservoirs of circum-galactic neutral
hydrogen (\HI) (although evidence for interactions can also be seen in
extended stellar envelopes; for an early example see
\citealt{ferguson02}).

The M81 triplet (with M81, M82 and NGC 3077 as the main galaxies) is
often presented as a prime example of the complexity of interactions,
their impact on the circum-galactic medium and, therefore, galaxy
evolution (see \citealt{yun94}).  The three main galaxies in the
triplet each highlight different aspects of galaxy evolution.  The
inner disk of the grand spiral galaxy M81 seems largely unaffected by
the interaction. Studies of the disk have been instrumental in
developing the theory of density waves and formation of (grand) spiral
structure (e.g., \citealt{rots75} for an early example).  The
starburst galaxy M82 is a unique target for studying
interaction-triggered star formation feedback processes in essentially
all wavelength bands (e.g., H$\alpha$, X-rays, dust, \HI, molecular
gas, see \citealt{walter02a,leroy15} and references therein).  The
third galaxy is NGC 3077, an optically smooth galaxy with an actively
starforming core, which has been stripped of most of its \HI (e.g.,
\citealt{walter02b}). Much of this \HI is now found immediately to
the east of the main galaxy as part of the ``Garland'' feature (e.g.,
\citealt{yun93a,walter11}).  The triplet is surrounded by
at least 20 dwarf galaxies (including a few tidal dwarfs) that
together form the greater `M81 group' (e.g.,
\citealt{karachentsev02}). One of the more prominent of these dwarf
galaxies is NGC 2976, an actively star-forming, gas-rich dwarf galaxy.

Very Large Array (VLA) D-array \HI observations of the M81 triplet,
obtained by \citet{yun94}, have played a
critical role in shaping our understanding of how interactions between
galaxies affect the distribution of the atomic gas (see also
\citealt{yun93a} and early work by \citealt{vanderhulst79} and
\citealt{appleton88}).  The 12 pointing ($\sim 2.8$ square degrees)
mosaic presented in \citet{yun94} (which was later extended to 24
pointings [$\sim 5.6$ square degrees], \citealt{yun00}) demonstrated
that the extended \HI emission in the triplet is dominated by
filamentary structures of many tens of kpc connecting the main
galaxies and containing most of the \HI in the system. These
structures are mostly due to the effects of the tidal interactions. No
such signs are visible in shallow optical imaging of the
triplet. However, recent star count analyses of the triplet, which are
sensitive to very low surface brightness emission, have revealed that
the stellar component is also affected by the interaction
\citep{okamoto15}.

Green Bank Telescope (GBT) \HI observations \citep{chynoweth08} of the
M81 triplet covered a larger area ($3^{\circ} \times 3^{\circ}$) down
to low column densities, albeit at a spatial resolution ($\sim 10$
kpc) that is insufficient to resolve the sub-kpc giant atomic
complexes that are present in nearby galaxies and that are key to our
understanding of star formation (e.g., \citealt{leroy08,
  bigiel08}). However, the GBT observations clearly demonstrated that
there is almost twice as much \HI present in the system than detected
in the \citet{yun94, yun00} observations.

Here, we use the dramatic increase in VLA capabilities over the last
two decades to completely map at high resolution (spatially and
spectrally) the \HI in the entirety of the M81 triplet and its
immediate surroundings.  We present a 105-pointing (7.6 square degrees
out to the 50 percent sensitivity level) C- and D-array mosaic,
covering the same area as the \citet{chynoweth08} GBT
observations. These new data form the most complete high-resolution
and high-sensitivity census of atomic gas in the M 81 triplet so far.
Our highest-resolution data set has a resolution of $\sim 24''$ (420
pc at $D = 3.63$ Mpc, the distance of the M81 triplet;
\citealt{karachentsev04}), or close to a factor of three higher
spatial resolution than the earlier 12-pointing data presented in
\citet{yun94}. These earlier data were limited by the capacities of
the correlator at the time, with different pointings observed over
different, fairly narrow velocity ranges.  In our new data, all
pointings cover the entire velocity range of the triplet, at a much
higher velocity resolution. These data thus form the currently most
complete and comprehensive view of the atomic gas in the M81 triplet
and its immediate surroundings.

A data set of this size with this level of detail has many
applications. Here, apart from presenting the data, we focus on the
following aspects. Our $\sim 400$ pc resolution observations
reach a limiting \HI mass of $\sim 10^4$ \msun and  constrain
the numbers of individual \HI clouds in the group down to very low
masses and sizes. This will provide a link to the missing satellite
problem, i.e., satellites with clumps of cool \HI, but no
star-formation.

A second important topic is the connection to high-redshift \HI
through measurements of the \HI probability distribution function.
Linking high-redshift \HI absorption measurements to local emission
properties is important as our high-$z$ \HI knowledge is based on
absorption measurements. If the M81 triplet were by chance observed in
the foreground against a high-redshift quasar, it would be classified
as either a Lyman Limit System (LLS), or a Damped Lyman-$\alpha$
absorber (DLA), depending on where exactly the quasar's sightline
would appear through the \HI distribution. Our covered area measures
about 0.2 Mpc $\times$ 0.2 Mpc, large enough to include typical impact
parameters between quasars and DLAs at high redshift (e.g.,
\citealt{rahmati14}). We can therefore directly compare the \HI
linewidths in the triplet with those seen in DLA systems at high
redshift.

Finally, we address the link between star formation and \HI in the
triplet.  Empirical descriptions of this link often treat star
formation as dependent on (amongst others) local conditions (such as
the gas column density; \citealt{skillman87}, or the cooling time;
\citealt{schaye01}) or assume a more global dependency on the galaxy
dynamics (e.g., Toomre-$Q$ or shear; \citealt{kennicutt89, hunter98}). Our high-resolution data will allow a direct comparison
with maps of the (resolved) stellar distribution of young stars.

In Sect.\ 2 we describe the observations, data reduction and data
products.  Section 3 compares the data with previous observations and
highlights the new aspects of this data set. We also compare our data
with the \citet{chynoweth08} deep Green Bank Telescope observations,
and discuss a number of low-mass \HI clouds visible in these data
sets.  In Sect.\ 4 we compare the \HI column densities with stellar
density maps and relate the profile velocity width in our data with
those found in DLAs. Finally, we summarise
our results in Sect.\ 5.

\section{Observations, data reduction and presentation}

The M81 triplet and its immediate surrounding were observed as part of
a large 105 pointing mosaic covering $3^{\circ} \times 3^{\circ}$ ($190 \times 190$ kpc),
centered on M81. This is the same area as observed by \citet{chynoweth08}
using the GBT.  The observations were done using the VLA in its C- and
D-configurations between October 2015 and April 2016.

The D-array observations took place in 10 separate observing sessions
between October and December 2015 (project 15B-122); the
C-configuration was used in 27 separate sessions during March and
April 2016 (project 16A-073).

\subsection{Mosaic layout and observations}

\begin{figure}
  \centering
\includegraphics[width=0.95\hsize]{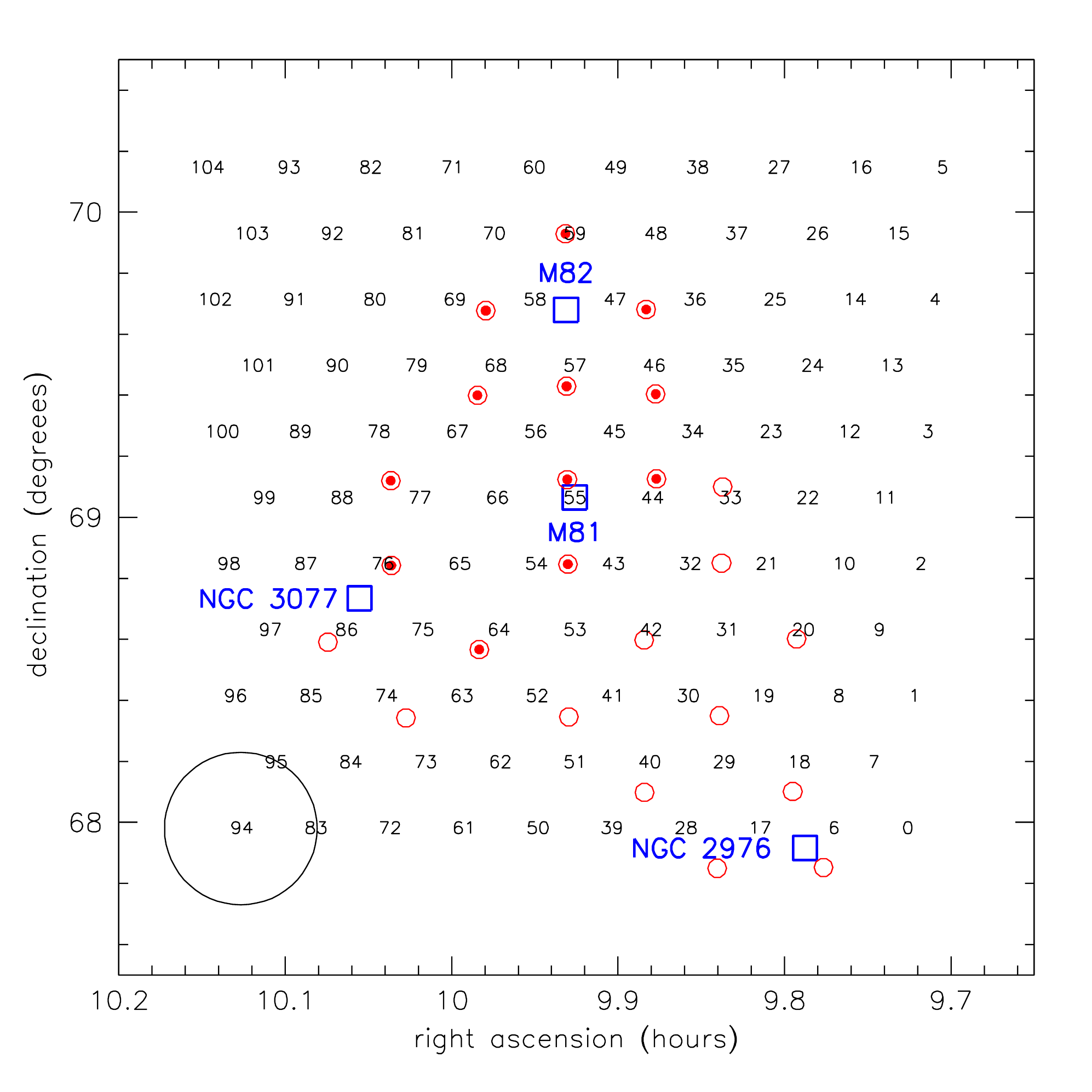}
\caption{Mosaic pointings. Numbers indicate the central positions of
  the pointings listed in Table \ref{tab:pointings}. Red, dotted
  circles indicate the positions of the mosaic presented in
  \citet{yun94}. Red, open circles show the additional mosaic
  pointings described in \citet{yun00} and also presented in
  \citet{chynoweth08}. The large circle in the lower left shows the
  size of the primary beam of a single pointing. The positions of M81,
  M82, NGC 3077 and NGC 2976 are indicated by blue squares.
\label{fig:fields}}
\end{figure}

We adopt a hexagonal Nyquist-sampled grid pattern of 105 separate
pointing as shown in Fig.\ \ref{fig:fields}. The horizontal and
vertical distance between pointings is $13'$ (half a primary beam
width at 21-cm wavelength), with each row offset horizontally by half
a spacing (one quarter primary beam width or $\sim 6.5'$). The center
positions of the pointings are listed in Table \ref{tab:pointings} and
shown in Fig.~\ref{fig:fields}. The total observing area measures
$\sim 3^{\circ} \times 3^{\circ}$, with the area inside the 50 percent
sensitivity level measuring $\sim 2.7^{\circ} \times 2.7^{\circ}$.

Each pointing was visited once per observing session. Pointings were
observed in turn where in both right ascension and declination
directions every second row or column was skipped. This meant
that to observe all pointings, the grid was covered a total of four
times, each visit starting with a different pointing. This strategy
ensures a more homogeneous $uv$-coverage and reduces the impact that
any intermittant radio frequency interference (RFI) may have on any
particular position.

Each pointing was visited for two minutes (including $\sim 10$ seconds
slewing and settling time). Every 15 pointings, including at the
beginning and the end of each observing session, the complex
gain/phase calibrator J0949+6614 was observed for one minute
on-source. At the start of each session the primary flux and bandpass
calibrator 3C147 was observed for two minutes on-source. The total
duration of each observing session was 4 hours.

Combining all observing sessions, this resulted in a total integration
time of 20 minutes per pointing for the D-array and 54 minutes for the
C-array, with a total integration time for the entire mosaic of 40 hours
for the D-array and 108 hours for the C-array.

The WIDAR correlator was used in combination with the L-band
receivers. We used the 8-bit correlator setup. An 8192-channel, dual
polarization subband was used to observe the \HI line at $0.4$ \kms
(1.953 kHz) channel width over a 3200 \kms (16 MHz) bandwidth.  In
addition, eight spectral windows were allocated to observe the full
polarization continuum at 1 MHz resolution.  A 4 MHz subband was used
to observe the OH and radio recombination lines. In this paper we
discuss the \HI observations only.

\begin{deluxetable*}{cccccc}
\tabletypesize{\scriptsize} \tablewidth{0pt} \tablecaption{Pointing
  center coordinates\label{tab:pointings}}
\tablehead{\colhead{Pointing} & \colhead{$\alpha (2000.0)$} &
  \colhead{$\delta (2000.0)$} & \colhead{Pointing} & \colhead{$\alpha
    (2000.0)$} & \colhead{$\delta (2000.0)$} \\ \colhead{} &
  \colhead{$^h\ \ ^m\ \ ^s$} & \colhead{$^\circ\ \ '\ \ ''$} &
  \colhead{} & \colhead{$^h\ \ ^m\ \ ^s$} &
  \colhead{$^\circ\ \ '\ \ ''$}}
\startdata
    0      &    09 43 32.98& +67 58 57.88 &       52    &     09 56 54.75& +68 24 56.73    \\
    1      &    09 43 19.24& +68 24 56.73 &       53    &     09 55 33.20& +68 37 56.15    \\
    2      &    09 43 04.92& +68 50 55.57 &       54    &     09 56 56.34& +68 50 55.57    \\
    3      &    09 42 49.99& +69 16 54.42 &       55    &     09 55 33.20& +69 03 54.99    \\
    4      &    09 42 34.41& +69 42 53.26 &       56    &     09 56 58.00& +69 16 54.42    \\
    5      &    09 42 18.13& +70 08 52.11 &       57    &     09 55 33.20& +69 29 53.84    \\
    6      &    09 46 13.02& +67 58 57.88 &       58    &     09 56 59.73& +69 42 53.26    \\
    7      &    09 44 46.96& +68 11 57.30 &       59    &     09 55 33.20& +69 55 52.69    \\
    8      &    09 46 02.34& +68 24 56.73 &       60    &     09 57 01.54& +70 08 52.11    \\
    9      &    09 44 34.49& +68 37 56.15 &       61    &     09 59 33.27& +67 58 57.88    \\
    10     &    09 45 51.21& +68 50 55.57 &       62    &     09 58 14.76& +68 11 57.30    \\
    11     &    09 44 21.50& +69 03 54.99 &       63    &     09 59 37.85& +68 24 56.73    \\
    12     &    09 45 39.59& +69 16 54.42 &       64    &     09 58 17.87& +68 37 56.15    \\
    13     &    09 44 07.94& +69 29 53.84 &       65    &     09 59 42.62& +68 50 55.57    \\
    14     &    09 45 27.47& +69 42 53.26 &       66    &     09 58 21.12& +69 03 54.99    \\
    15     &    09 43 53.79& +69 55 52.69 &       67    &     09 59 47.60& +69 16 54.42    \\
    16     &    09 45 14.81& +70 08 52.11 &       68    &     09 58 24.51& +69 29 53.84    \\
    17     &    09 48 53.07& +67 58 57.88 &       69    &     09 59 52.79& +69 42 53.26    \\
    18     &    09 47 28.52& +68 11 57.30 &       70    &     09 58 28.05& +69 55 52.69    \\
    19     &    09 48 45.44& +68 24 56.73 &       71    &     09 59 58.22& +70 08 52.11    \\
    20     &    09 47 19.17& +68 37 56.15 &       72    &     10 02 13.32& +67 58 57.88    \\
    21     &    09 48 37.49& +68 50 55.57 &       73    &     10 00 56.31& +68 11 57.30    \\
    22     &    09 47 09.42& +69 03 54.99 &       74    &     10 02 20.95& +68 24 56.73    \\
    23     &    09 48 29.19& +69 16 54.42 &       75    &     10 01 02.55& +68 37 56.15    \\
    24     &    09 46 59.26& +69 29 53.84 &       76    &     10 02 28.90& +68 50 55.57    \\
    25     &    09 48 20.54& +69 42 53.26 &       77    &     10 01 09.04& +69 03 54.99    \\
    26     &    09 46 48.64& +69 55 52.69 &       78    &     10 02 37.20& +69 16 54.42    \\
    27     &    09 48 11.49& +70 08 52.11 &       79    &     10 01 15.82& +69 29 53.84    \\
    28     &    09 51 33.12& +67 58 57.88 &       80    &     10 02 45.85& +69 42 53.26    \\
    29     &    09 50 10.08& +68 11 57.30 &       81    &     10 01 22.90& +69 55 52.69    \\
    30     &    09 51 28.54& +68 24 56.73 &       82    &     10 02 54.90& +70 08 52.11    \\
    31     &    09 50 03.84& +68 37 56.15 &       83    &     10 04 53.37& +67 58 57.88    \\
    32     &    09 51 23.77& +68 50 55.57 &       84    &     10 03 37.87& +68 11 57.30    \\
    33     &    09 49 57.35& +69 03 54.99 &       85    &     10 05 04.05& +68 24 56.73    \\
    34     &    09 51 18.79& +69 16 54.42 &       86    &     10 03 47.22& +68 37 56.15    \\
    35     &    09 49 50.57& +69 29 53.84 &       87    &     10 05 15.19& +68 50 55.57    \\
    36     &    09 51 13.60& +69 42 53.26 &       88    &     10 03 56.97& +69 03 54.99    \\
    37     &    09 49 43.49& +69 55 52.69 &       89    &     10 05 26.80& +69 16 54.42    \\
    38     &    09 51 08.17& +70 08 52.11 &       90    &     10 04 07.14& +69 29 53.84    \\
    39     &    09 54 13.17& +67 58 57.88 &       91    &     10 05 38.92& +69 42 53.26    \\
    40     &    09 52 51.64& +68 11 57.30 &       92    &     10 04 17.75& +69 55 52.69    \\
    41     &    09 54 11.64& +68 24 56.73 &       93    &     10 05 51.58& +70 08 52.11    \\
    42     &    09 52 48.52& +68 37 56.15 &       94    &     10 07 33.41& +67 58 57.88    \\
    43     &    09 54 10.05& +68 50 55.57 &       95    &     10 06 19.43& +68 11 57.30    \\
    44     &    09 52 45.27& +69 03 54.99 &       96    &     10 07 47.15& +68 24 56.73    \\
    45     &    09 54 08.39& +69 16 54.42 &       97    &     10 06 31.90& +68 37 56.15    \\
    46     &    09 52 41.88& +69 29 53.84 &       98    &     10 08 01.47& +68 50 55.57    \\
    47     &    09 54 06.66& +69 42 53.26 &       99    &     10 06 44.89& +69 03 54.99    \\
    48     &    09 52 38.34& +69 55 52.69 &       100   &     10 08 16.40& +69 16 54.42    \\
    49     &    09 54 04.85& +70 08 52.11 &       101   &     10 06 58.45& +69 29 53.84    \\
    50     &    09 56 53.22& +67 58 57.88 &       102   &     10 08 31.98& +69 42 53.26    \\
    51     &    09 55 33.20& +68 11 57.30 &       103   &     10 07 12.61& +69 55 52.69    \\
         &                &                 &       104   &     10 08 48.26& +70 08 52.11   \\
\enddata
\end{deluxetable*}

\subsection{Calibration and flagging}

We extracted the \HI data from each observing session's measurement
set and averaged the $uv$ samples to an integration time of 10
seconds. We ran the standard scripted VLA calibration pipeline
(version 1.3.8) using CASA \citep{mcmullin07} version 4.6.0. The
pipeline was modified slightly to interpolate over $\sim 30$ \kms of
Galactic absorption in the 3C147 observations. The standard flagging
set-up in the pipeline removed some of the bright \HI target emission,
so at this stage only calibrator pointings were automatically flagged.

The data were Hanning-smoothed prior to calibration. After calibration,
every second channel was discarded, resulting in 4096 independent
channels with a velocity spacing and resolution of $0.8$ \kms.  


Due to issues with the CASA mosaicking routines, we exported all
calibrated data to the Miriad \citep{sault95} reduction package and
used this for the rest of our data reduction. The conversion to Miriad
format corrected all velocities from topocentric to barycentric, with
the full velocity range of the \HI data sets going from $-1735$ \kms
to $+1645$ \kms (radio definition of velocities). Self-calibration was
used, which improved the ratio of peak flux to noise by a factor
three.

We used the {\tt uvlin} task to subtract the continuum using a
second-order polynomial. We experimented with lower-order polynomials
but found that these gave less satisfactory results for the bright and
resolved emission of M82.  To fit the continuum emission, we used two
ranges spanning $\sim 1000$ channels each, covering velocities from
$-1325$ \kms to $-415$ \kms and from $+490$ to $+1315$ \kms,
respectively.  This is well away from the velocity range where \HI is
expected: the deep GBT observations by \citet{chynoweth08} detect \HI
in the velocity range between $-250$ and $+340$ \kms. Later inspection
of our data validated our choice.

As the target data were not flagged during the calibration stage, some
flagging was done at this stage using the continuum-subtracted
data. We used the task {\tt pgflag} with its default settings to do a
conservative {\tt SumThreshold} flagging \citep{offringa10} at
7$\sigma$, followed by flagging of visibilities with fewer than three
unflagged neighbour visibilities. We then flagged time intervals or
channels with less than 20 per cent good data.  We checked that no \HI
line visibilities were flagged. Inspection of the data showed that
this procedure removed most of the artefacts from the $uv$-data, but
some additional flagging was necessary to remove a number of more
localised artefacts. Specifically, we flagged all visibilities with an
amplitude $>8$ Jy; all visibilities from baseline ea03--ea24; and the LL
polarisation of baseline ea18--ea24 below a $uv$-distance of 2 k$\lambda$.

A remaining subtle  large-scale spatial ripple over the entire mosaic could
not be readily identified in the $uv$-data. We therefore created an
average $uv$-data set by averaging spectral channels 1000 to 1500 and
produced a single-channel image.  Inspection of the Fourier transform
of this image showed the ripple to originate in high amplitude
visibilities between 200$\lambda$ and 400$\lambda$. The spatial scales
with which these correspond make it conceivable that the ripples are
due to some residual solar interference.  We therefore flagged in the
averaged $uv$ data set all visibilities in this range with amplitudes
$>0.4$ Jy. This flagged, averaged data set was used as a mask to flag
the corresponding visibilities in the full data set.  The resulting
$uv$ data set was used to produce image cubes.

\subsection{Deconvolution and clean masks\label{sec:deconv}}

The {\tt invert} task was used to produce the dirty beam and data
cubes of the mosaic, combining all pointings of the C- and D-array. We
adopt a pixel size of $5''$ and a channel width of 2 \kms over the
velocity range $-400$ km/s to $+450$ km/s. This results in a data cube
of $2406 \times 2370 \times 425$ pixels. We produced cubes using
natural weighting ({\tt robust=2.0} in the Miriad definition) and a
higher-resolution version using {\tt robust=0.5}. For convenience, we
refer to these as the ``natural-weighted'' and the ``robust-weighted''
cubes, respectively.

We used the {\tt mossdi} task to deconvolve the mosaicked cube. The
aim was to clean the data cube deeply, to avoid residual-scaling
effects (see \citealt{walter08, jorsater95}).  Due to the large area
involved, the desire to not clean large amounts of noise, and to
minimise the possibility of clean-bias, we created masks to indicate
areas where deconvolution is allowed (``clean masks'').  Due to the
complex, extended and fragmented nature of the \HI emission in the
cube, we used a source-finding algorithm, specifically the smooth and
clip algorithm implemented in the SoFiA software \citep{serra15}.

This applies a number of user-defined
convolution combinations (Gaussian along the spatial axes and boxcar
along the velocity axis) and, for each of these combinations defines a
binary mask by selecting all pixels above a user-defined threshold,
expressed in multiples of the noise in the convolved data cube.
The final mask is then the union of the masks belonging to each filter
combination. Using a slight variation on the procedure described in
\citet{serra12}, we then apply a size filter to the SoFiA output mask
by smoothing it spatially with the largest Gaussian beam used in the
previous step.  In the output mask, values $>0.5$ are then selected
for the final mask.

This procedure creates masks objectively, but it is important that the
final mask correctly isolates real signal. We have therefore
extensively tested the algorithm described above on THINGS (The \HI
Nearby Galaxy Survey; \citealt{walter08}) data, where cubes, masks and
moment maps are readily available.  THINGS is a multi-configuration
VLA survey with a similar spatial and velocity resolution as our data
set and should thus be representative. We used the source finding
algorithm and the THINGS data to create masked cubes and moment maps
where the parameters were tweaked to most closely resemble the
published THINGS results.  We find that the optimum final mask is
produced by a combination of masks using a $3\sigma$ level at
resolutions $(1,1,1)$, $(1,1,2)$, $(2,2,1)$ and $(2,2,2)$, where each
triplet of numbers indicates the two spatial resolutions and the
velocity resolution, in multiples of the original resolution.

To create the final clean masks, we cleaned the natural-weighted data
cube down to 2$\sigma$ (without any pre-defined masks) and used the resulting cube
as input for SoFiA. The task {\tt mossen} was used to produce cubes
showing the expected noise and gain (mosaic ``primary beam'' correction)
values. We used the inverse of the noise cube as a weights cube in
SoFiA to de-emphasize the higher noise values towards the edge of the
mosaic and to prevent an excessive number of these noise peaks from
entering the mask. The data set contains a number of channels with
Galactic foreground \HI. This foreground emission was included as part
of the clean mask.

We then used {\tt mossdi} again for a final deconvolution, using the
clean mask, and this time cleaning down to $1\sigma$ within the
mask. The task {\tt restor} was used to create the final, restored
data cubes.  For the robust-weighted data we used the same procedure
and the same clean mask as for the natural-weighted data.

As noted, we use a deep cleaning limit to make residual-scaling
corrections (see \citealt{walter08, jorsater95}) negligible.
\citet{ianja17} have shown that (for THINGS data) cleaning to below
$1.5\sigma$ results in virtually identical fluxes from both
residual-scaled and ``standard'', non-residual-scaled data.  We
checked this for our data by cleaning two representative channel maps
to different depths of $1\sigma$, $0.75\sigma$ and $0.5\sigma$,
creating a standard version as well as a residual-scaled version at
each clean depth (using the residual scaling parameters given in
\citealt{walter08} and \citealt{ianja17}).

We chose channel maps at $v=-92$ \kms and at $v=+172$ \kms which
contain a significant amount of extended low-level emission. The first
map is characteristic of the structures seen near M81, the second of
those near M82. We determine the fluxes within the respective clean
masks. For the M81 (M82) channel map we find that the residual-scaled
fluxes derived using the three different depths agree to within 0.5
(1.8) percent. The three ``standard'' fluxes show a variation of 0.8
(0.4) percent. More importantly, we find that the $1\sigma$
residual-scaled and standard fluxes agree to within 1.7 (2.8) percent,
with the ratio decreasing to 0.4 (0.6) percent for the $0.5\sigma$
clean depth maps.  In addition to the three depths just discussed, we
checked the difference for a more shallow $2\sigma$ limit, and find a
difference of 6.2 (10.7) percent between standard and residual-scaled
fluxes, consistent with the increasing relevance of residual-scaling
for shallow clean limits. We therefore conclude that the standard
fluxes derived here using a $1\sigma$ clean are virtually identical to
the residual-scaled fluxes, in agreement with \citet{ianja17}, and can
be used in our further analysis.

\subsection{Beam size and sensitivity\label{sec:noise}}

The natural-weighted C+D data gives a synthesised beam size of $38.1''
\times 30.9''$, and a beam position angle (PA) of
$75.5^{\circ}$. Assuming a distance of 3.63 Mpc
\citep{karachentsev04}, this corresponds to a linear resolution of
$0.67 \times 0.54$ kpc. The beam size for the robust-weighted data is
$24.3'' \times 20.0''$ with a beam PA of $80.7^{\circ}$. The
corresponding linear resolution is $0.43 \times 0.35$ kpc.  The noise
in a 2 \kms channel is 1.17 mJy beam$^{-1}$ for the natural-weighted
cube and 1.25 mJy beam$^{-1}$ for the robust data. These values are
close to the theoretical noise.  The corresponding column density
sensitivities are $2.2 \cdot 10^{18}$ cm$^{-2}$ (natural) and $5.7
\cdot 10^{18}$ cm$^{-2}$ (robust). These are 1$\sigma$ values over a
single 2 \kms channel. More representative sensitivies are given by
$3\sigma$ and 16 \kms (8 unaveraged channels) limits. These are $5.3
\cdot 10^{19}$ cm$^{-2}$ (natural) and $1.4 \cdot 10^{20}$ cm$^{-2}$
(robust).  For unresolved sources, these noise levels imply an
5$\sigma$ \HI mass limit of $3.4 \cdot 10^3\, (W/[10\ {\rm
    km\,s}^{-1}])$ \msun for the natural weighting, and a limit of
$8.9 \cdot 10^3\, (W/[10\ {\rm km\,s}^{-1}])$ \msun for the robust
weighting. Here $W$ is the width of the \HI profile in \kms.

Selected channel maps of the natural-weighted cube are shown in
Fig.\ \ref{fig:chanmap} to give an overview of the M81 triplet data
set.  The channel maps clearly show the regular rotation of the inner
parts of M81, the streaming motions in the outer arms and the
connection with NGC 3077 in the southern part. This is in great
contrast with the more chaotic and extended distribution of the \HI in
the northern part, including the connection with M82.  The diffuse gas
around M82 is visible over a large range in velocity. We can also
clearly see the presence of Galactic foreground emission in a number
of channels around velocities of $\sim -60$ \kms and $\sim 0$ \kms.

\begin{figure*}
  \centering
\includegraphics[width=0.9\hsize]{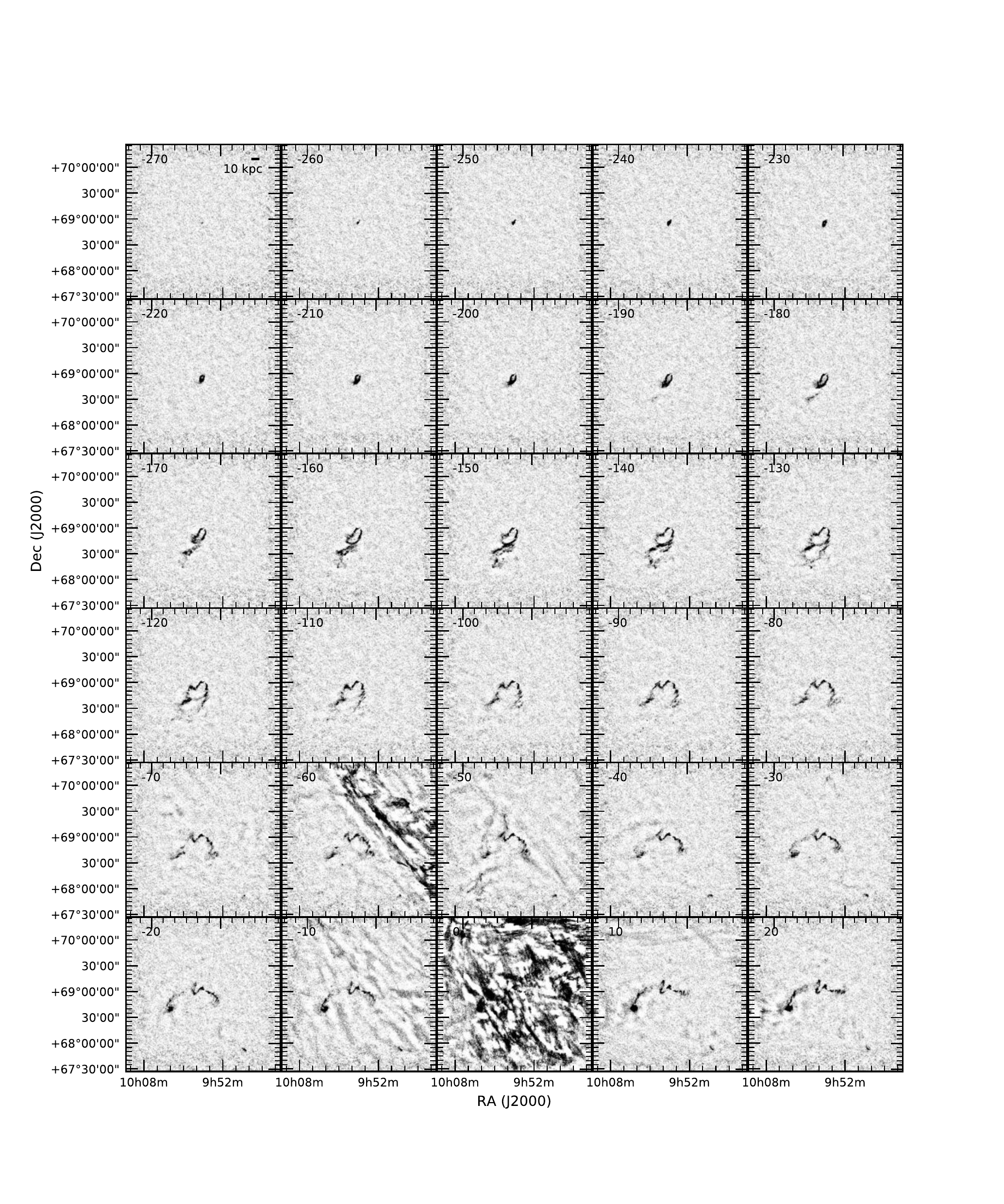}
\caption{Selected channel maps from the natural-weighted data. Every
  fifth channel is shown. The grayscale runs from $-0.5$ mJy beam$^{-1}$
  (white) to $+7$ mJy beam$^{-1}$ (black). The velocity of the
  channel in \kms is shown in the top-left corner of each sub-panel.
  Only the full-sensitivity area of the mosaic is shown. The scale-bar in the
top-left panel indicates 10 kpc.}
\end{figure*}

\addtocounter{figure}{-1}
\begin{figure*}
\centering
\includegraphics[width=0.9\hsize]{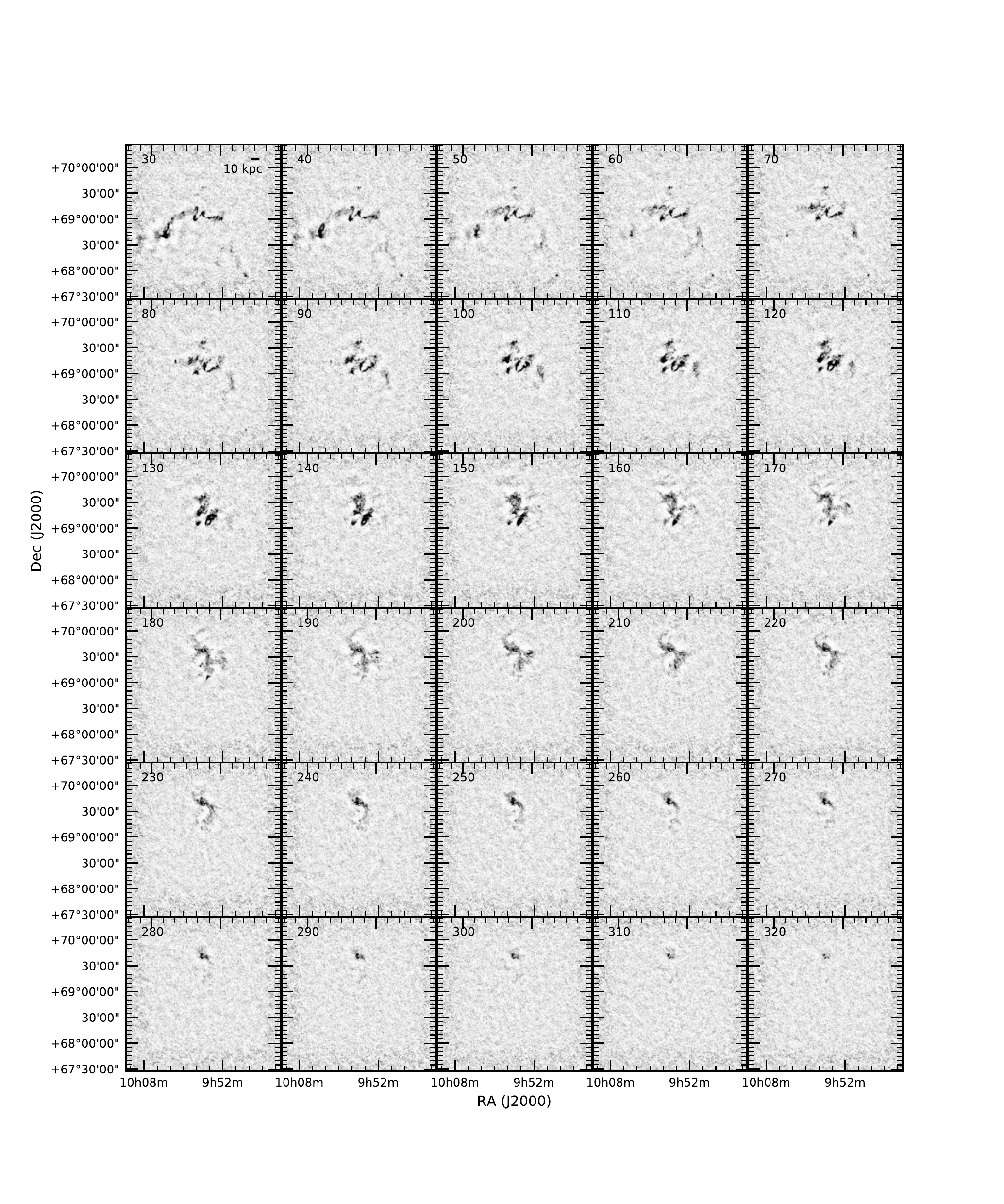}
\caption{\emph{Continued.} Selected channel maps from the
  natural-weighted data. Every fifth channel is shown. The grayscale
  runs from $-0.5$ mJy beam$^{-1}$ (white) to $+7$ mJy beam$^{-1}$
  (black). The velocity of the channel in \kms is shown in the
  top-left corner of each sub-panel.  Only the full-sensitivity area
  of the mosaic is shown. The emission from M82 extends to about 350
  \kms (not shown here). The scale-bar in the top-left panel indicates 10 kpc.
\label{fig:chanmap}}
\end{figure*}

\subsection{Moment maps and Galactic foreground\label{sec:mom}}

To create moment maps we use a modified version of the clean mask with
the Galactic foreground emission removed.  Figure \ref{fig:chanmap}
shows that this emission is present at two distinct velocities. The
main component is at $\sim 0$ \kms, with a second, fainter component
at $\sim -58$ \kms.  Specifically, from $-64$ to $-60$ \kms, Galactic
emission is present in one corner of the image, without overlapping
the M81 triplet area. We masked the Galactic signal manually for these
channels. From $-58$ to $-48$ \kms, and again from $-10$ to $+8$ \kms,
the Galactic emission is bright and overlaps the M81 triplet
area. These channels were masked completely. The channels in between
these components, from $-46$ to $-12$ \kms, were not affected.  We did
not attempt to interpolate the emission in the blanked channel maps
due to the relatively small velocity range that was affected, and the
complexity of the structures in these maps. In the rest of the cube no
Galactic contamination is present and the remaining channels were not
affected.

We use the updated mask to create the moment maps of both the
natural-weighted and robust-weighted data using the {\tt moment} task
in Miriad.  For both weightings we create zeroth (total intensity),
first (intensity-weighted mean velocity), and second (velocity
dispersion) moment maps. We used the gain cube produced by {\tt
  mossen} to mask spurious noise peaks towards the edges of the
mosaic. We only retained signal that was present in more than 3
channels at each spatial position. In addition, we create total
intensity maps using the original clean masks (i.e., with the Galactic
emission still included). These can be used to gauge the effect of
blanking the Galactic emission channels.

Due to different numbers of channels contributing to each of the
pixels in the moment maps, the noise in an integrated intensity
(zeroth moment) map is not constant. We derive the noise as
follows. For a zeroth moment map based on independent channels (as is
the case here), the noise in a pixel $\sigma_{\rm mom0}$ is defined as
$\sigma_{\rm mom0} = \sigma_{\rm chan} \cdot \sqrt{N_{\rm chan}}$,
where $\sigma_{\rm chan}$ is the noise in a single channel and $N_{\rm
  chan}$ the number of channels contributing to a moment-map pixel. (A
zeroth moment is a sum, not an average, which is why the noise
increases.) The signal-to-noise S/N of each pixel can be derived by
dividing the zeroth-moment map by a map of $\sigma_{\rm chan}$ (taking
care to treat the units consistently, i.e., the $\sigma_{\rm chan}$
map should also have Jy beam$^{-1}$ \kms units, to avoid introducing
extra factors equal to the channel separation). We then selected all
pixels in the S/N map with values between 2.5 and 3.5. We take the
mean value of the corresponding pixels in the zeroth-moment map to
represent the average S/N $= 3$ column density sensitivity.  We find a
value of $15.0 \pm 5.5$ mJy beam \kms, corresponding to a column
density sensitivity of $1.42 \cdot 10^{19}$ cm$^{-2}$ for the
natural-weighted data. A similar procedure for the robust-weighted
data gives a S/N $= 3$ value of $24.2 \pm 6.1$ mJy beam \kms,
corresponding to a column density of $5.54 \cdot 10^{19}$ cm$^{-2}$.

To ensure a homogeneous column density limit across the maps, we apply
a zeroth-moment value cutoff of $1.5 \cdot 10^{19}$ cm$^{-2}$ to the
natural-weighted maps and $5.5 \cdot 10^{19}$ cm$^{-2}$ to the robust
map. The resulting maps are also applied as masks to the respective
first and second-moment maps.

\begin{deluxetable*}{lccrcrr}
  \tabletypesize{\scriptsize} \tablewidth{0pt} \tablecaption{Properties of the four main galaxies\label{tab:props}}
  \tablehead{\colhead{Galaxy} & \colhead{$\alpha(2000.0)$} & \colhead{$\delta(2000.0)$} & \colhead{$D_{\rm Holmberg}$} & \colhead{$i_{\rm opt}$}& \colhead{PA$_{\rm opt}$} & $V_{\rm sys}^{\rm hel}$ \\
\colhead{} & \colhead{$(^h\ ^m\ ^s)$} & $({}^{\circ}\ '\ '')$& \colhead{($'$)}  & \colhead{(${}^{\circ}$)} & \colhead{(${}^{\circ}$)} & \colhead{(\kms)}\\
 \colhead{(1)} & \colhead{(2)} & \colhead{(3)} & \colhead{(4)} & \colhead{(5)} & \colhead{(6)} & \colhead{(7)}}
\startdata
M81      & 09 55 33.2  & +69 03 55  & $35.0 \times 14.4$ &  57 & 157 &  $-34$ \\ 
M82      & 09 55 52.7  & +69 40 46  & $13.4 \times 8.5$ &  82 & 65  & 203  \\
NGC 3077 & 10 03 19.1  & +68 44 02  & $8.8 \times 8.0$  &  38 & 45  & 14   \\
NGC 2976 & 09 47 15.4  & +67 54 59  & $9.7 \times 5.7$  &  61 & 143 &  3   \\
\enddata
\tablecomments{(1) Name of galaxy. (2) Right Ascension (J2000.0). (3) Declination (2000.0). (4) Major and minor axis Holmberg diameter from \citet{appleton81}. (5) Optical inclination from \citet{appleton81}. (6) Optical position angle from UGC \citep{ugc}. (7) Heliocentric systemic velocity from RC3 \citep{rc3}.}
\end{deluxetable*}

\begin{figure*}
  \centering
  \includegraphics[width=0.48\hsize]{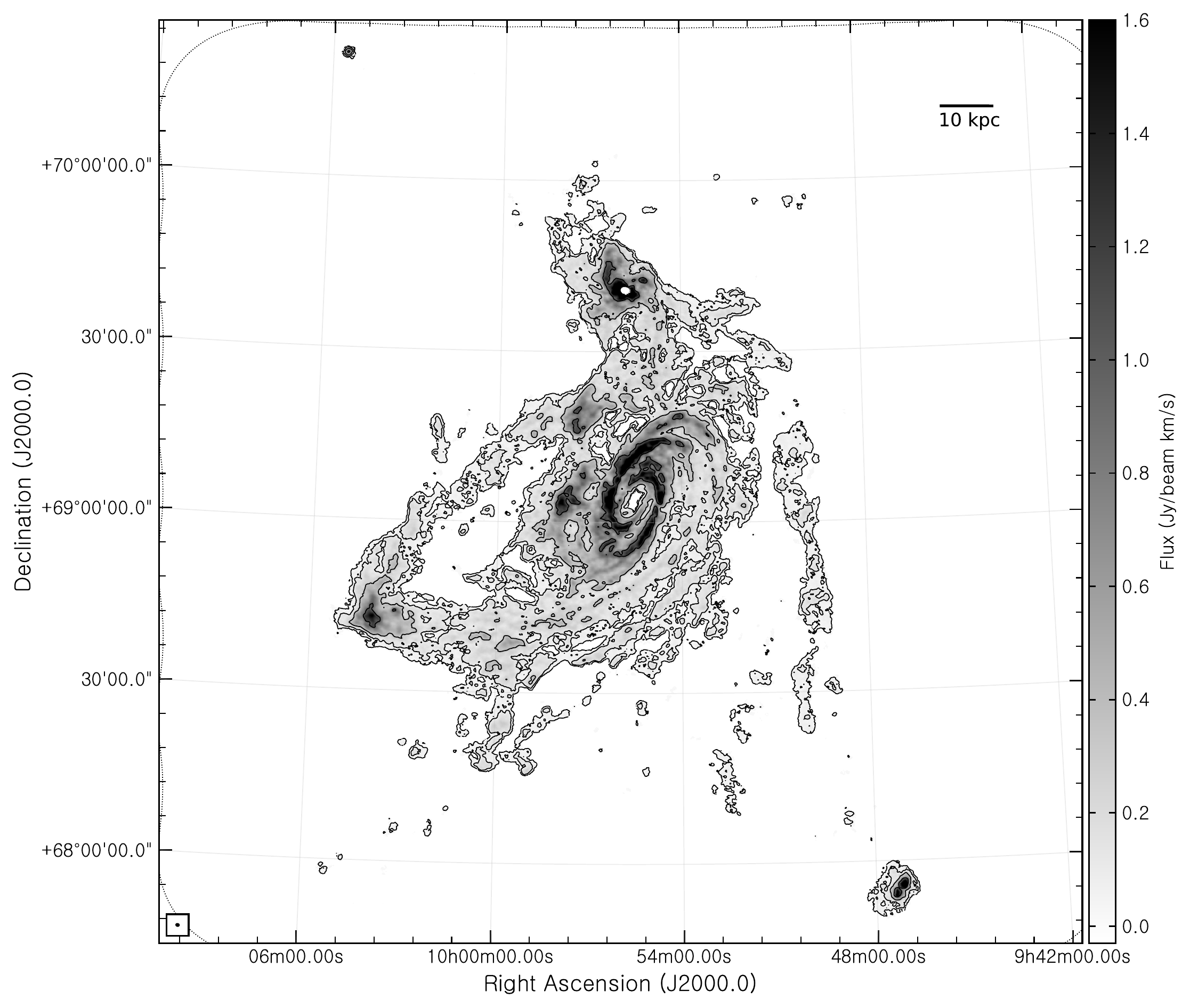}
  \includegraphics[width=0.48\hsize]{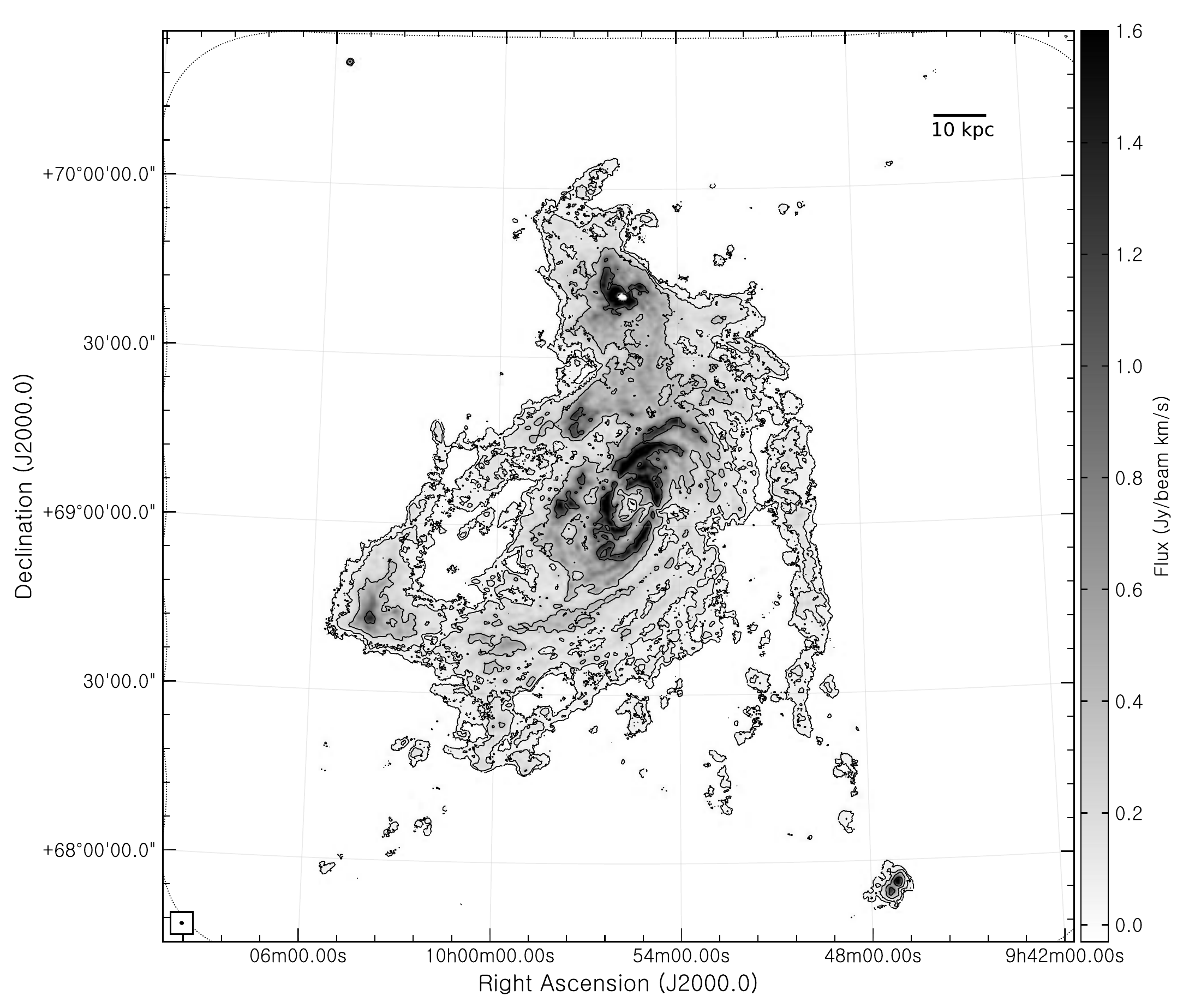}
\caption{Left: Natural-weighted integrated intensity (zeroth moment)
  map derived using the C+D data. The grayscale runs from 0 (white) to
  1.6 (black) Jy beam$^{-1}$ \kms. Contours levels are $0.0316 \cdot
  10^x$ Jy beam$^{-1}$ \kms where $x=(0, 0.5, 1, 1.5)$. This
  corresponds to column densities of $(0.3, 0.95, 3, 9.5) \cdot
  10^{20}$ \cm.  Only the area inside the 50 percent sensitivity
  contour (dotted curve visible in the corners) is shown. Right:
  Zero-spacing corrected zeroth moment map based on the
  natural-weighted C+D VLA and GBT data. Contours and grayscales as in
  left panel. \label{fig:mom0}}
\end{figure*}

\begin{figure*}
\centering
\includegraphics[width=0.9\hsize]{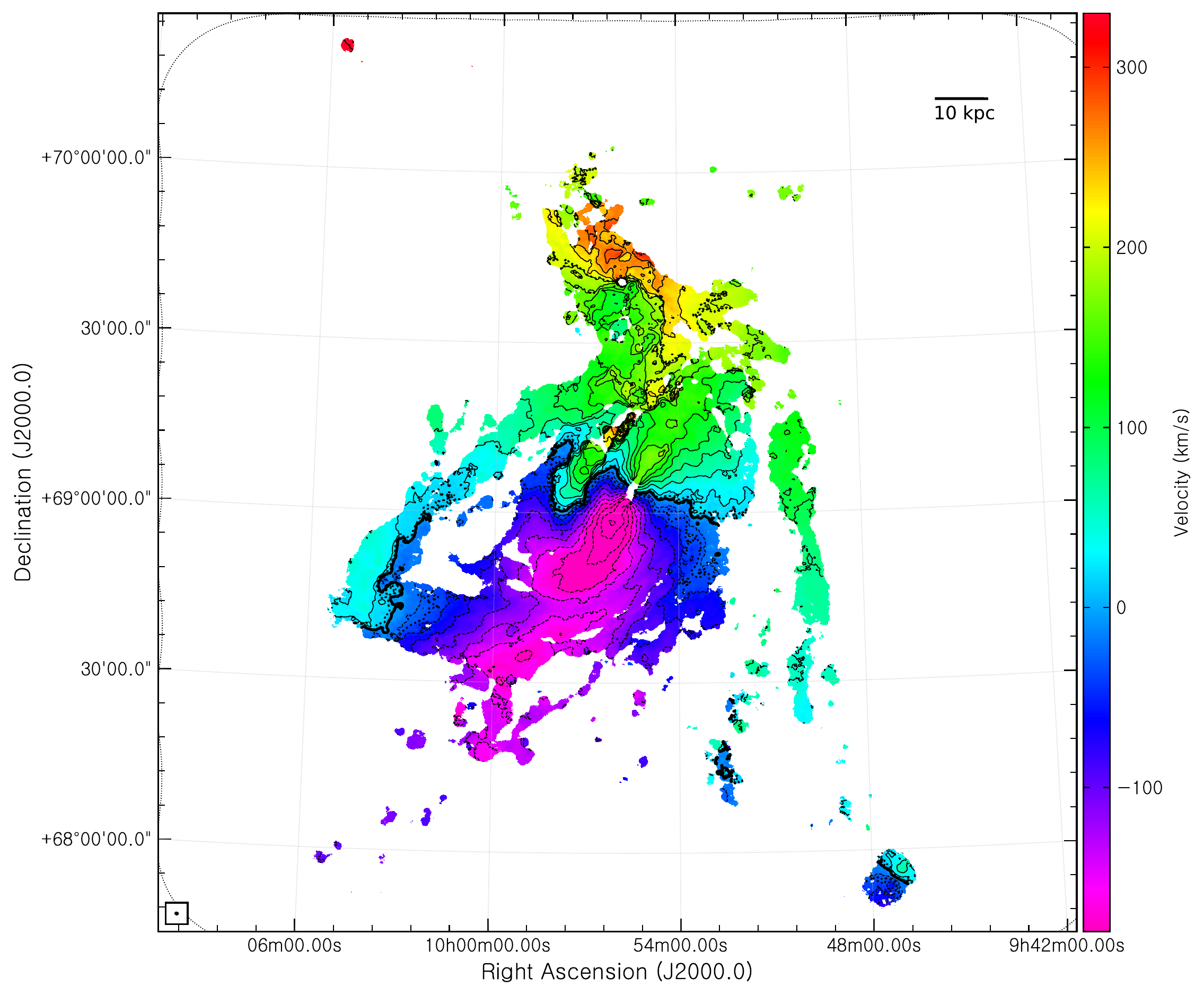}
\caption{Natural-weighted intensity-weighted mean velocity (first moment) map derived using the C+D
  data. The color-scale runs from $-180$ to $330$ \kms, as indicated
  by the color bar.  Contour levels run from $-250$ to $+400$ \kms and
  are spaced by 25 \kms. Negative contours are dashed. The thick
  contour is at 0 \kms. Only the area inside the 50 percent
  sensitivity contour (dotted curve visible in the corners) is shown. \label{fig:mom1}}
\end{figure*}

\begin{figure*}
\centering
\includegraphics[width=0.9\hsize]{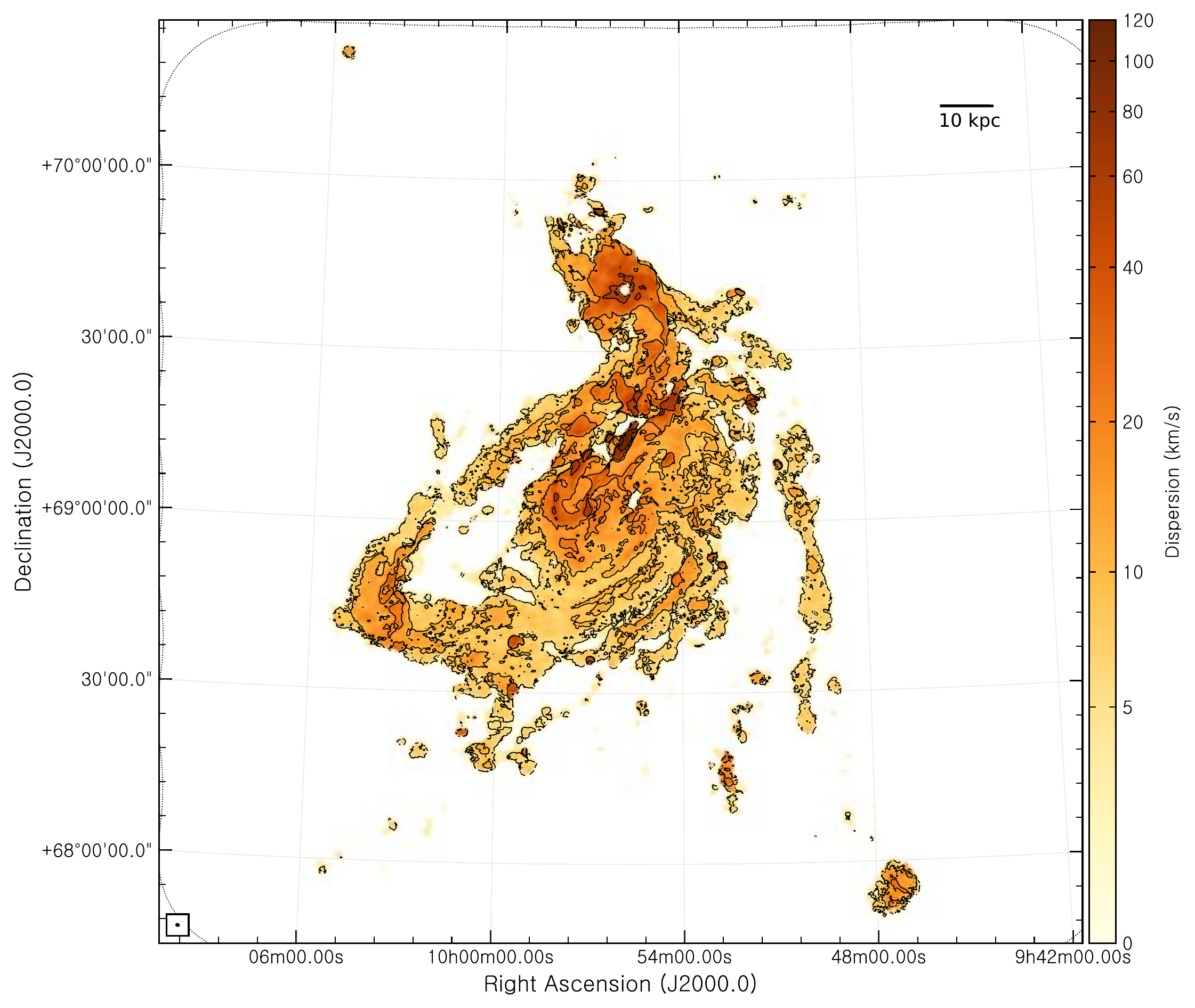}
\caption{Natural-weighted velocity dispersion (second moment) map using the C+D
  configurations. The color-scale uses an arcsinh stretch, running
  from 0 (light) to 120 (dark) \kms. Contour levels are at 5, 10, 20,
  50 and 100 \kms. Only the area inside the 50 percent sensitivity
  contour (dotted curve visible in the corners) is shown.\label{fig:mom2}}
\end{figure*}

\begin{figure*}
\centering
\includegraphics[width=0.9\hsize]{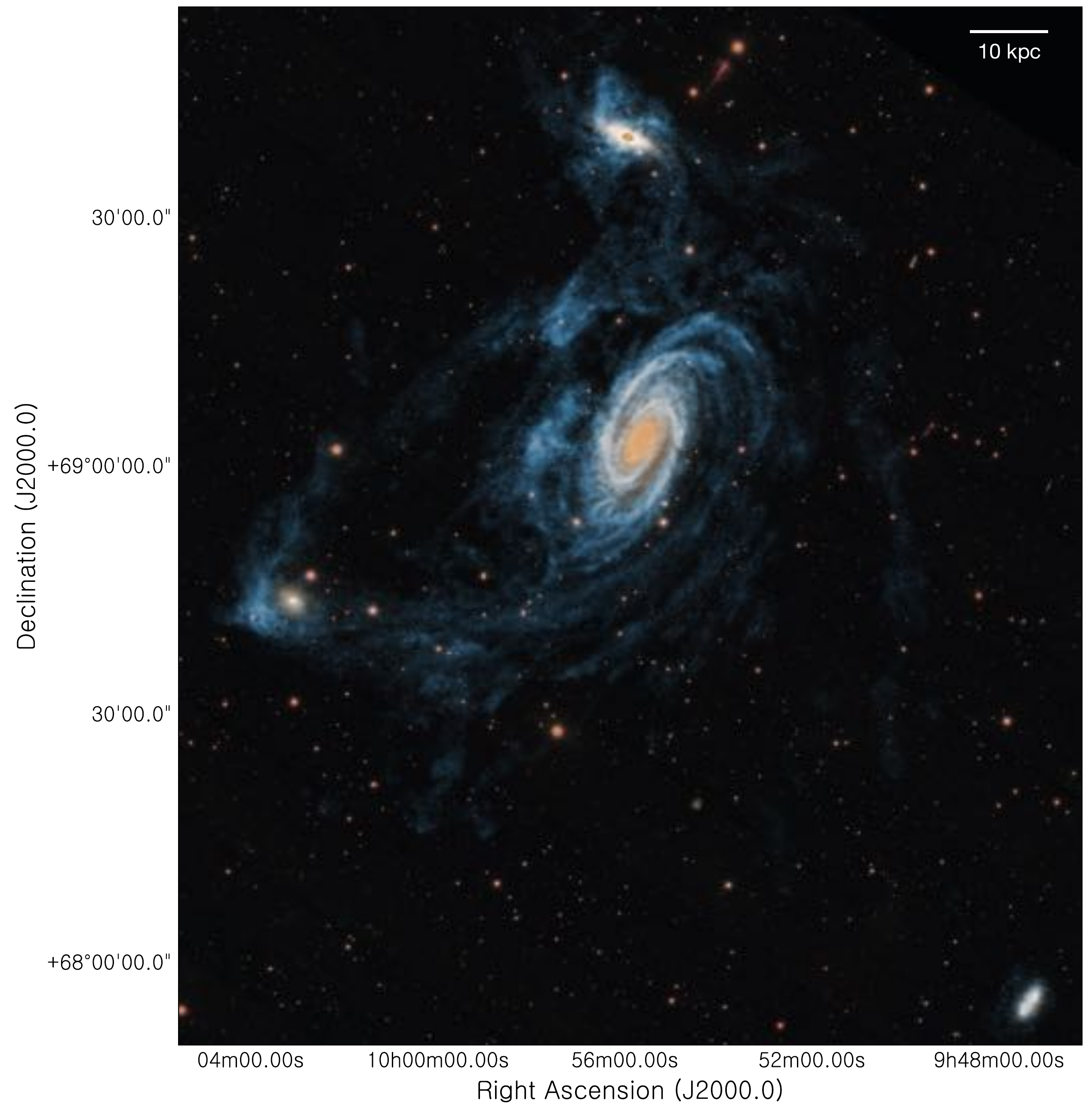}
\caption{False-color overlay of the robust-weighted zeroth-moment map
  (in blue) on an color SDSS image of the M81 triplet. The area shown
  in slightly smaller than in Fig.~\ref{fig:mom0}.\label{fig:sdss}}
\end{figure*}

The natural-weighted moment maps are presented in
Figs.\ \ref{fig:mom0} (left) (zeroth-moment map), \ref{fig:mom1}
(first-moment map) and \ref{fig:mom2} (second-moment map). These maps
are further discussed in Sect.\ \ref{sec:moms}.
In Fig.~\ref{fig:sdss} we show a false-color
representation of the robust-weighted zeroth-moment map overlaid on a
SDSS optical image. A summary of the optical positions and sizes is given in
Table \ref{tab:props}.

\begin{figure*}
  \centering
  \includegraphics[width=0.49\hsize]{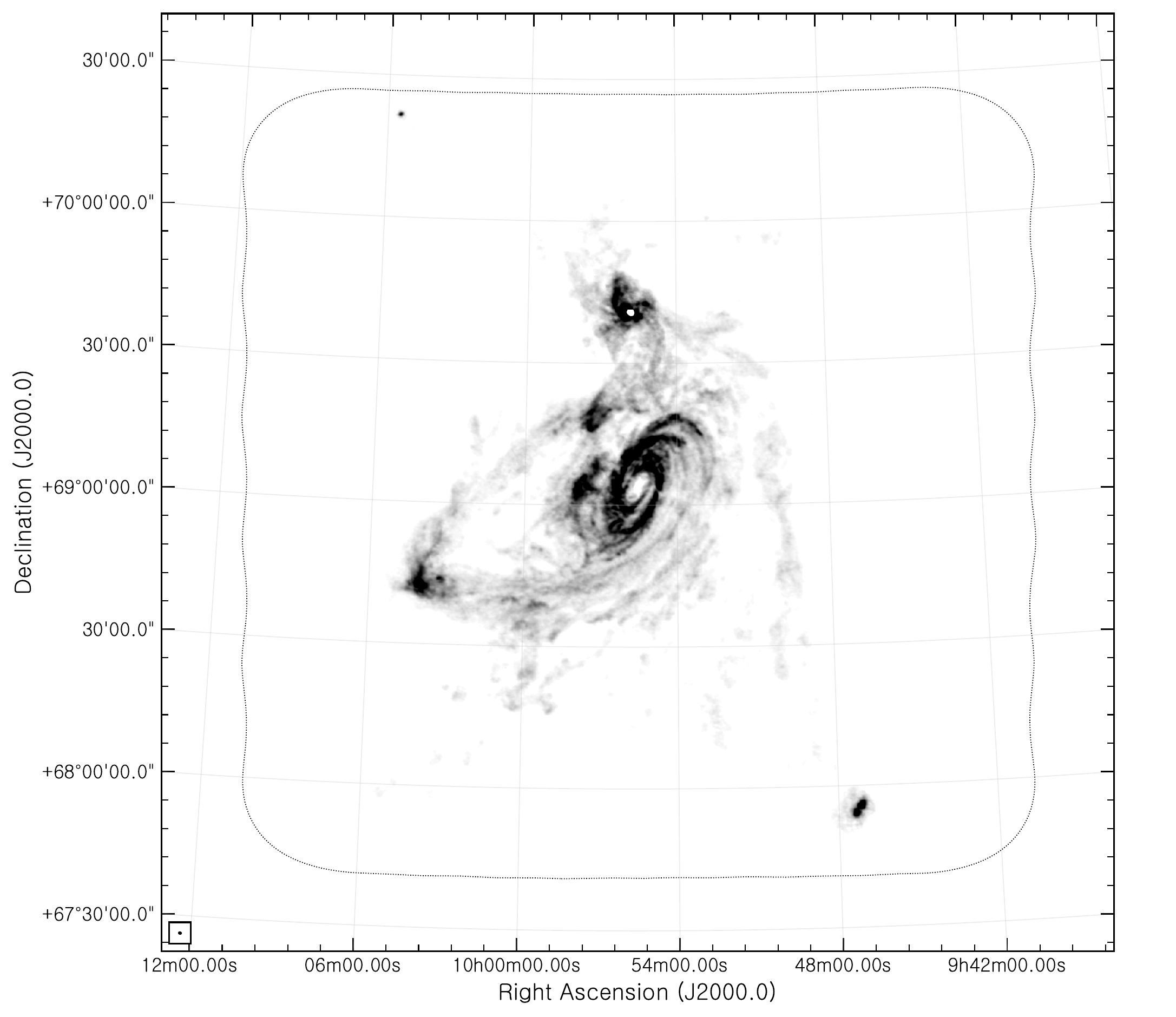}
  \includegraphics[width=0.49\hsize]{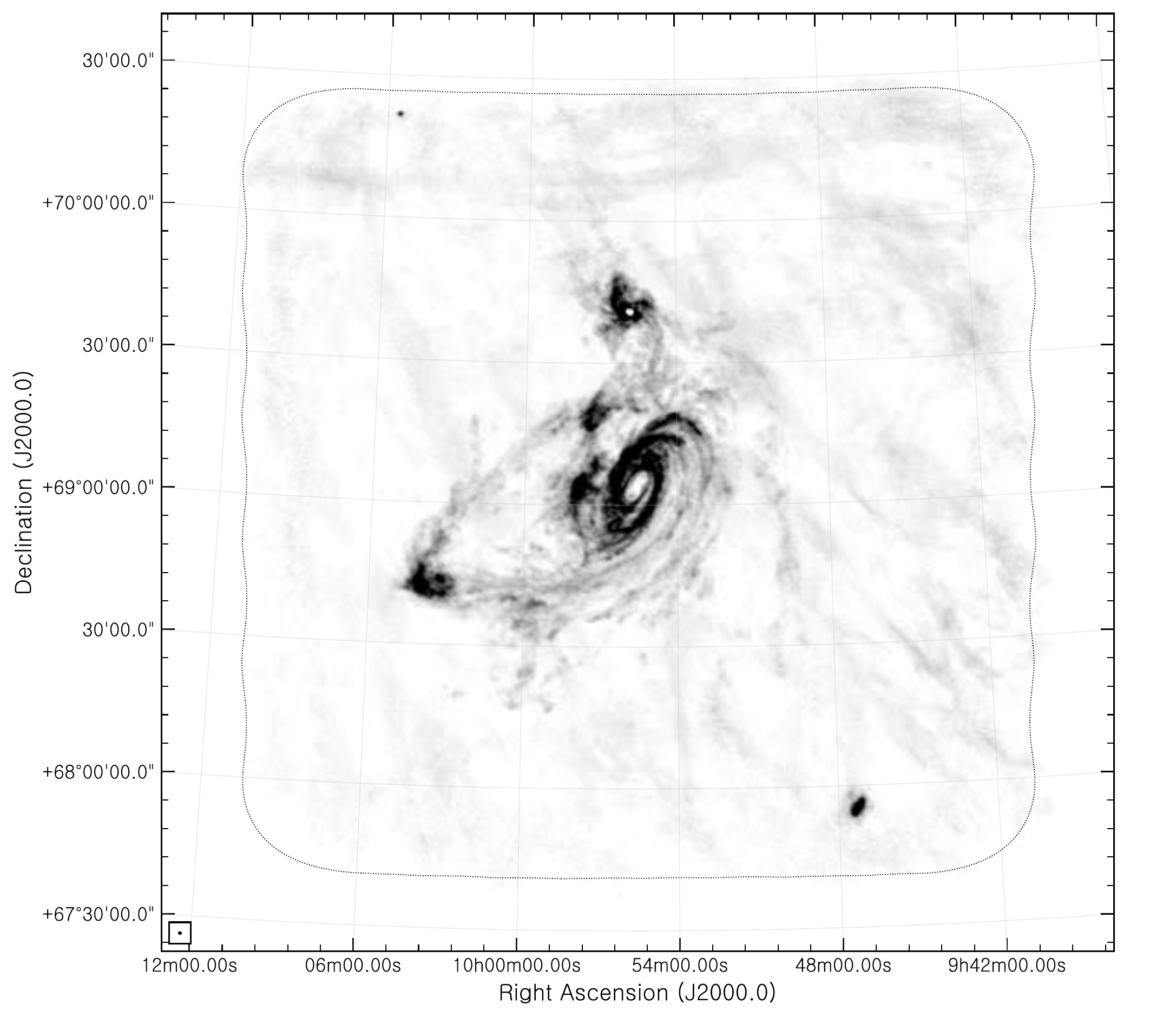}
\caption{Comparison of C+D natural-weighted zeroth-moment maps without
  (left panel) and with (right panel) the Galactic emission
  channels. In both panels the grayscale runs from 0 (white) to 1.0
  (black) Jy beam$^{-1}$ \kms. The beam is indicated in the lower-left
  corner. The dotted curve indicates the 50 percent sensitivity level
  of the mosaic area. \label{fig:momcomp}}
\end{figure*}

In Fig.\ \ref{fig:momcomp}, we compare the two zeroth-moment maps
created using the masks with and without the Galactic foreground
emission.  Comparison of the two moment maps shows the effect of the
Galactic masking. For example, the appearance of NGC 2976 seemingly
having two separate \HI components is due to the masking used, and
similarly, some emission is missing along the minor axis
of M81.  The final effect of this masking on properties like total \HI
masses is, however, small compared to other uncertainties, as
discussed in more detail in Sect.\ \ref{sec:masses}.


\subsection{D-array cubes and moment maps}

We also produced more sensitive, lower resolution versions of
the cubes and maps using the D-array data and the shorter baselines
from the C-array data by selecting all data with a $uv$-distance $< 5$
k$\lambda$. For convenience, we refer to this data set as the
``D-array'' data.

We used the same procedure as for the C+D data described above.  The
resulting natural-weighted deconvolved cube has a beam size of $94.5''
\times 76.0''$ with a beam PA of 78.2$^{\circ}$. At the distance of
M81, this corresponds to a resolution of $1.67 \times 1.34$ kpc.  For
the D-array data we did not consider a robust weighting.

The noise in the natural-weighted data set is 1.81 mJy beam$^{-1}$ for
a single 2 \kms channel. This corresponds to a $1\sigma$, 1 channel
column density sensitivity of $5.6 \cdot 10^{17}$ cm$^{-2}$, or a more
representative $3\sigma$, 16 \kms (8 independent, unaveraged channels)
limit of $1.3 \cdot 10^{19}$ cm$^{-2}$.

For the moment maps calculation, we created new masks due to the
increased prominence of the Galactic emission. From $-66$ to $-58$
\kms, Galactic emission covered part the field, without affecting the
main M81 triplet emission, and this Galactic emission was masked by
hand. The channels from $-56$ to $-50$ and from $-8$ to $+6$ \kms were
blanked completely. In between these ranges, from $-48$ to $-10$ \kms,
no blanking was needed. A small amount of manual blanking was needed
for the channels at $+8$ and $+10$ \kms, where prominent Galactic
emission covered a small part of the field. No further blanking was
needed for the remaining channels, except for a small area immediately
to the east of NGC 3077, where from $+12$ to $+34$ \kms some aliasing
and mosaicking artefacts were removed.

In the zeroth-moment map, the average value of S/N $=3$ pixels is
$18.3 \pm 8.0$ mJy beam \kms, corresponding to an average column
density of $2.83 \cdot 10^{18}$ cm$^{-2}$. To achieve a homogeneous
sensitivity, we blanked the zeroth-moment map, and the corresponding
pixels in the first- and second-moment maps, at a column density value
of $3.0 \cdot 10^{18}$ cm$^{-2}$.

\subsection{Zero-spacing corrections using GBT data\label{sec:zero}}

Interferometers are limited in their ability to recover the total
  fluxes of objects, especially if these are extended compared to the
  size scale corresponding to the shortest baseline. Single-dish data
  are often used to correct the fluxes in the interferometric data and
  enhance extended structures. Here we use GBT data to apply this
  zero-spacing correction to our data.

As noted in the Introduction, GBT observations of the survey area are
published in \citet{chynoweth08}.  We could, however, not use the
\citet{chynoweth08} data cube as an unflagged version (still including
Galactic emission) was not available. For the zero-spacing correction
we therefore use the GBT data set covering the M81/M82 and NGC 2403
groups as published in \citet{chynoweth11} (though this data set
incorporates the \citealt{chynoweth08} data).

The channel spacing of the data set is 5.2 \kms, with a noise level
between $\sim 8$ and $\sim 14$ mJy beam$^{-1}$. The variations in noise level are
due to the patching together of many different observations
(cf.\ Fig.\ 3 in \citealt{chynoweth11}). In the area covering
the triplet, the data set is for all practical purposes equal to the
\citet{chynoweth08} data, resulting in a noise level of
$\sim 8$ mJy beam$^{-1}$.  For the GBT beam size of $9.4'$, this corresponds to a
$1\sigma$, 1 channel (5.2 \kms) column density sensitivity of $2.5
\cdot 10^{17}$ cm$^{-2}$.

We extracted the region corresponding to our VLA mosaic from this data
set and regridded it to the spatial and spectral pixel size of the VLA
mosaic. Note that this meant oversampling the GBT velocity channels by
a factor $\sim 2.5$ to achieve a 2 \kms channel spacing. We combined
the natural-weighted C+D VLA data and the GBT data using the Miriad
task {\tt immerge}. This task combines the two image cubes in the
Fourier plane, and optionally uses the range in spatial frequencies
where the single-dish and interferometer data overlap to determine a
scale factor to bring the single dish flux scale in agreement with the
interferometer one. For our data we used a $uv$-range between 35 and
90 meters for the overlap.

Comparing fluxes in the velocity range between $-252$ and $-102$ \kms
we found an optimal scale factor of 1.08 for the GBT data. Tests using
different velocity ranges (excluding that of the Galactic emission)
yielded similar values.  The final, combined cube as produced by {\tt
  immerge} has a noise level and resolution equal to that of the VLA
C+D data cube.

The increased prominence of Galactic foreground emission and presence
of additional features introduced in the combined cube, meant we
created a new mask to produce moment maps. As before, we used SoFiA,
using the same settings, and applied the same size filter.

The velocity range from $-450$ to $-78$ \kms, and from $+22$ to $+450$
\kms needed no additional blanking. Galactic emission dominated the
velocity range from $-62$ to $-44$ \kms, and from $-10$ to $+20$
\kms. These channels were completely blanked. Finally, from $-76$ to
$-64$ and from $-42$ to $-12$ \kms Galactic emission was present but
did not overlap with the triplet emission. Here the Galactic emission
was identified and blanked by hand. In addition, a small aliasing
effect towards the edge of the mosaic east of NGC 3077 was also
removed by hand. Comparison with Sect.\ \ref{sec:mom} shows that in
the combined cube a substantially larger range in velocity is affected
by Galactic emission.

This mask was then used to create moment maps, applying the same S/N
$= 3$ column density cut, and retaining only signal occuring over more
than 3 consecutive channels.

The zeroth moment map is shown in Fig.\ \ref{fig:mom0} (right panel). A
comparison with the VLA-only map shown in the left panel of the same
Figure clearly shows that the artefact running along the minor axis
due to the blanking of Galactic emission is more prominent in the
zero-spacing corrected cube. Note that the increased blanking, along
with the lower velocity resolution of the GBT data, affects the
corrected first- and second-moment maps and in the rest of this paper
we therefore only consider the VLA-only first- and second-moment
maps as shown in Figs.\ \ref{fig:mom1} and \ref{fig:mom2}.

\subsection{Position-velocity slices}

The moment maps presented here give a concise description of the
morphology and kinematics of the \HI in the M81 triplet.  A disadvantage
of these moment maps is that much information on the detailed velocity
structure of the gas is lost.  Moment maps along the two other
(spatial) axes of the cube can show some of the global velocity
structure of the emission, but due to the projection, detailed
information on smaller-scale structures is lost here as well.

An alternative solution is to make use of position-velocity
slices. These show the velocity structure of the gas along a spatial
slice. In principle, these slices can be extracted from the data cube
at any arbitrary position and position angle, fine-tuned to highlight
a particular feature. Here, we want to produce a general overview of
the velocity structure of the triplet. Using the zero-spacing
corrected natural-weighted C+D data, we extract a number of slices parallel
to the major axis of M81, covering the full extent of the triplet
along each slice. We assume a major axis PA of 330$^{\circ}$
\citep{deblok08}, which is also a reasonable approximation for the
orientation on the sky of the entire triplet.

To keep the number of slices manageable and increase the
signal-to-noise in each slice, we extract slices with a perpendicular
thickness of $140''$ ($\sim 4$ natural-weighted beams). We tested
several slide thicknesses and found that the value of $140''$ gives a
good compromise between increasing the signal-to-noise and preserving
the visibility of small-scale features.

\begin{figure}
  \centering
  \includegraphics[width=\hsize]{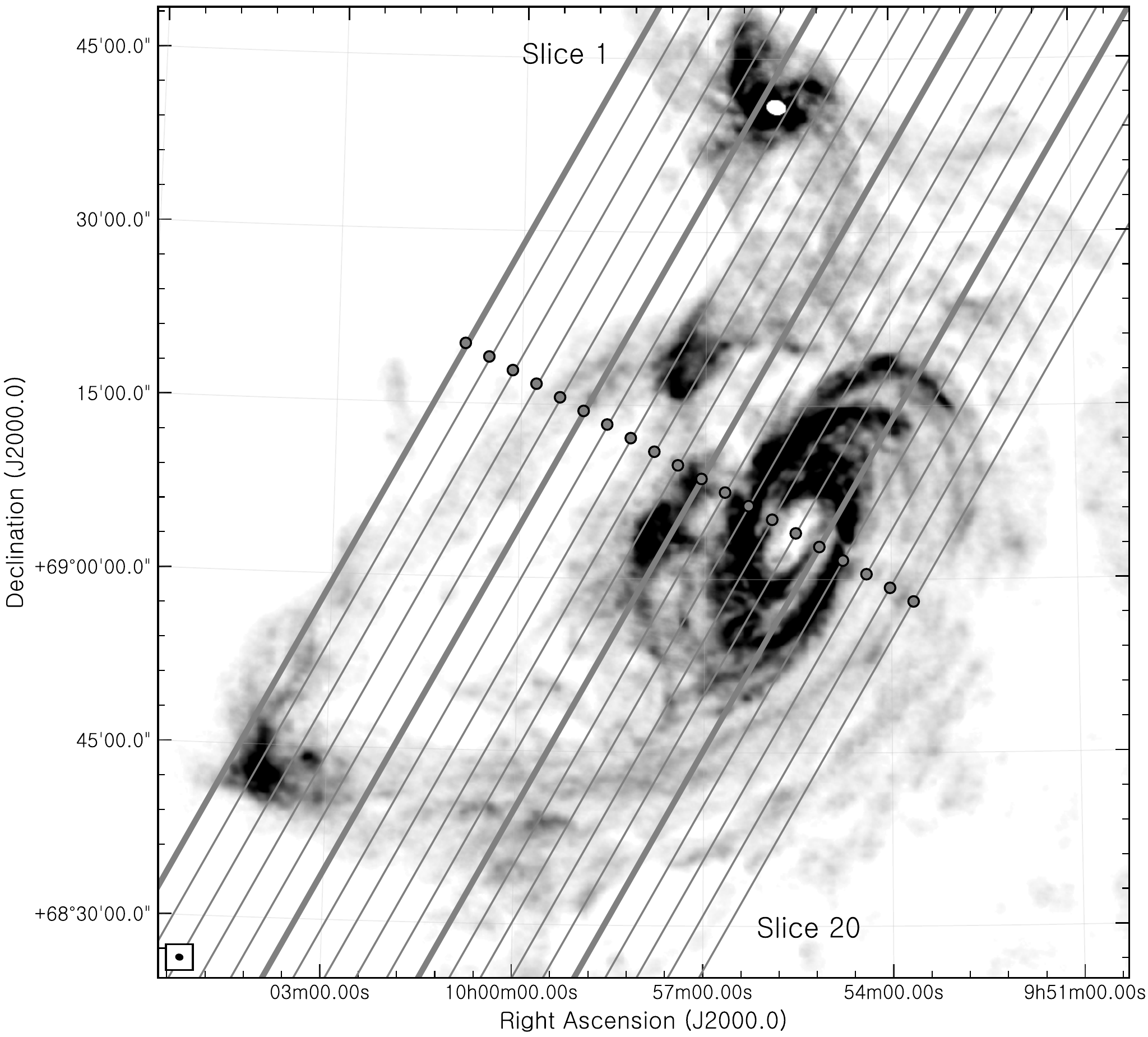}
  \caption{Central positions of the position-velocity slices presented
    in Fig.\ \ref{fig:slices}, superimposed on a zeroth-moment
    map. The circles indicate the zero-points for the offsets along
    the slices. For ease of reference, every fifth slice is shown
    using a thick line. Slice 1 is the easternmost slice, slice 20 the
    westernmost slice.  Slice 15 is centered on the center of M81.
    Every slice is $140''$ thick and separated by the same amount from
    the adjacent slices. The lines shown here indicate the centres of
    the slices. The position angle of the slices is $150.3^{\circ}$,
    corresponding to the position angle of the major axis of M81.
  \label{fig:mom0slice}}
\end{figure}

Most of the prominent velocity features in the triplet can be covered
by 20 contiguous slices covering most of the eastern part of the
triplet and a smaller fraction of the western part. Figure
\ref{fig:mom0slice} shows the positions of the slices superimposed on
the zeroth-moment map of the triplet. Slices are numbered from 1 to
20, with slice 1 the easternmost slice, and slice number increasing to
the west. Slice 15 is centered on the center of M81 and is located on
the M81 major axis.  Slices 6 and 7 pass close to the center of M82.

The slices' position-velocity diagrams are shown in
Fig.\ \ref{fig:slices}. The Galactic emission is clearly visible in
all slices at velocities of $\sim 0$ and $\sim -50$ \kms. The leftmost
part (negative velocites and negative offsets) of the panels
corresponds to the southern part of the mosaic, the rightmost part
(positive velocities and positive offsets) to the northern part. The
increased noise in the very leftmost part of the slices is due to the
decreased sensitivity at the southern edge of the mosaic. The
rightmost (northern) edge is not shown due to a lack of features
there. The position-velocity slices are discussed in more detail in
Sect.\ \ref{sec:moms}.

\begin{figure*}
  \centering
  \includegraphics[width=0.45\hsize]{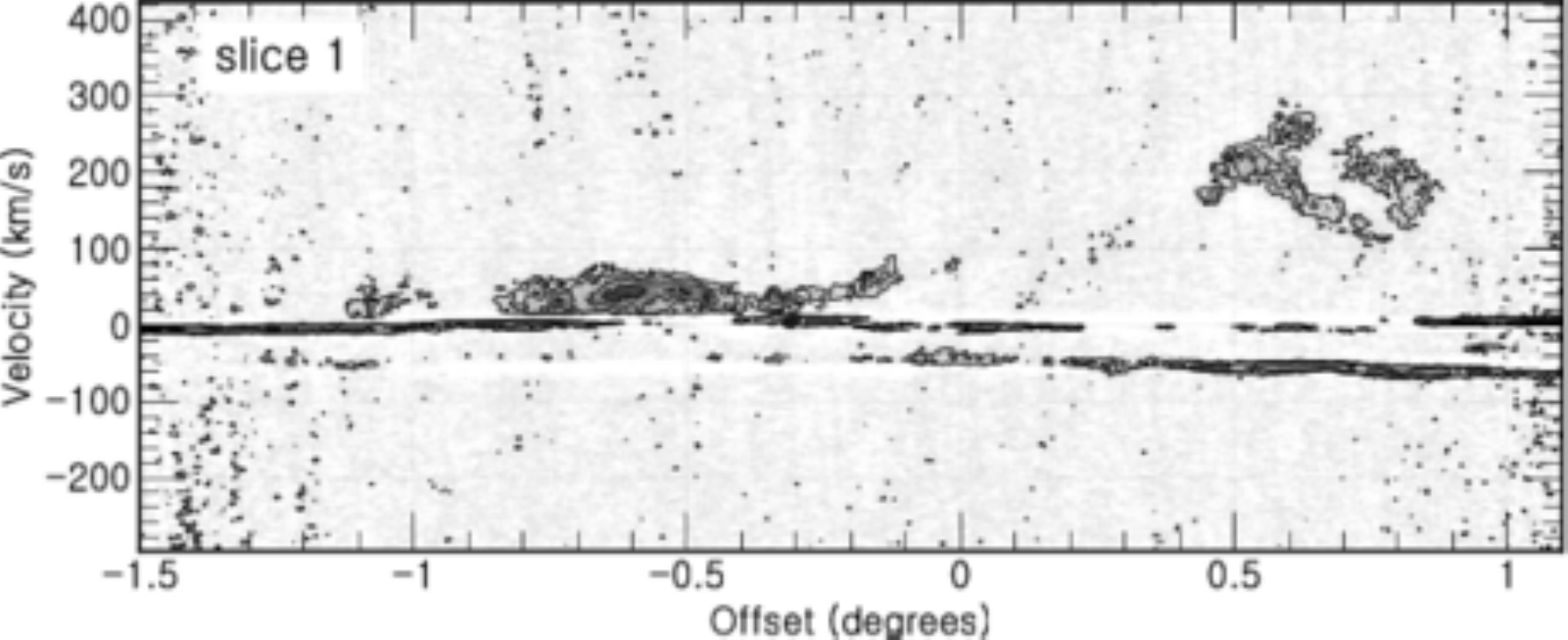}
  \includegraphics[width=0.45\hsize]{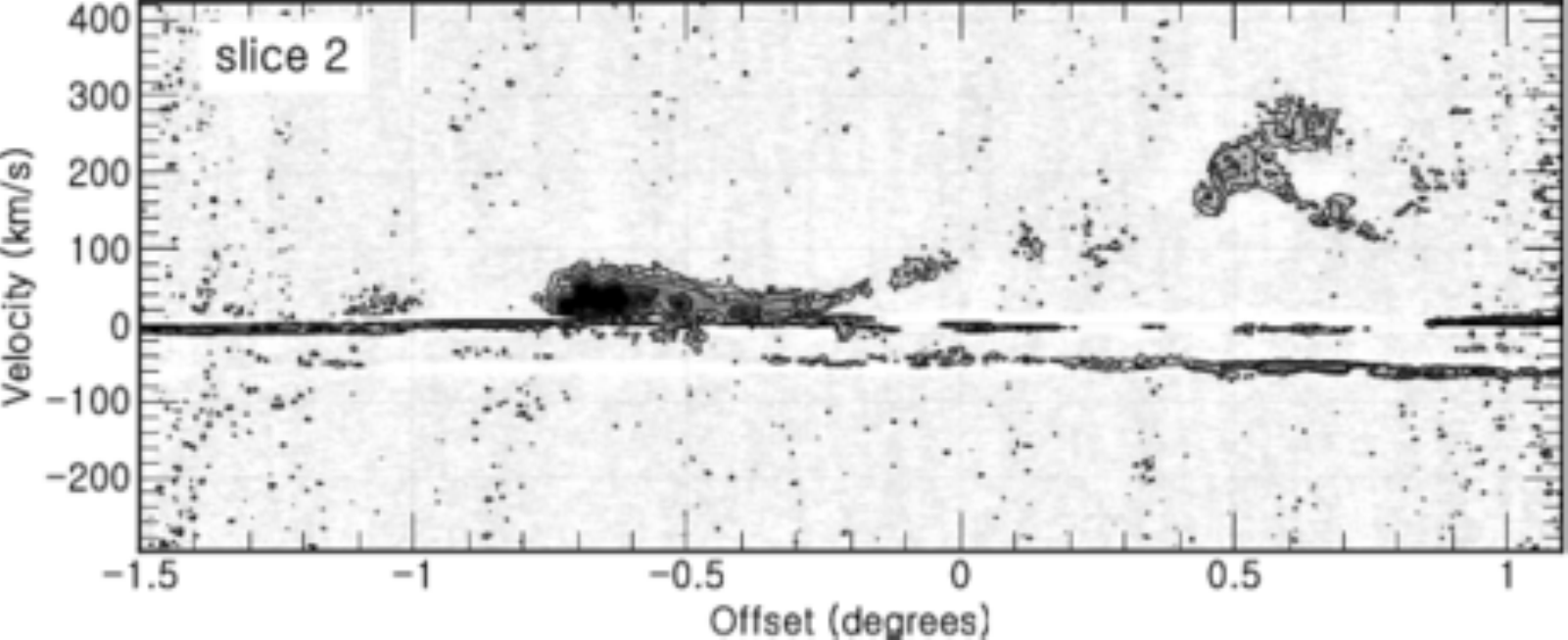}\\
  \includegraphics[width=0.45\hsize]{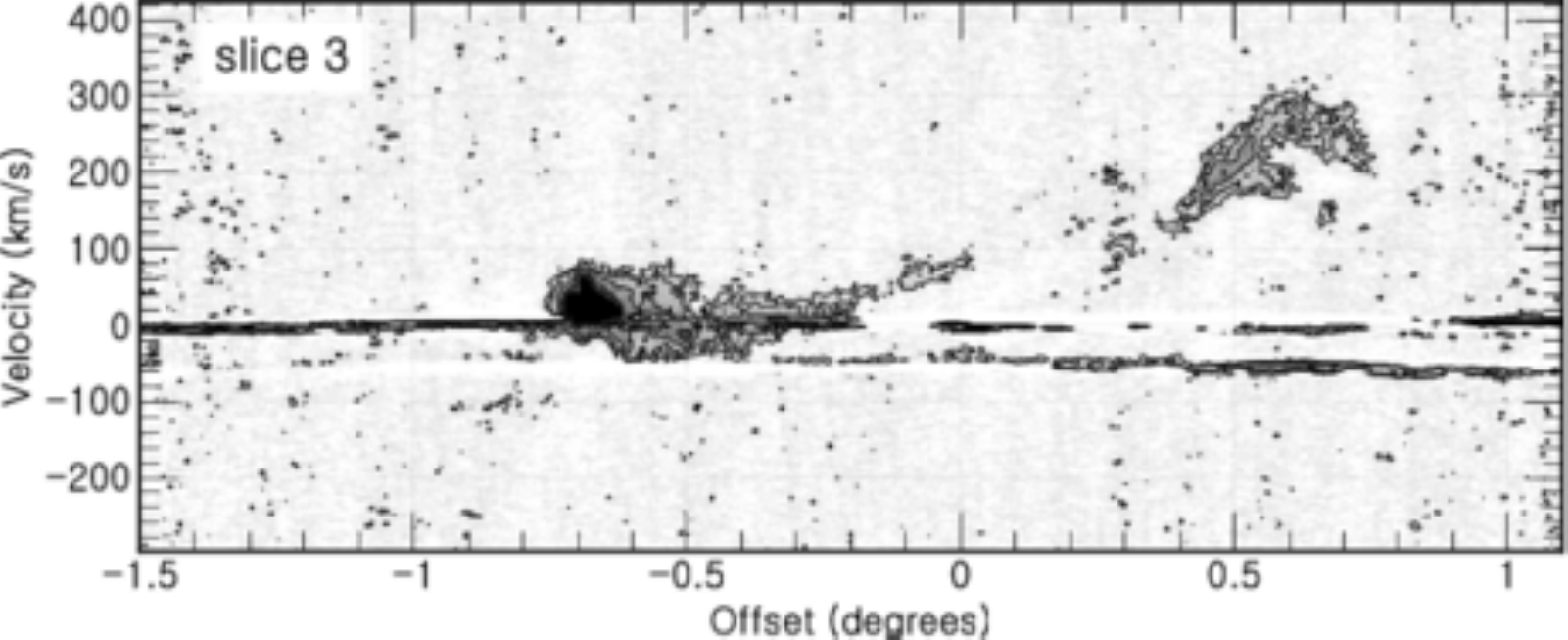}
  \includegraphics[width=0.45\hsize]{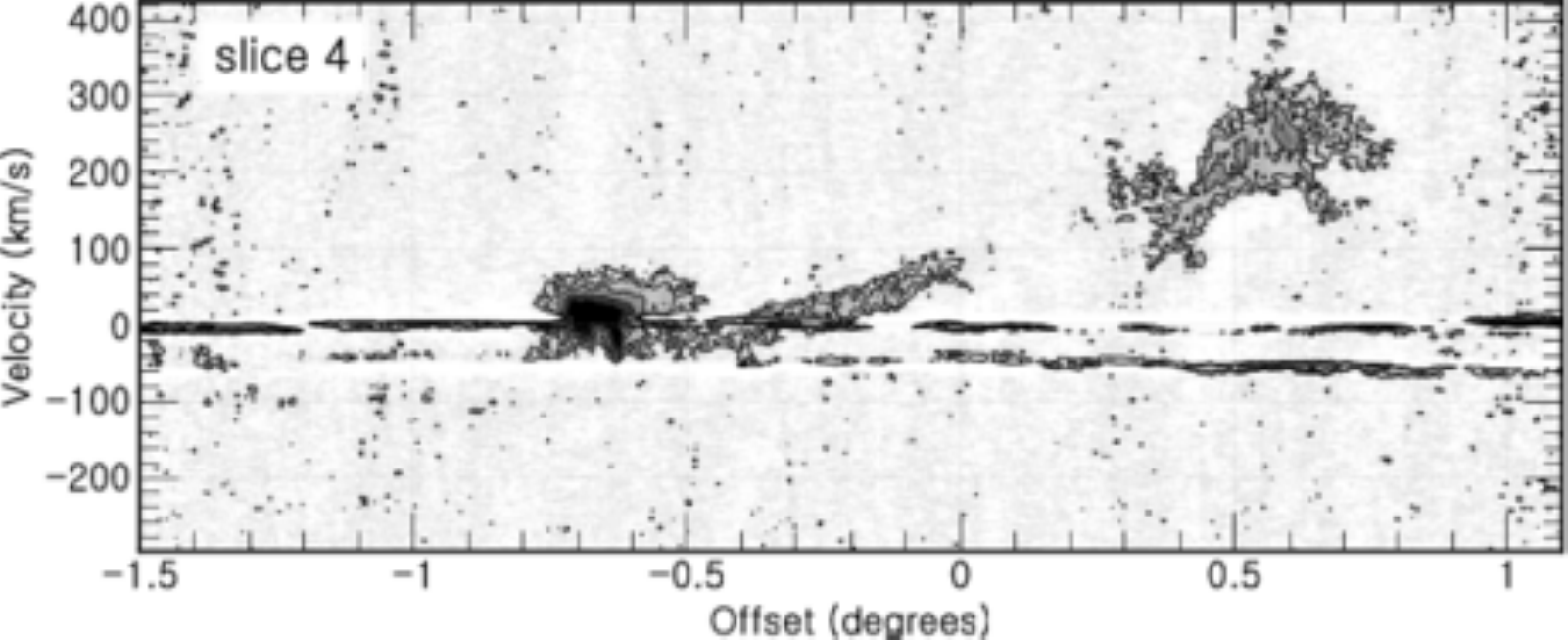}\\
  \includegraphics[width=0.45\hsize]{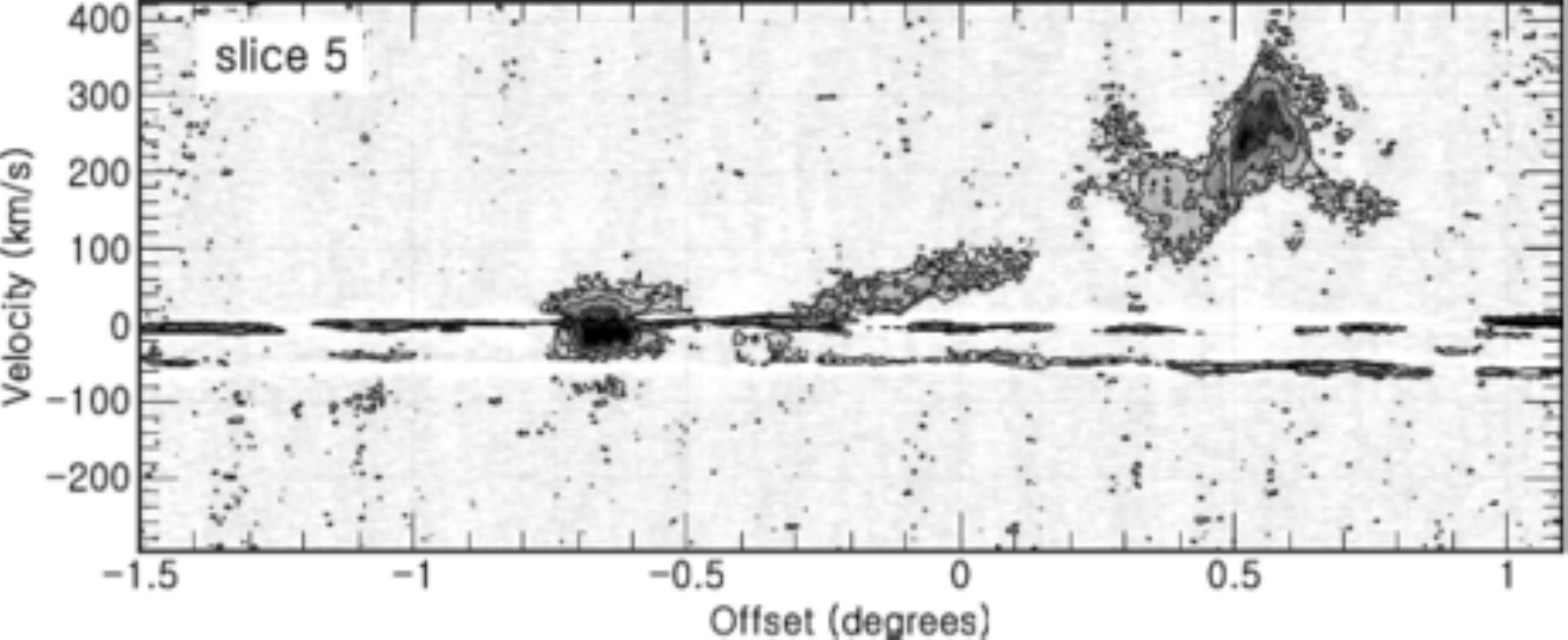}
  \includegraphics[width=0.45\hsize]{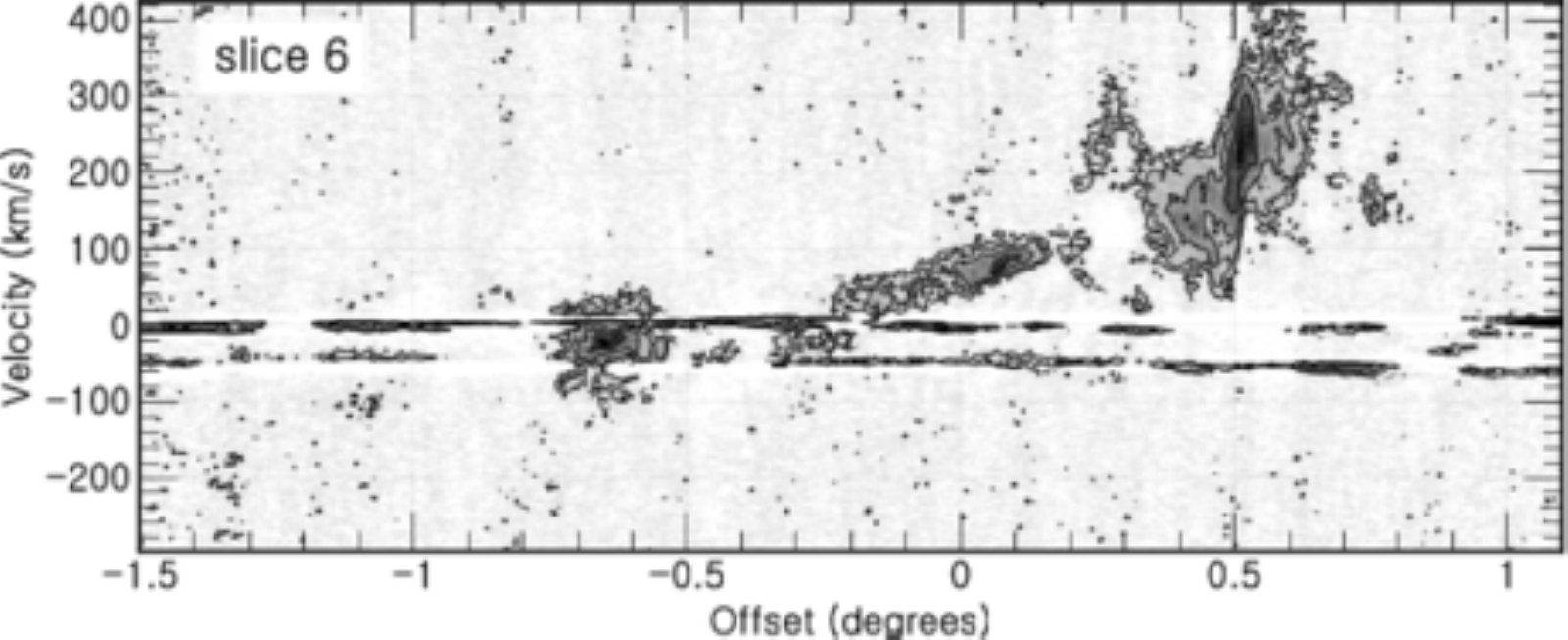}\\
  \includegraphics[width=0.45\hsize]{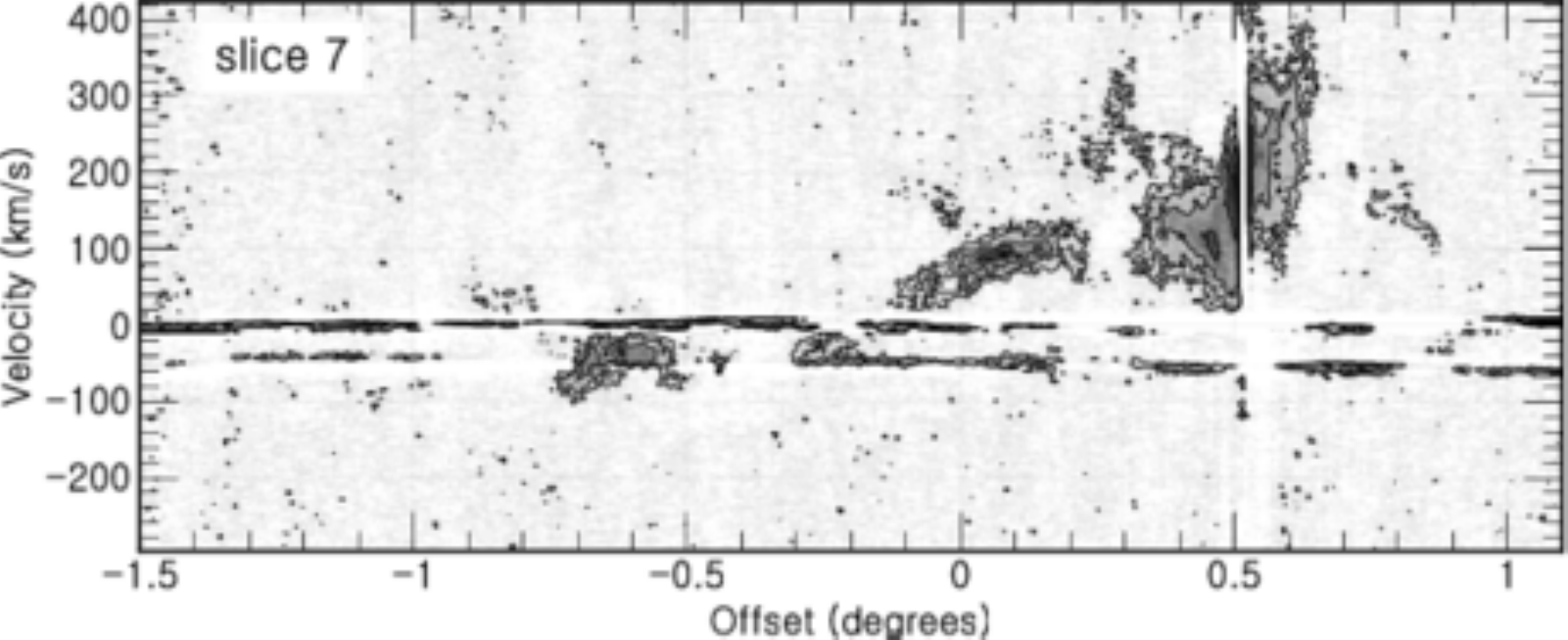}
  \includegraphics[width=0.45\hsize]{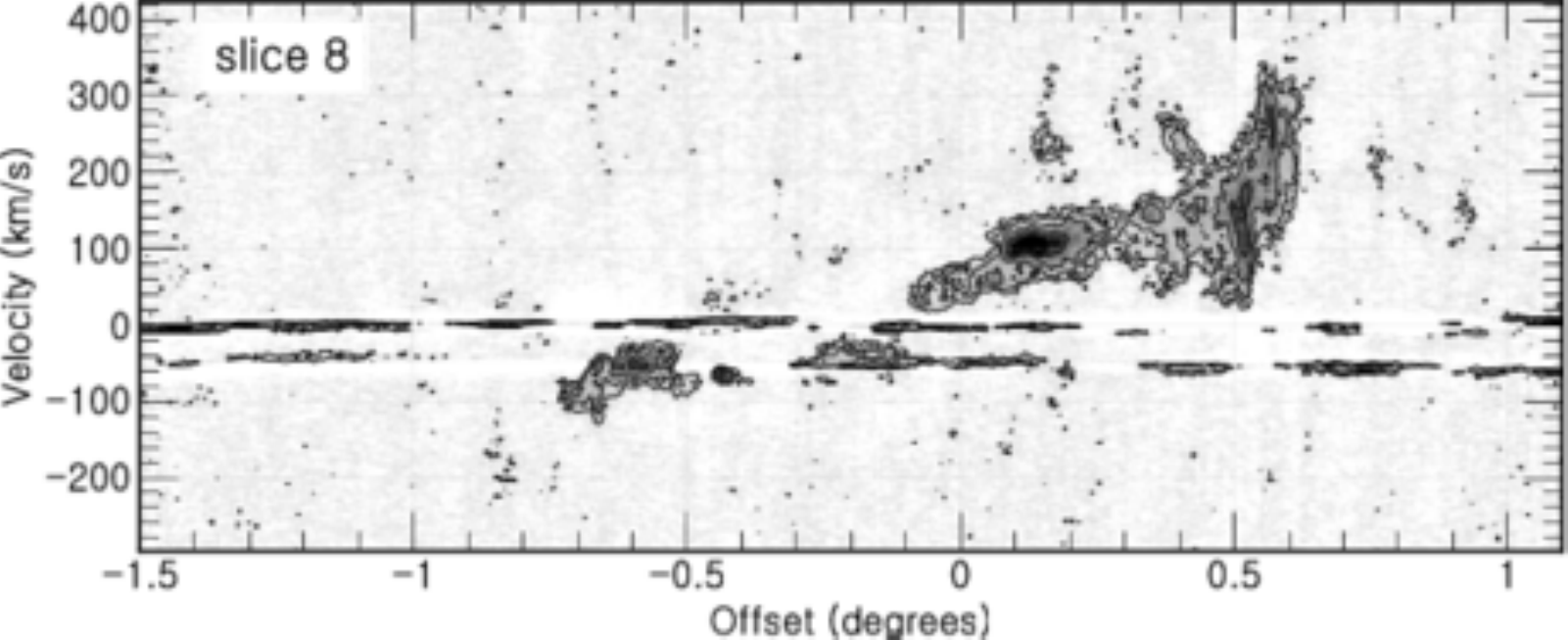}\\
  \includegraphics[width=0.45\hsize]{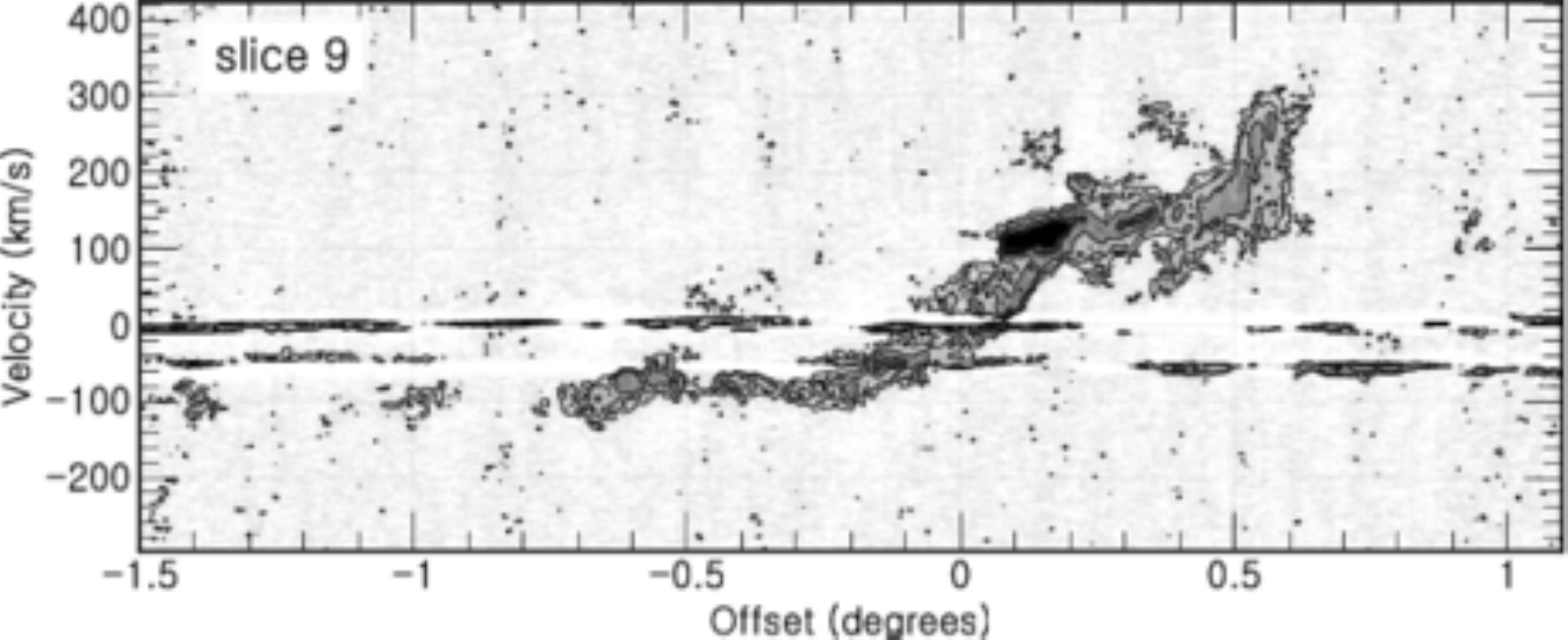}
  \includegraphics[width=0.45\hsize]{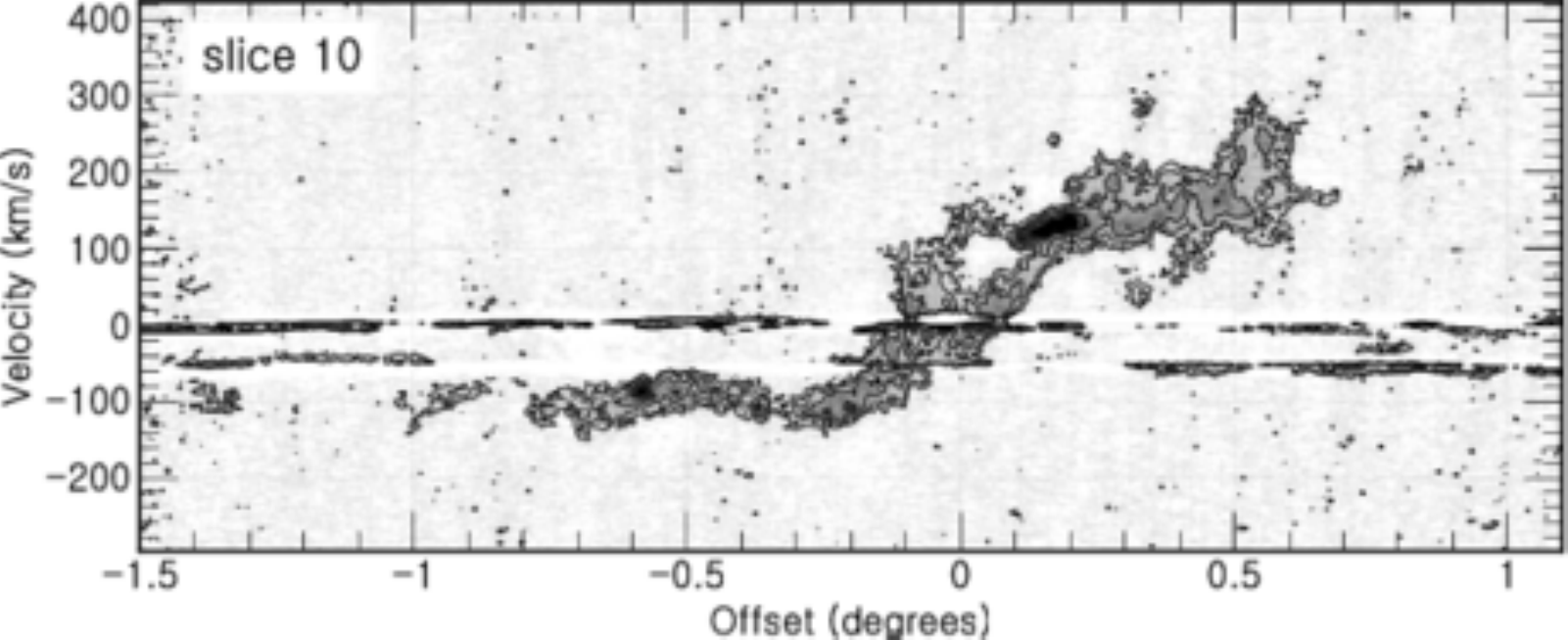}\\
  \caption{Position-velocity slices covering part of the M81 triplet,
    as shown in Fig.\ \ref{fig:mom0slice}. Numbering of the slices is
    as shown in that Figure. Negative offsets are towards the south,
    positive offsets to the north. The zero-point corresponds with the
    respective circles indicated in Fig.\ \ref{fig:mom0slice}. The
    slices are $140''$ thick, and emission is summed perpendicularly
    to each slice. The lowest contour shown is 0.015 Jy beam$^{-1}$,
    corresponding to $3\sigma$ in these summed slices. Contour levels
    then increase by factors of two. The grayscale runs from $-0.01$
    Jy beam$^{-1}$ (white) to $+0.2$ Jy beam$^{-1}$ (black). Galactic
    emission is visible in all slices at $0$ and $-50$ \kms. Increased
    noise in the left-most part of the slices is due to decreased
    sensitivity due to the edge of the mosaic.}
\end{figure*}

\addtocounter{figure}{-1} 
\begin{figure*}
  \centering
  \includegraphics[width=0.45\hsize]{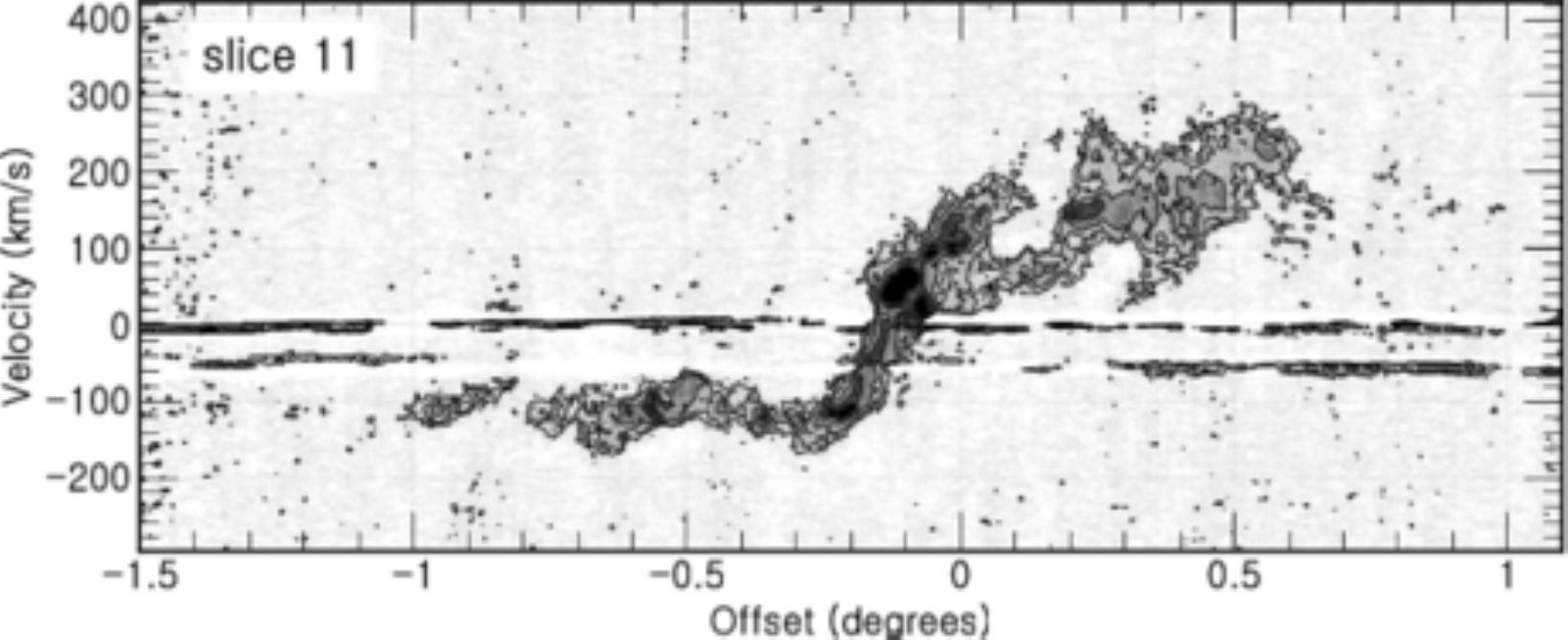}
  \includegraphics[width=0.45\hsize]{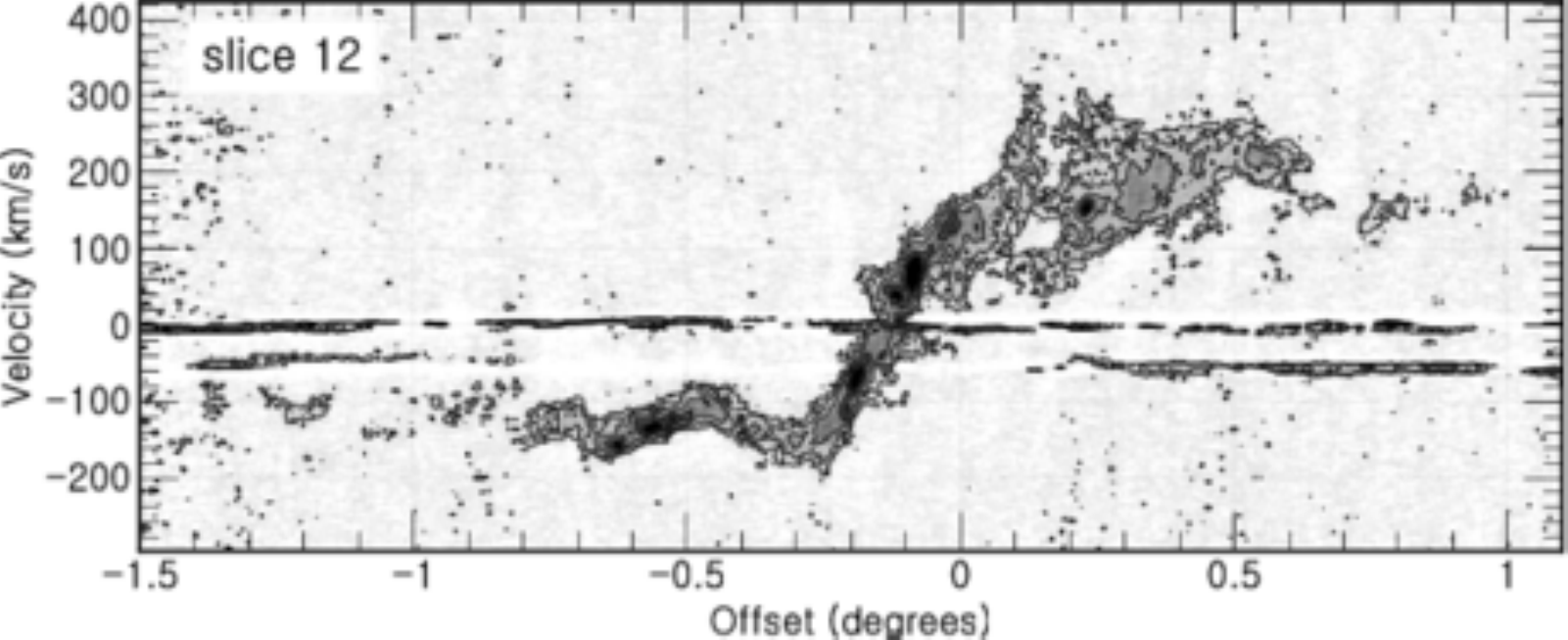}\\
  \includegraphics[width=0.45\hsize]{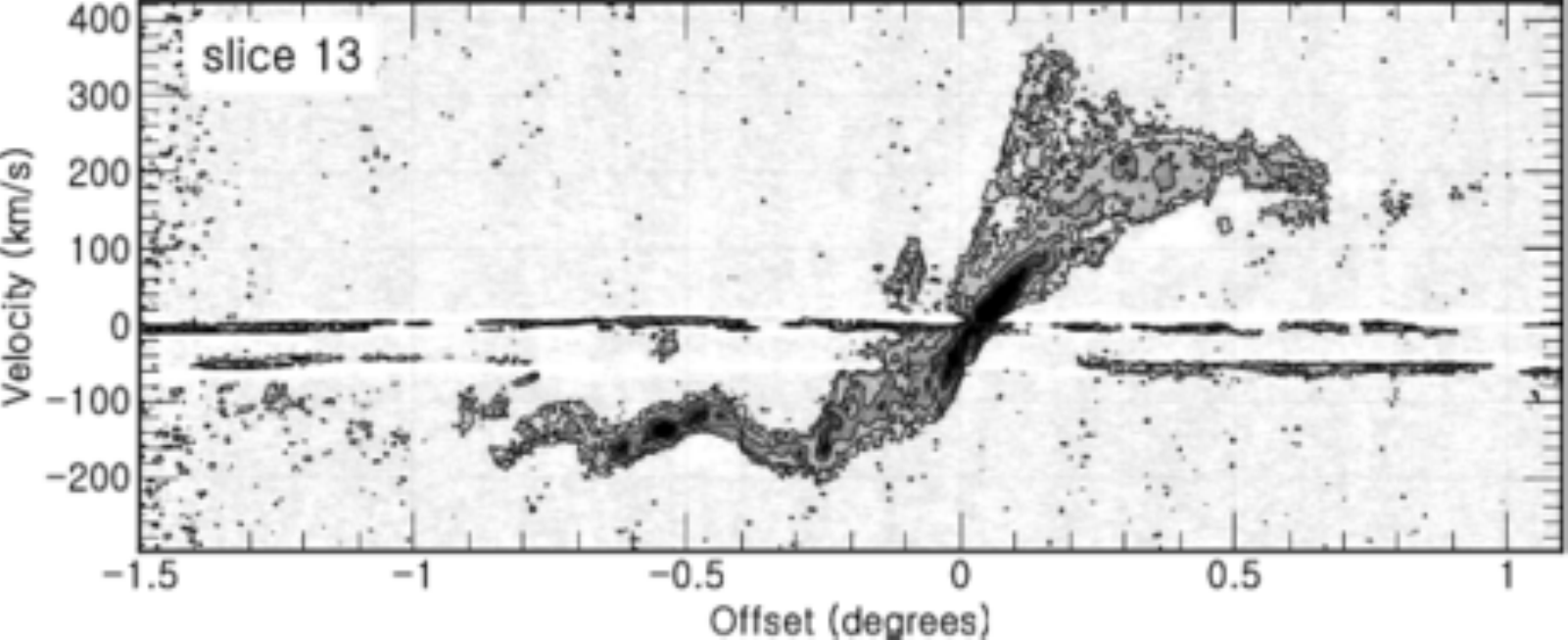}
  \includegraphics[width=0.45\hsize]{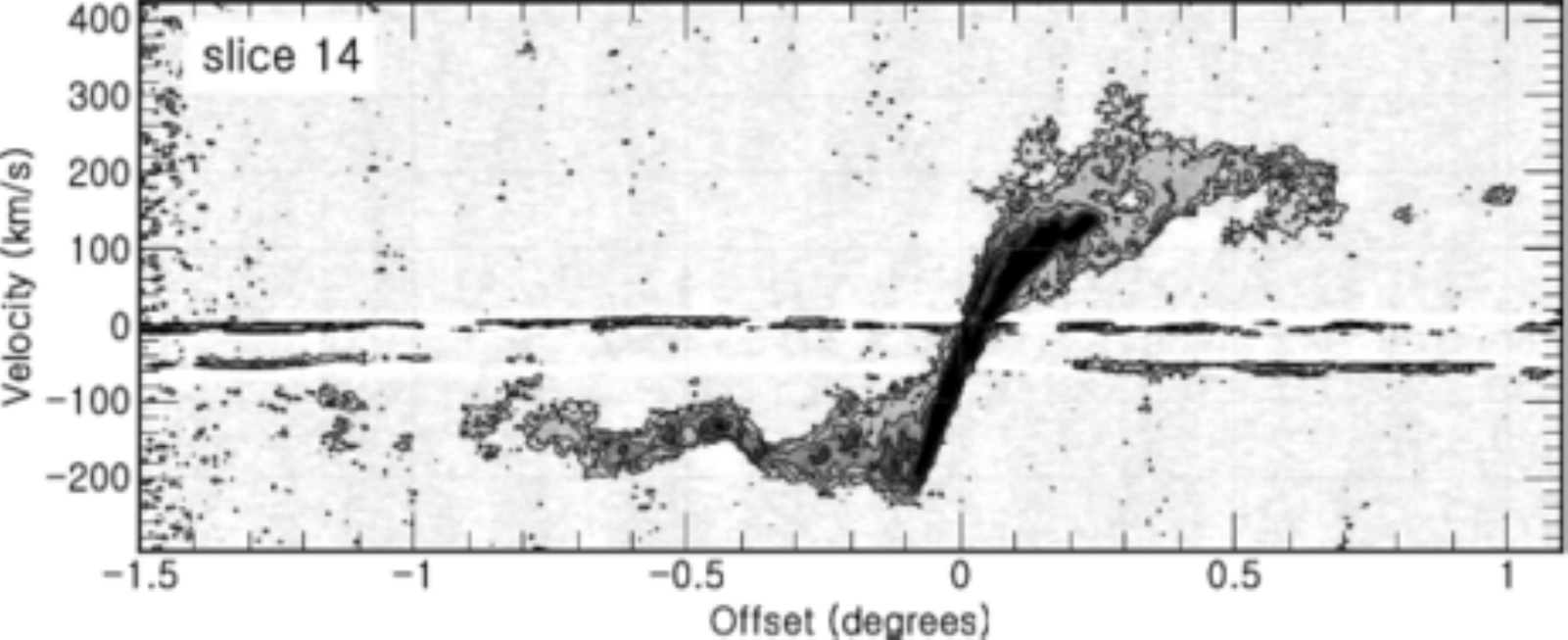}\\
  \includegraphics[width=0.45\hsize]{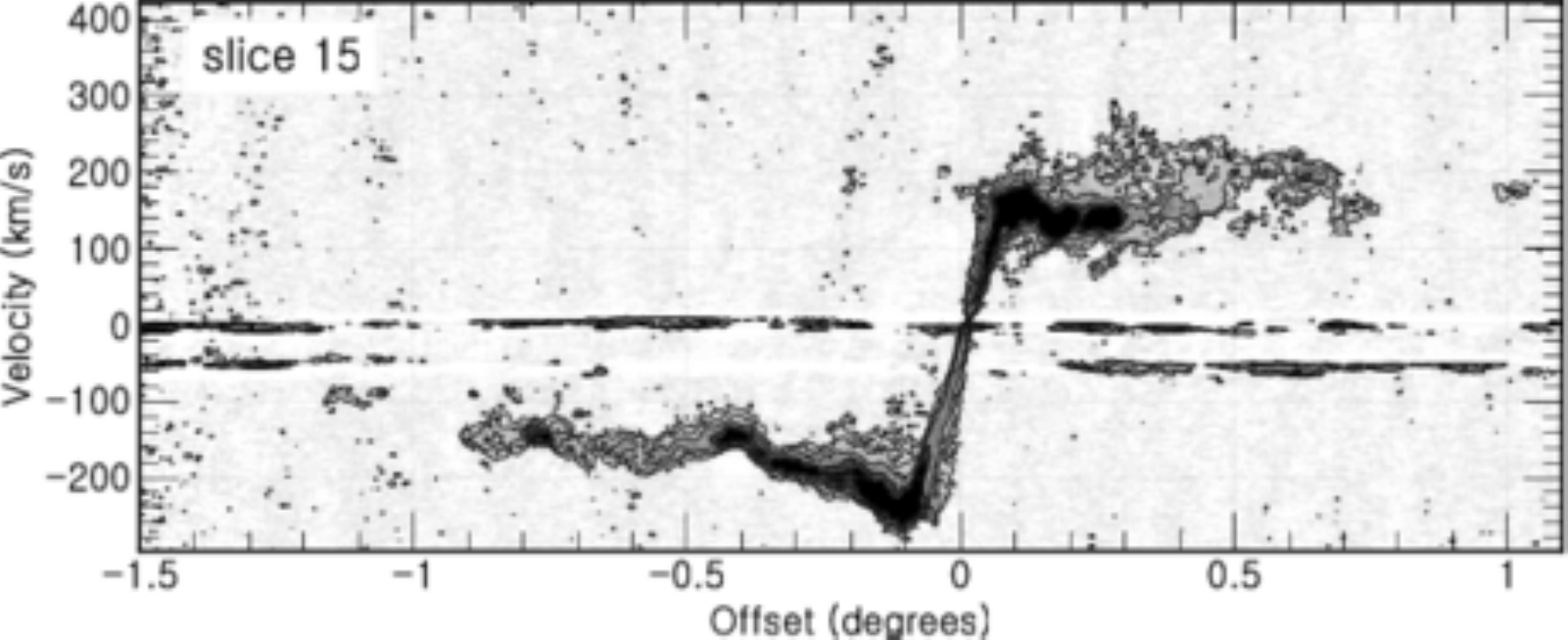}
  \includegraphics[width=0.45\hsize]{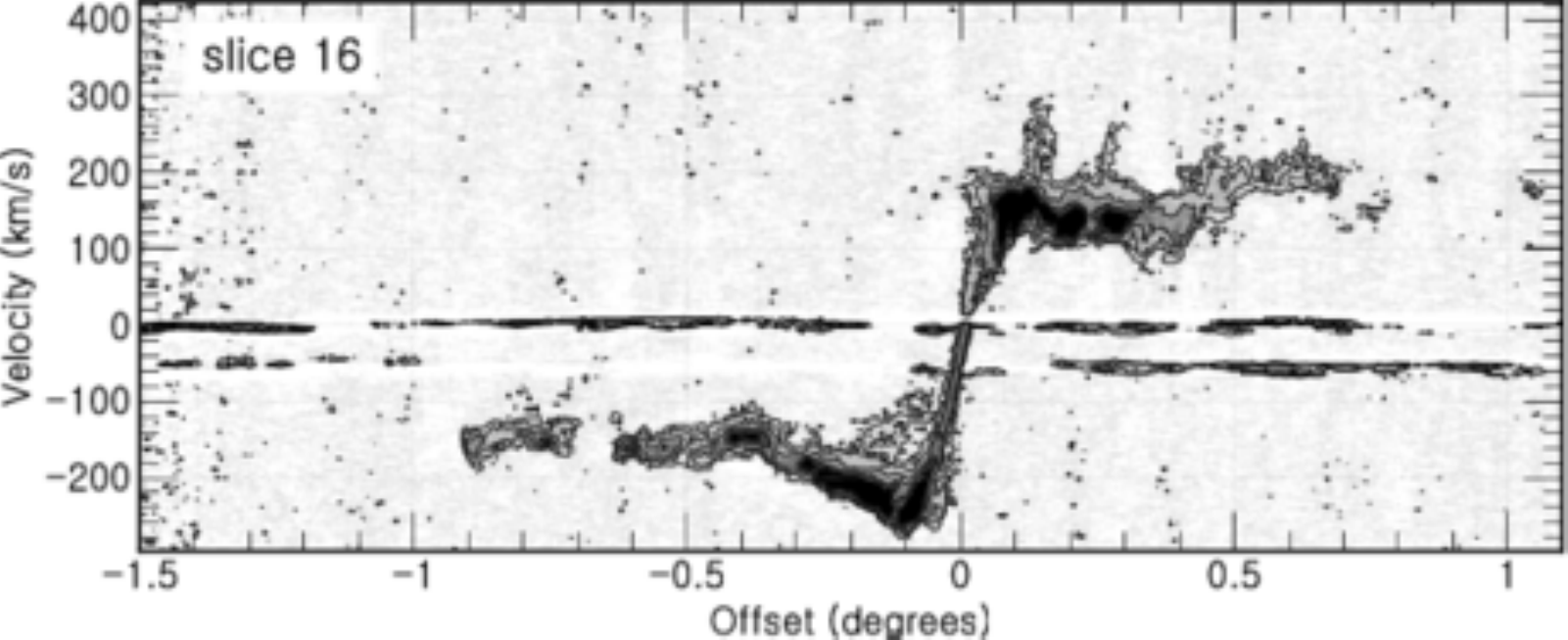}\\
  \includegraphics[width=0.45\hsize]{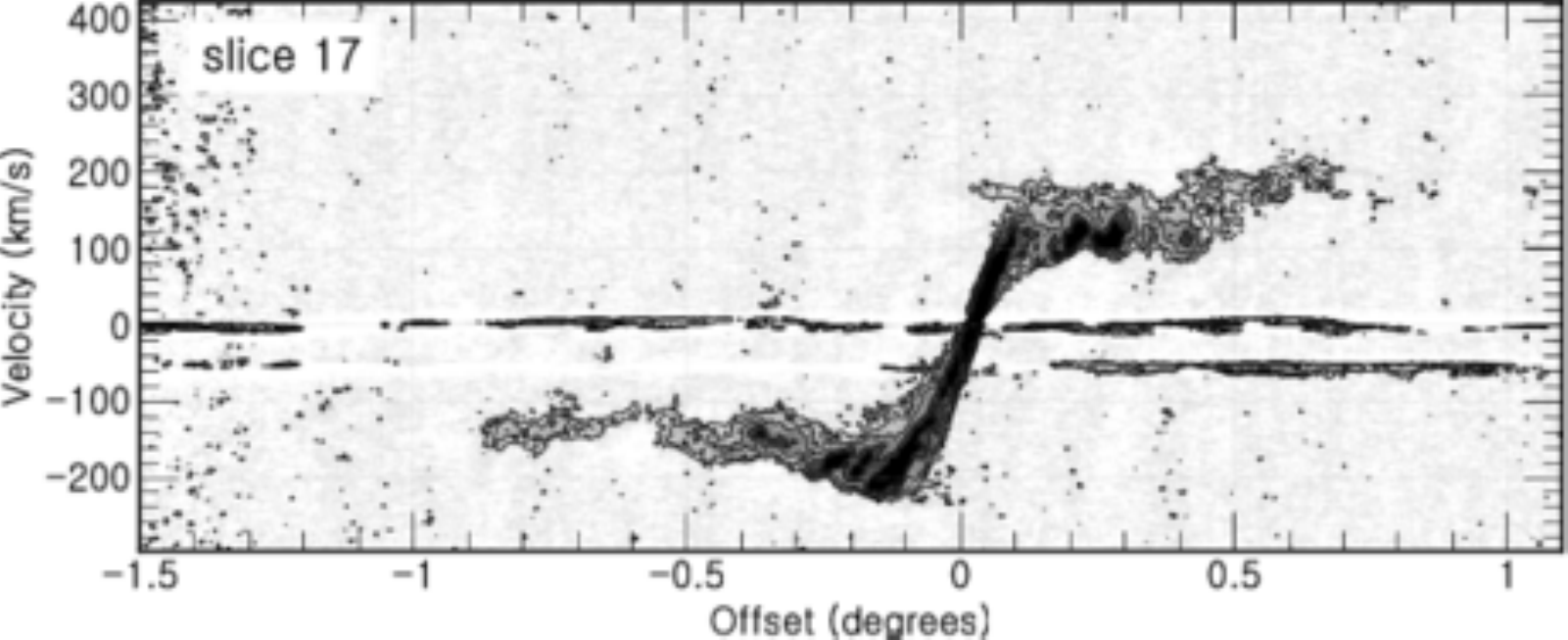}
  \includegraphics[width=0.45\hsize]{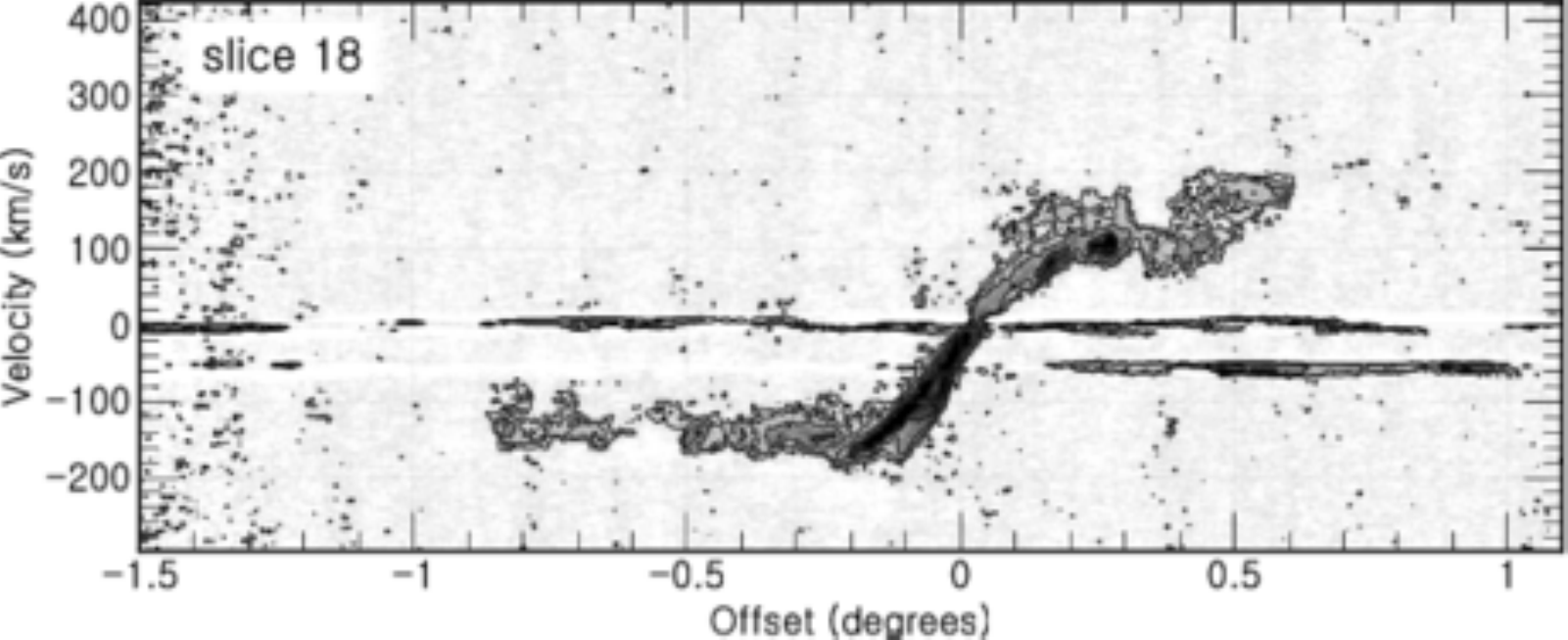}\\
  \includegraphics[width=0.45\hsize]{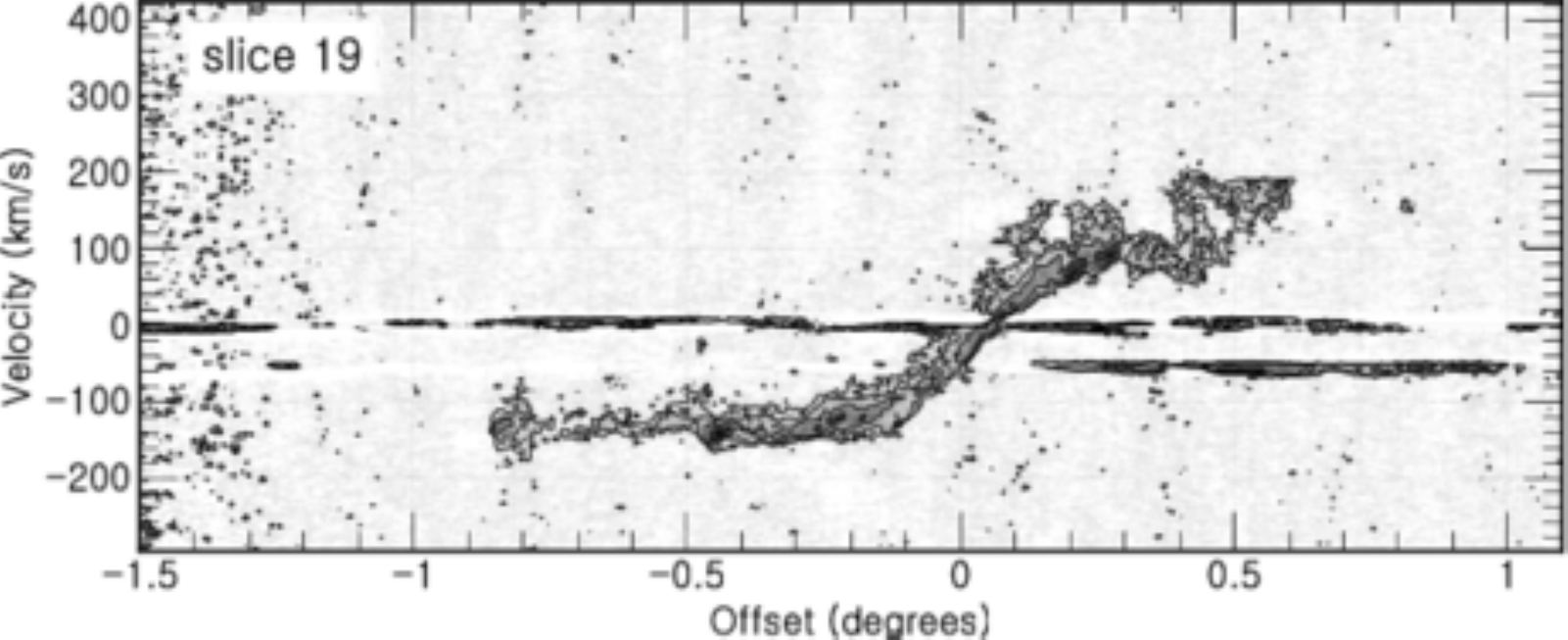}
  \includegraphics[width=0.45\hsize]{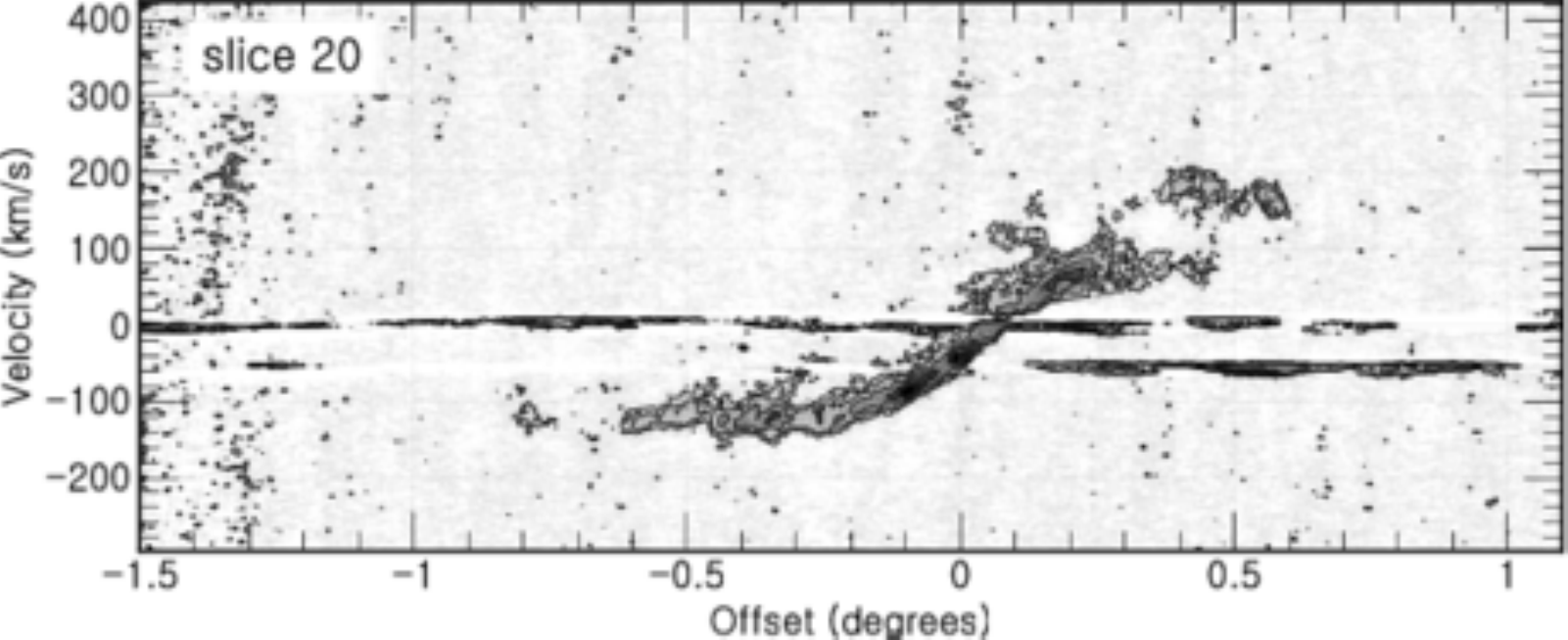}
  \caption{\emph{Continued.} Position-velocity slices covering part of
    the M81 triplet, as shown in Fig.\ \ref{fig:mom0slice}. Numbering
    of the slices is also as shown in that Figure. Negative offsets
    are towards the south, positive offsets to the north. The slices
    are $140''$ thick, and emission is summed perpendicularly to each
    slice. The lowest contour shown is 0.015 Jy beam$^{-1}$,
    corresponding to $3\sigma$ in these summed slices. Contour levels
    then increase by factors of two. The grayscale runs from $-0.01$
    Jy beam$^{-1}$ (white) to $+0.2$ Jy beam$^{-1}$ (black). Galactic
    emission is visible in all slices at $0$ and $-50$ \kms. Increased
    noise in the left-most part of the slices is due to decreased
    sensitivity due to the edge of the mosaic.
    \label{fig:slices}}
\end{figure*}

\subsection{Additional south-east mosaic pointings\label{sec:se-data}}

\begin{figure}
  \centering
\includegraphics[width=0.9\hsize]{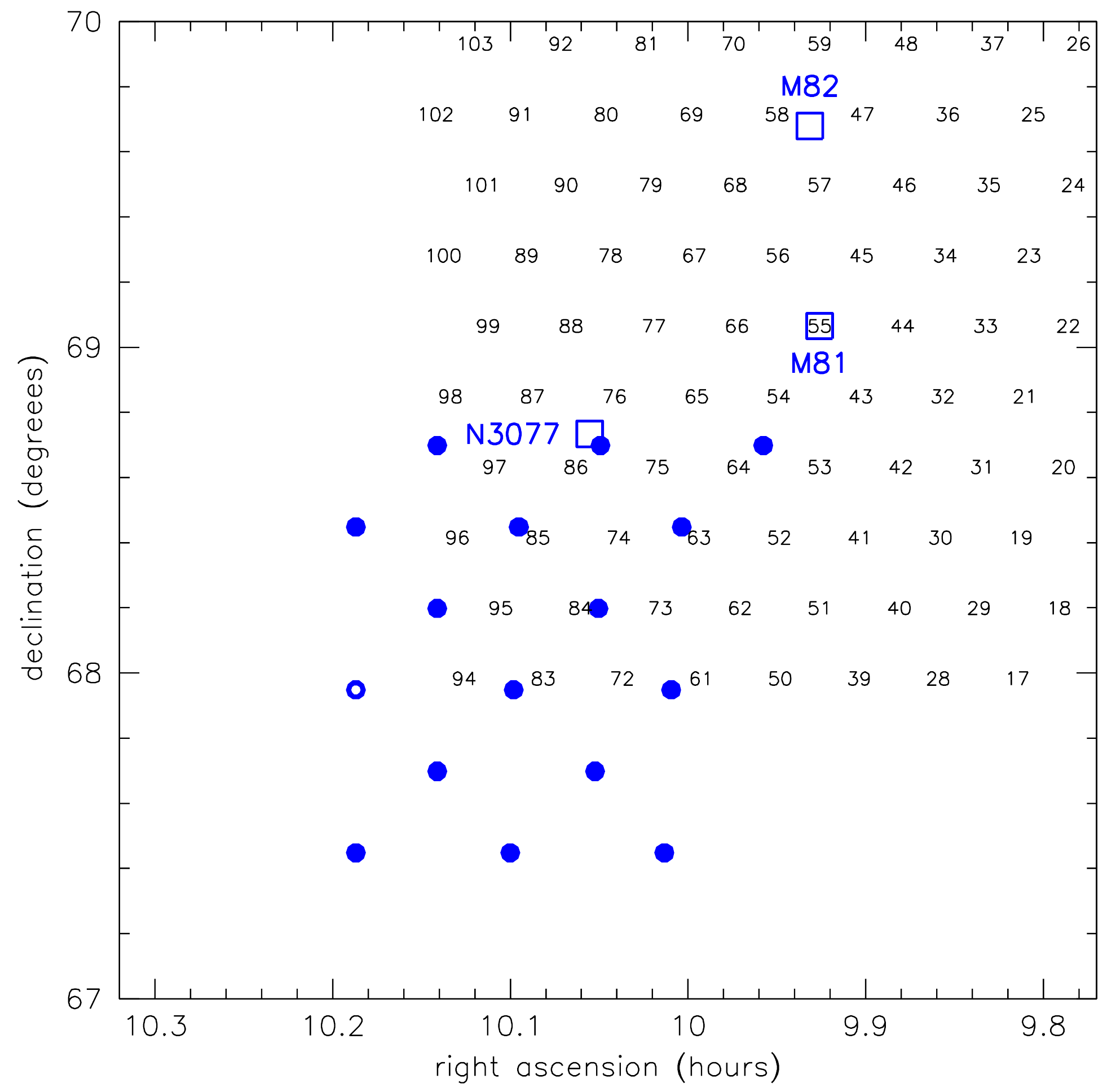}
\caption{Central positions of the pointings of the AW683
  mosaic compared with our mosaic. Filled blue circles indicate the
  pointings used here. Blue open circle shows the position of a
  pointing observed but not used. Other numbers and symbols are as in
  Fig.\ \ref{fig:fields}.
  \label{fig:flufffields}}
\end{figure}

In addition to the main mosaic data, we also use additional data from
project AW683 to extend the mosaic coverage further towards the
south-east.  These data consist of a 16-pointing mosaic observed in C-
and D-array and partly overlapping with the SE corner of the main
mosaic (see Fig.\ \ref{fig:flufffields}).

These data were taken in December 2006 (C-array) and April 2007
(D-array), when the VLA/EVLA transition was underway, meaning not all
baselines were usable. The integration time was about 50 minutes per
pointing in each of the two configurations. The observations were done
with a channel spacing of 5 \kms between $-355$ and $+210$ \kms. The
C-array data did not significantly improve the signal-to-noise of the
final data set, so we do not consider these data any further.  The
central of the three easternmost pointings was severely affected by
RFI, and we discard that pointing.

We subtracted a zeroth-order continuum fit, and produced a
natural-weighted, D-array-only datacube using the remaining pointings
and a channel spacing of 10 \kms. The noise per 10 \kms channel is 1.1
mJy beam$^{-1}$. We cleaned the cube down to 1.5$\sigma$ using {\tt
  mossdi} in Miriad. The synthesized beam is $80.2'' \times 69.2''$,
with a beam position angle of 30.0$^{\circ}$. The column density limit
of these data is $1.95 \cdot 10^{18}$ \cm ($1\sigma$, 1 channel of 10
\kms), or, more representative, $1.17 \cdot 10^{19}$ \cm ($3\sigma$,
20 \kms or 2 channels).

The integration time per pointing is approximately equal to those of
the D-array observations of our mosaic, however, the noise level in
the AW683 data is $\sim 40$ percent higher (taking into account the
different channel widths used). This is due to a combination of the
smaller number of baselines available (1/3 of the telescopes had
already transitioned to EVLA status and were not used) and the
relatively large amount of RFI which necessitated a significant amount
of flagging. We tried combining these data with our VLA mosaic to
produce one combined data set, but this produced inferior results due
to the irregular pointing grid and varying noise levels in the overlap
region.

The higher noise level and presence of residual RFI artefacts in the
data means we use an alternative method to create an unbiased
integrated intensity map.  All \HI in the observed region is
constrained to the velocity range from $-120$ to $-80$ \kms and we
therefore only consider the channels in this velocity range.  These
were spatially smoothed to twice the original beam size.  We selected
all signal above $3\sigma$ (smoothed) per channel and also present in
at least 2 consecutive channels. The resulting mask was applied to the
original resolution data cube, and from the latter a zeroth-moment map
was created. These data are discussed further in
Sect.\ \ref{sec:clouds}.

\section{Discussion of the data}

\subsection{Moment maps and position-velocity slices\label{sec:moms}}


The zeroth moment map (Fig.\ \ref{fig:mom0}) shows features not
visible in the \citet{yun94} and \citet{yun00} data, such as the full
length of the arm between M81 and NGC 2976, emission between NGC 2976
and M81 and the presence of clouds to the SE of the triplet. The
existence of the northern part of the NGC 2976 arm was already known
from observations by \citet{appleton81} and \citet{appleton88}, as
well as from the 24-pointing mosaic by \citet{yun00}. The zero-spacing
corrected moment map convincingly shows that this arm forms with one
part extending down to NGC 2976, as was also shown in the GBT
observations in \citet{chynoweth08}. Also visible close to the
northernmost edge of the mosaic is dwarf galaxy M81dwB (UGC 5423) at
$10^h05^m30^s, +70^{\circ}21'52''$.

One striking result is that the observed area away from the triplet is
mostly empty.  We do not find a large population of small \HI clouds
that are not associated with the tidal features, even though the
5$\sigma$ \HI mass limit for an unresolved cloud is $\sim 10^4$ \msun
for a velocity width of $\sim 10$ \kms. Even taking into account that
clouds may be resolved by a few beams, or have velocity widths that
are a factor of few larger, this still implies upper limits below
$\sim 10^5$ \msun for a hypothetical population of free-floating \HI
clouds. It is often thought that these free-floating clouds could be
embedded in mini-dark-matter halos, with implications for cosmological
problems such as the ``missing satellites'' problem (e.g.,
\citealt{kauffmann93}).  A more extensive discussion on cloud masses
is given in Sect.\ \ref{sec:clouds}.

The velocity field of M81 (Fig.\ \ref{fig:mom1}) shows a regularly
rotating inner disk.  The outer disk is more disturbed. The
  transition occurs at approximately the Holmberg radius. The largest
deviations from regular rotation occur to the east of the center,
along the minor axis, and are visible as strong kinks in the velocity
contours. This region corresponds to the location of dwarf galaxy
Holmberg IX.  This is also visible in the position-velocity slices in
Fig.\ \ref{fig:slices}.  Slice 10 and 11 cross this location, and the
presence of the extra \HI is clearly visible at an offset of $\sim
-0.1^{\circ}$.

The orientation of the kinematical minor axis of M82 seems to be
almost perpendicular to its optical minor axis. It is likely that this
is caused by the gas outflows in M82 (e.g., \citealt{yun93b,
  walter02a,leroy15,martini18}) affecting the velocity field. Slices 6
and 7 in Fig.\ \ref{fig:slices} show that in these regions \HI is
present with a velocity spread of close to 400 \kms.

NGC 3077 is hardly visible kinematically, and the dynamics of the gas
in that region are dominated by the interaction. It also does not
stand out in slices 1--5 (Fig.\ \ref{fig:slices}) which cross this
area.

Note that the smaller clumps and stream fragments surrounding the main
body of the triplet all have velocities close to those of the nearby
parts of the triplet, indicating they are probably all associated with
the observed tidal features.

The second-moment map (Fig.\ \ref{fig:mom2}) shows a north-south
gradient in velocity dispersion, with lower values of around 5--10
\kms mainly found towards the south-west, while high values of 20 \kms
and higher are found towards the north-east. Many of these high values
are associated with M82, and inspection of the data cube shows that
this is indeed diffuse gas that is spread over a large range in
velocity, as shown by slices 6--8 (Fig.\ \ref{fig:slices}).

The situation is different in the northern part of M81 and the
connection with M82. Here the high values indicate the presence of
multiple components at different velocities along the line of sight.
This is explains the extremely high second-moment values of $>100$
\kms found about $10'$ to the north of the center of M81. Here,
multiple, separate components with a maximum separation of $\sim 260$
\kms are present. Slices 12--14 (Fig.\ \ref{fig:slices}) show this
region at offsets between $\sim +0.1^{\circ}$ and $\sim +0.3^{\circ}$.

To disentangle these multiple components, most likely different
physical structures along the same line of sight, requires a full 3D
structural and kinematic model of all the \HI, both the rotating disk
of M81 and the various tidal filaments wrapping around M81 and its
satellite galaxies. Athough the features just discussed are the most
prominent, similar structures can be found at many places within the
group; see, e.g., slice 9 at $-0.4^{\circ}$ and slice 7 at
$0.0^{\circ}$.

Some of the high second-moment value clumps seen in the bridge between
M81 and NGC 3077 are caused by \HI clouds at different velocities from
the main \HI bridge features. These clouds are in the tidal
structures, well away from the main galaxies. In contrast, the high
values in the immediate proximity of NGC 3077 are intrinsic again, and
indicate the presence of a gas component spread over a large range in
velocity, as shown by the feature in slice 4 (Fig.\ \ref{fig:slices})
at $\sim -0.55^{\circ}$.

In addition to the larger-scale phenomena described above, several
interesting individual smaller-scale features can be made out in the
position-velocity slices. One example is the high-velocity feature
visible in slice 16 at an offset of $+0.14^{\circ}$ with anomalous
velocities of up to $\sim 100$ \kms. It is located in the interarm
region just south of the inner of the two prominent northern \HI
spiral arms of M81.  Ultra-violet GALEX \citep{gildepaz07} and
H$\alpha$ \citep{greenawalt98} data, as well as the stellar density map
discussed in Sect.\ \ref{sec:sf}, show the presence of star formation in the
area, and it is likely that the feature is associated with a recent
star formation event.  Several similar, but less prominent
features are visible in the same area.

\subsection{Comparison with GBT data\label{sec:GBT}}

As noted in the Introduction, the survey area presented here was
  also observed with the GBT, as published in \citet{chynoweth08}. In
  Sect.\ \ref{sec:zero} we described using the GBT data to correct for
  the missing spacings in the VLA data.  As the zero-spacing corrected
  cube is a combined data set with the resolution of the
  interferometry data and the flux of the single-dish data, it in
  principle contains no new information that is not already present in
  the two source data sets.  It is therefore instructive to compare
  these original data sets to get a understanding of where the
  various features visible in the moment maps originate.

\begin{figure*}
  \centering
\includegraphics[width=0.9\hsize]{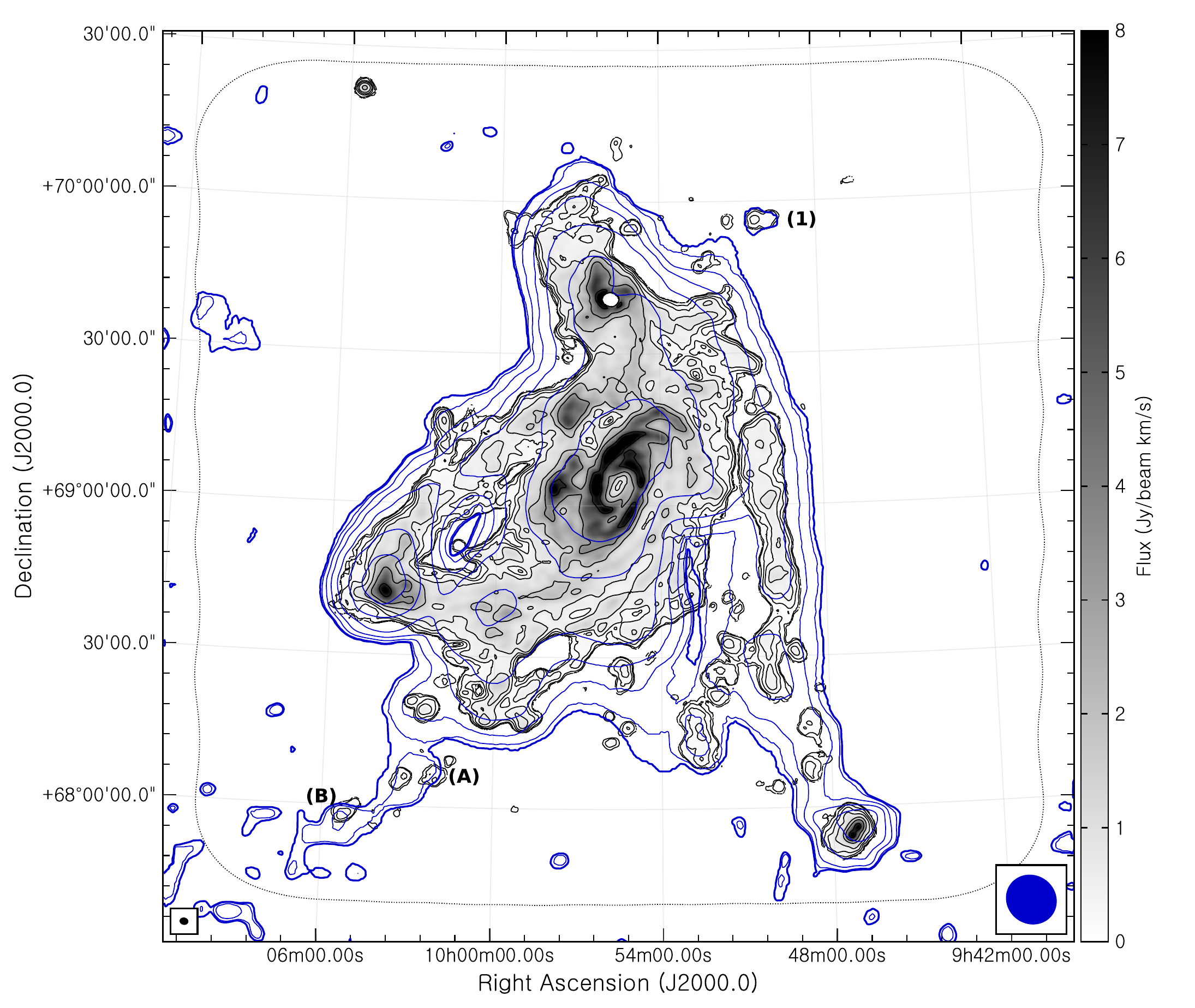}
\caption{Comparison of our natural-weighted D-array zeroth-moment map
  with the GBT zeroth-moment map from \citet{chynoweth08}.  The
  D-array data are shown as grayscale and black contours, the GBT data
  as dark-blue contours. The grayscale runs from 0 (white) to 8
  (black) Jy beam$^{-1}$ \kms.  The GBT contour levels are shown at
  1500 (thick contour), 3000, 7500, 15000, 30000, 75000, 150000 and
  300000 kJy beam$^{-1}$ \kms which corresponds to $(4.5, 9.0, 22.5,
  45, 90, 225, 450, 900) \cdot 10^{18}$ \cm. The D-array mosaic
  contour values were chosen to have the same column densities, and
  are shown at $0.0329 \cdot (1, 2, 5, 10, 20, 50, 100, 200)$ Jy
  beam$^{-1}$ \kms. The full GBT survey area is shown. The mosaic 50
  percent sensitivity contour is shown as the dotted curve.  The VLA
  beam is indicated in the bottom-left corner, the GBT beam in blue in
  the bottom-right corner.  Numbers and letters indicate the cloud
  complexes described in Sect.\ \ref{sec:GBT} and
  Sect.\ \ref{sec:clouds}. \label{fig:GBT}}
\end{figure*}

Figure \ref{fig:GBT} displays an overlay of the \citet{chynoweth08}
data on top of our D-array mosaic. The GBT beam size is $10.1' \times
9.4'$, with a major axis position angle of $53^{\circ}$. This
translates to a physical size of $10.7 \times 9.9$ kpc.

The column density sensitivities of both data sets are
similar. \citet{chynoweth08} quote a $1\sigma$, 1 channel (5.2 \kms)
sensitivity of $2.5 \cdot 10^{17}$ cm$^{-2}$. Smoothing our D-array
data to the same velocity resolution yields a sensitivity of $3.5
\cdot 10^{17}$ cm$^{-2}$.  In Fig.\ \ref{fig:GBT}, we therefore chose
identical contour levels for both data sets. We see a good
correspondence between the \HI distribution as observed by the VLA and
the GBT. The only major discrepancy is immediately to the south-west
of M81, where the GBT data show an extended north-south trough that is
not visible in the VLA data. This trough is artificial and entirely
due to the interpolation over the blanked Galactic emission that was
used in the \citet{chynoweth08} paper to construct the moment map.

The low-column density filament seen in the GBT data near $10^h06^m$,
$+68^{\circ}00'$, which is resolved into clumps with the VLA, extends
to the edge of the GBT survey area, suggesting there may be additional
HI clouds beyond the VLA survey area. We will return to this in
Sect.\ \ref{sec:clouds}.

The feature in the GBT data located near $10^h11^m$, $+69^{\circ}30'$
has a velocity of $\sim -110$ \kms as detected in the original GBT
data cube. At this position and velocity it is also marginally visible
in the VLA mosaic. It is not included in the VLA moment map as its peak
flux is below $3\sigma$ and its location close to the 50 percent
sensitivity contour makes identification more uncertain based on the
VLA data alone.

The reverse situation is true for M81 Dw B (UGC 5423), a dwarf
galaxy which is clearly detected in the VLA mosaic (at $10^h05^m30^s,
+70^{\circ}21'52''$), but is not visible in the GBT moment
map. Inspection of the GBT data cube shows a marginal detection
at the correct position and velocity, but it is located in the edge
region of the GBT map where the noise is enhanced and many
artificial features of similar extent and brightness are present.

It is striking that, especially towards the south, the low-column
density arms and streams detected by the GBT break up in clouds and
clumps as observed by the VLA. An interesting question is whether
these clouds represent all the \HI seen in the lower-resolution GBT
data, or whether they form the high column density tip of the iceberg
in a surrounding lower column density component.

To address this, we compare the \HI masses of a number of these
  clouds, selecting only objects that are far enough away spatially
  and spectrally from bright \HI emission that may affect the object
  fluxes.  As noted above, we consider the VLA and GBT data sets
  separately to better trace the origin of emission features.  The
  zero-spacing corrected data is (for individual low-flux objects)
  less suited due to the various contributions from, amongst others,
  flux scale factors, masking and difference in velocity resolution
  that are difficult to quantify.
 
One example of a low-mass \HI cloud is the isolated cloud to the NW
of M82, which \citet{chynoweth08} denote as ``Cloud 1'' (indicated as
``1'' in Fig.\ \ref{fig:GBT}).  We find an \HI mass of $3.2 \cdot
10^6$ \msun, which is a factor 4.6 less than found by
\citet{chynoweth08}. (The other clouds discussed in that paper are affected
by Galactic emission and therefore not discussed here.)

Other examples can be found to the south of the triplet.  These are
indicated in Fig.\ \ref{fig:GBT} as ``A'' and ``B''. Complex A consist
of two small clouds in the VLA data, and corresponds to single
overdensity in the GBT map. Cloud B is a single cloud in the VLA data,
corresponding with a single overdensity in the GBT map.

The two clouds A have a total mass of $4.8 \cdot 10^6$ \msun. The mass
of the corresponding GBT peak is $1.2 \cdot 10^7$ \msun, or a factor
2.5 higher. Cloud B has a mass of $3.2 \cdot 10^6$ \msun in the VLA
data, and a mass of $1.2 \cdot 10^7$ \msun in the GBT map. This is a
factor 3.6 different. For completeness, we did check the combined
data, and for the \HI clouds discussed here found masses intermediate
to the GBT and VLA masses.

In these particular comparisons we can be confident that the GBT is
detecting excess \HI not seen in the VLA data. This indicates that the
low-column density filaments seen in the GBT data are not simply the
VLA \HI clouds observed at low resolution, but that they consist of
substantial amounts of low-column density \HI in which the clouds are
embedded.

\subsection{Comparison of HI masses: GBT vs.\ VLA\label{sec:masses}}

\subsubsection{Total \HI mass}
The previous section established that some of the isolated clouds seen
in the VLA data are embedded in a low-column density \HI component
detected by the GBT.  We can check if this is more generally the case
by comparing the respective \emph{total} \HI masses found in both data
sets. As discussed above, we compare the individual VLA and GBT
  sets, rather than the zero-spacing-corrected data.

We use the moment maps to determine the total \HI mass detected in the
mosaic area. For the VLA C+D natural-weighted data we find a total flux of
2234.4 Jy \kms. Using the assumed distance of 3.63 Mpc, this gives a
total HI mass of $6.94 \cdot 10^9$ \msun.

The D-array data gives a slightly higher value of 2489.0 Jy \kms. This
translates into an HI mass of $7.74 \cdot 10^9$ \msun. These values
are $\sim 35$ percent higher than the total \HI masses given in
\citet{yun99} and \citet{appleton81}. This discrepancy is likely due to
a combination of different survey volumes, column density
sensitivities and Galactic foreground corrections.  We show below that
the latter alone can already amount to differences of $\sim 30$
percent in the total fluxes.

For the GBT data of the M81 triplet, \citet{chynoweth08} report a
total \HI mass of $10.46 \cdot 10^9$ \msun. This is substantially
higher than the previous literature values, but also $\sim 35$ percent
higher than the value derived from our D-array data.

\begin{figure}
  \centering
\includegraphics[width=0.9\hsize]{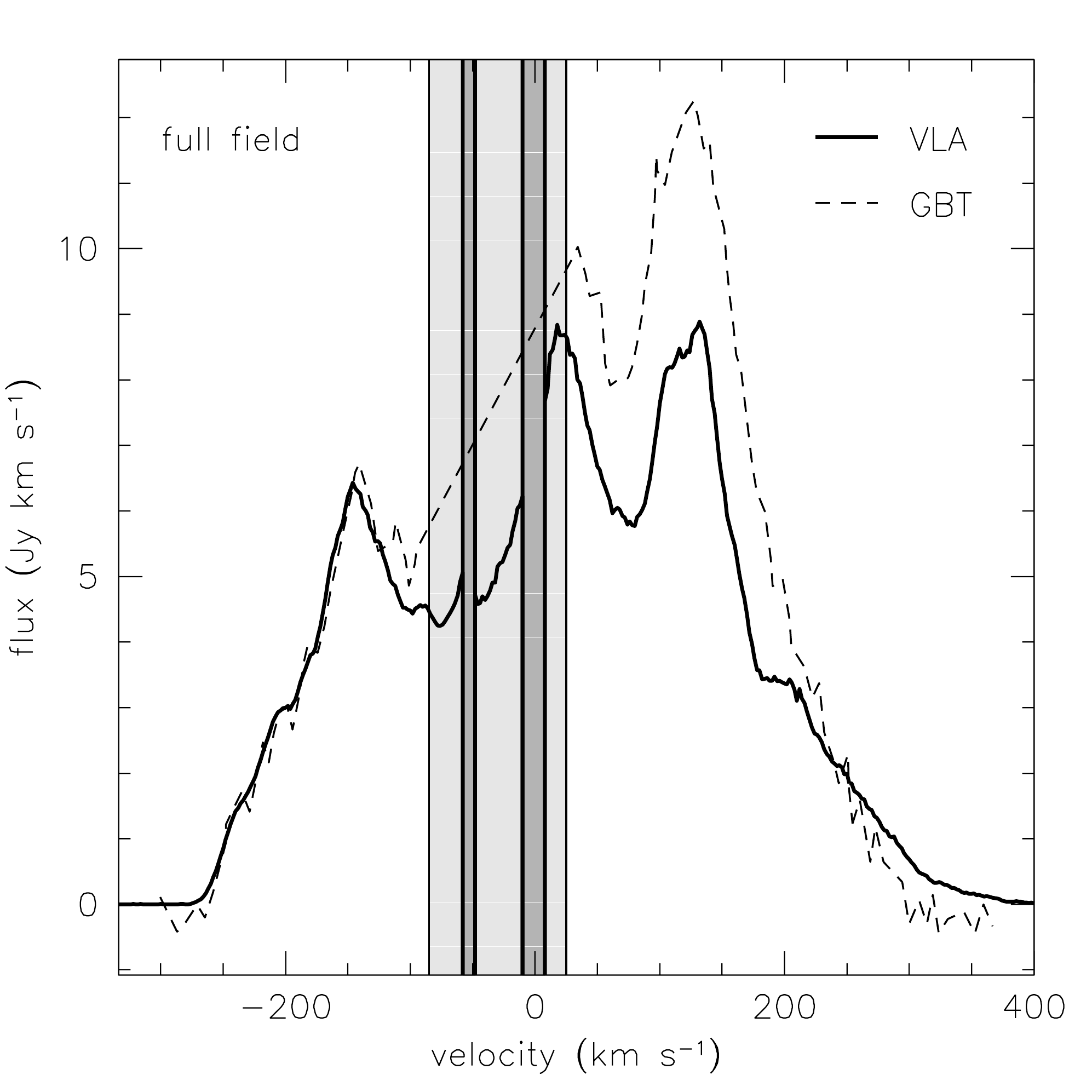}
\caption{Comparison of integrated intensity profiles of the observed
  area.  The thick full profile shows the integrated flux based on our
  D-array mosaic. The thin dashed profile show the integrated flux derived
  from the \citet{chynoweth08} GBT observations. The light-gray area
  indicates the velocity range over which \citet{chynoweth08} have
  interpolated their data. The dark-gray areas indicate the velocity
  ranges which we omitted from our data due to the Galactic emission.
  Note the different behavior of the profiles at positive velocities,
  probably indicating the presence of diffuse gas associated with M82.
  \label{fig:fieldflux}}
\end{figure}

\citet{chynoweth08} note that their data were affected by Galactic
foreground emission between $-85$ and $+25$ \kms. They replaced the
data in these velocity channels with a linear interpolation based on
the channels immediately adjacent to this range. From the global \HI
profile of the full area as shown in Fig.\ 2 of \citet{chynoweth08},
and also reproduced in Fig.\ \ref{fig:fieldflux}, we find that this
interpolated part of the spectrum constitutes 29 percent of total flux
they report.

The higher velocity resolution of our data allows us to gauge the
accuracy of this correction. We overplot the global profiles of the
full mosaic area in Fig.\ \ref{fig:fieldflux}.  The D-array fluxes in
the interpolated region of the GBT spectrum are $\sim 30$ percent
lower than the GBT interpolations. It is, however, not trivial to
correct the GBT \HI mass on the basis of
this. Fig.\ \ref{fig:fieldflux} shows that at negative velocities the
GBT and D-array fluxes agree very well with each other, whereas at
positive velocities the GBT has detected substantially more flux than
the D-array. Note that, as discussed in Sect.\ \ref{sec:deconv}, the
behaviour of detected flux as a function of clean depth is identical
for channel maps with positive and negative velocities, so that the
difference is not due to different relative importance of uncleaned
flux in these channel maps.

In other words, at negative velocities the D-array observations have
managed to detect almost all of the \HI flux (mostly associated with
the southern part of M81), while the extra GBT flux at positive
velocities (associated with the very northern part of M81, with M82,
and with the transition region in between) indicates the presence of
an extended low-column density HI component that is not present in the
southern part of the triplet.  Fig.\ \ref{fig:fieldflux} shows that
the difference between the integrated spectra is largest around the
peak at $\sim 125$ \kms, and the ``missing'' gas is thus most likely
associated with the already detected diffuse \HI around M82.  For a
full synthesis observation, the largest angular scale the VLA is
sensitive to at 1.4 GHz is $\sim 16'$, while for a single snapshot
observation this is $\sim 8'$. This range of scales is mostly larger
than the GBT beam. So while the length of the integration time per
pointing may have some influence on the recovery of structures larger
than $\sim 10'$, it is more likely that the difference between the
integrated spectra is due to surface brightness sensitivity
limitations.

We also considered the integrated spectrum of the zero-spacing
  corrected C+D data, and found a good match with the GBT profile at
  positive velocities. However, at negative velocities this profile
  overestimates significantly overestimates the flux compared to the
  GBT profile. The situation is reverse when using D-array corrected
  data. We did test the combination using various different velocity
  ranges to determine the scale factor, but found this did not affect
  the outcomes. Due to the uncertainty in relative flux scales of these
  combined data, we therefore do not consider the zero-spacing
  corrected data further in this context. A full study of the relative
  fluxes found in the VLA and GBT data as a function of resolution is
  beyond the scope of this paper.

The presence of extra \HI at positive velocities and its absence at
negative velocities, with the transition happening exactly in the
region affected by Galactic emission, makes deriving a more accurate
correction for Galactic foreground correction
difficult. Fig.\ \ref{fig:fieldflux} suggests that, with various extra
components and corrections cancelling each other, the \emph{total} \HI
mass estimate given in \citet{chynoweth08} is probably an
overestimate, but likely by not more than $\sim 5-10$ percent.  Taking
all this into account, we can therefore conclude that the GBT data
shows the presence $\sim 25$--30 percent more \HI than our VLA D-array
mosaic.

\subsubsection{\HI masses of the triplet galaxies}

The \HI masses of the major triplet galaxies are more difficult to
determine and compare, as the extent of their \HI disks cannot be well
determined due to the presence of the tidal \HI component.
\citet{chynoweth08} compare the \HI masses of the three major triplet
galaxies as derived from the GBT data, the \citet{yun99} data and the
\citet{appleton81} data.  The latter are also based on single-dish
data.

\citet{appleton81} define the \HI masses of the galaxies as the mass
measured within the Holmberg ellipse of the respective objects and
\citet{chynoweth08} follow that definition. As
noted by \citet{appleton81}, this choice of radius likely
underestimates the \HI masses.  In M81 the Holmberg radius only
encompasses the inner, high-density spiral arms; in M82 it misses much
of the extra-planar gas, while in NGC 3077 the main \HI component
falls outside the Holmberg radius. Nevertheless, in the absence of any
clear physical indicators, other choices would be equally arbitrary.

We here apply the same procedure to our D-array data, using the
parameters given in Table 1 of \citet{appleton81}. To get an estimate
of the uncertainty in the masses, we also derive the \HI masses within
a radius of $2R_{25}$, with the $R_{25}$ values taken from HyperLEDA,
but still adopting the orientations given in
\citet{appleton81}. Larger radii are impractical as the disks of M81
and M82 start overlapping at $\sim 2.5R_{25}$. The masses are listed
in Table \ref{tab:masses} and compared with the \citet{chynoweth08},
\citet{yun99} and \citet{appleton81} masses.  For M81,
  an alternative definition for the HI mass could be made by using the
  transition radius between the ordered motion of the inner disk and
  the more disturbed motion beyond that (cf.\ the velocity field in
  Fig.\ \ref{fig:mom1}). This radius turns out to be almost exactly
  equal to the Holmberg radius, so this mass is equal to the one
  already listed in Table \ref{tab:masses}.

There is a large spread in \HI mass values for each galaxy. We have
already established that Galactic foreground corrections introduce an
extra uncertainty in the \citet{chynoweth08} data, mostly due to the
lower velocity resolution. It is likely that a similar uncertainty
applies to the \citet{yun99} and \citet{appleton81} masses as well. A
full and proper determination of the ``true'' \HI masses of the three
main galaxies would thus require a further in-depth analysis and
comparison of all these effects. 

\begin{deluxetable}{lrrrrr}
  \tabletypesize{\scriptsize} \tablewidth{0pt} \tablecaption{Comparison of \HI masses\label{tab:masses}}
  \tablehead{\colhead{Galaxy} & \colhead{$M_{\rm HI}$} & \colhead{$M_{\rm HI}$} & \colhead{$M_{\rm HI}$} & \colhead{$M_{\rm HI}$}& \colhead{$M_{\rm HI}$} \\
\colhead{} & \colhead{(Ho)} & \colhead{($2R_{25}$)}  & \colhead{(Ch08)} & \colhead{(Y99)} & \colhead{(Ap81)}\\[2pt]
\colhead{} & \multicolumn{4}{c}{$(\times 10^9\,M_{\odot})$}}
\startdata
M81 & 2.29      & 2.79 & 2.67 & 2.81 & 2.19 \\ 
M82 & 0.44      & 0.75 & 0.75 & 0.80 & 0.72 \\ 
NGC 3077 & 0.23 & 0.31 & 1.01 & 0.69 & 1.00 \\ 
Total\tablenotemark{a} & 7.74 & --- & 10.46 & 5.6\phantom{0} & 5.4\phantom{0}\\
\enddata
\tablenotetext{a}{Value refers entire observed area.}
\tablecomments{(Ho): \HI mass within Holmberg radius from D-array mosaic; ($2R_{25}$): \HI mass within $2R_{25}$ radius from D-array mosaic; (Ch08): \HI mass from \citet{chynoweth08}; (Y99): \HI mass from \citet{yun99}; (Ap81): \HI mass from \citet{appleton81}.} 
\end{deluxetable} 

\subsection{The South-East clouds\label{sec:clouds}}

In Sect.\ \ref{sec:GBT}, we compared the \HI masses of the
overdensities ``A'' and ``B'' seen in the GBT map in
Fig.\ \ref{fig:GBT}. These overdensities correspond with a number of
more compact clumps as observed with the VLA.

The GBT map also shows that the \HI filament containing the
overdensities extends all the way to the south-east corner of the
observed area.  Our VLA mosaic does not extend this far, but we can
use additional observations to explore this area at higher
resolution. We use the data from project AW683 which consists of a
16-pointing mosaic observed in C- and D-array and partly overlapping
with the SE corner of our mosaic. A description of these data is given
in Sect.\ \ref{sec:se-data}.

\begin{figure*}
\centering
\includegraphics[width=0.9\hsize]{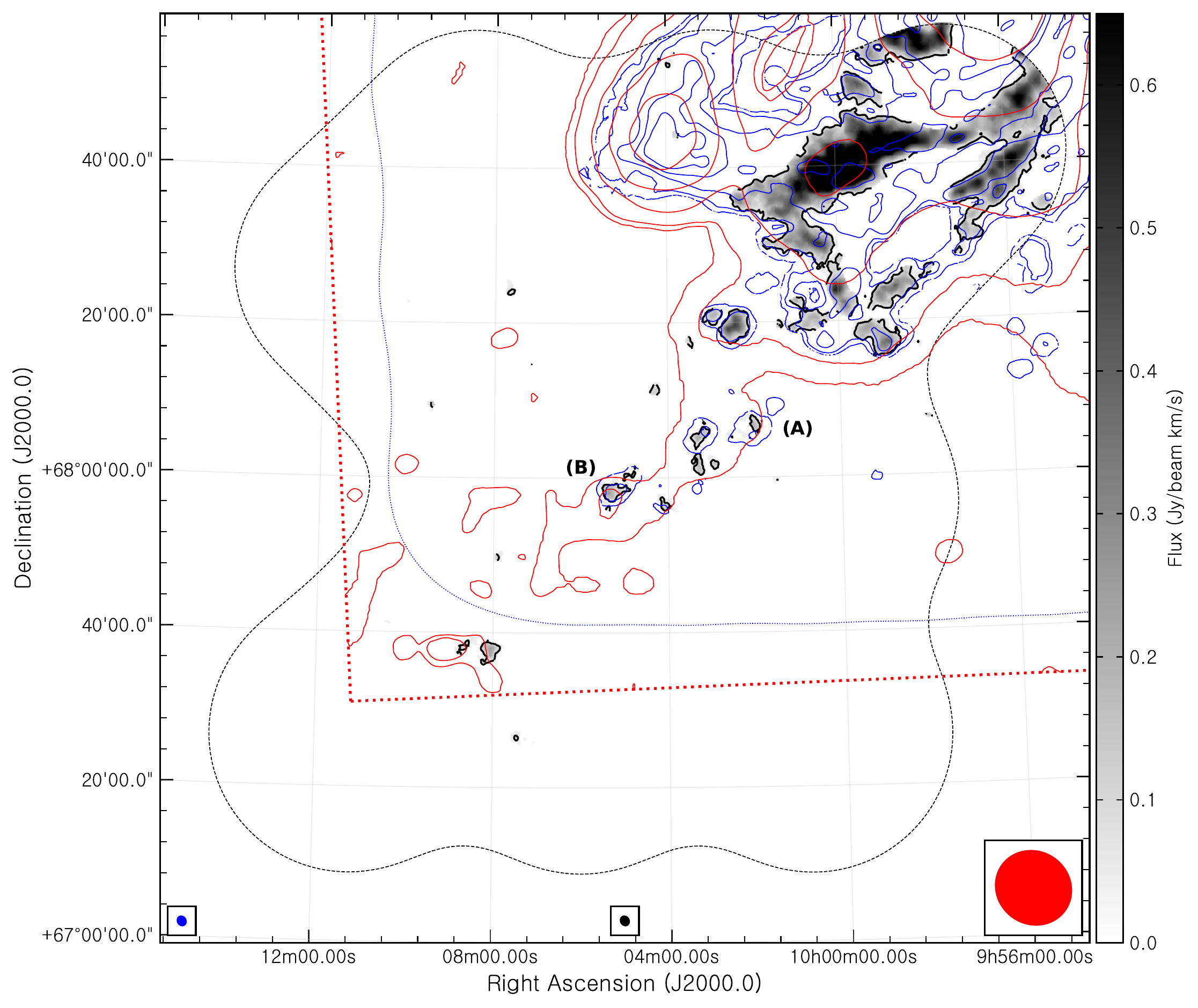}
\caption{Comparison of emission detected in the D-array AW683 mosaic,
  the GBT observations and our D-array mosaic. Grayscale shows the
  emission detected between $-120$ and $-80$ \kms in the AW683
  mosaic. The grayscale runs from 0 (white) to 0.65 (black) Jy
  beam$^{-1}$ \kms.  The corresponding black contour shows the $1.2
  \cdot 10^{19}$ \cm column density level. Red and blue contours show
  the GBT and VLA emission at identical repective column density
  levels of $(4.5,22.5,90,225,450)\cdot 10^{18}$ cm$^{-2}$
  (cf.\ Fig.\ \ref{fig:GBT}). The different \HI distribution in the
  M81-NGC 3077 bridge is due to the different velocity ranges
  displayed. The blue dotted curve indicated the 50 percent sensitivity
  level of our mosaic, the black curve that of the 15-pointing AW683
  mosaic. The red dashed lines show the extent of the GBT survey
  area. Beams are indicated at the bottom using the respective contour
  colours.  The letters indicate the cloud complexes described in
  Sect.\ \ref{sec:clouds}. \label{fig:clouds}}
\end{figure*}

Inspection of the AW683 data cube clearly shows the presence of clouds
A and B. We show the zeroth-moment map derived from these data in
Fig.\ \ref{fig:clouds}, in combination with the corresponding maps
from the GBT and our mosaic. It is clear that the low-column density
structure detected by the GBT at the edge of the survey area coincides
with a clump of \HI in the extended area mosaic. For this clump, we
find a mass of $6.3 \cdot 10^6$ \msun, comparable to that of the A and
B clumps.

The velocities of these clumps are all close to that of the more
prominent \HI features in this general area, suggesting that they are
tidal debris from the triplet interactions.  There are no other new
\HI clumps of comparable flux in this area.  We find a number of
marginal detections of smaller clumps, but deeper observations will be
needed to confirm whether these are real.

The \HI masses of the SE clumps are larger than those of the smallest
dwarf galaxies that have been detected in \HI in the Local Volume. An
example is Leo P, a low-mass, gas-dominated galaxy with an \HI mass of
$8.1 \cdot 10^5$ \msun \citep{mcquinn15}. It, and other galaxies like
it, are known to contain dark matter \citep{bernstein15}, and
potentially could help solve some of the problems that exist in
small-scale $\Lambda$CDM, such as the ``missing satellites'' problem
(see, e.g., \citealt{kauffmann93} for an early discussion). Similarly,
ultra-compact high-velocity clouds (UHVCs), which have similar \HI
masses and sizes as the SE clouds, are thought to be candidate
low-mass galaxies harbouring dark matter halos \citep{adams16}.

\begin{figure}
\includegraphics[width=0.8\hsize]{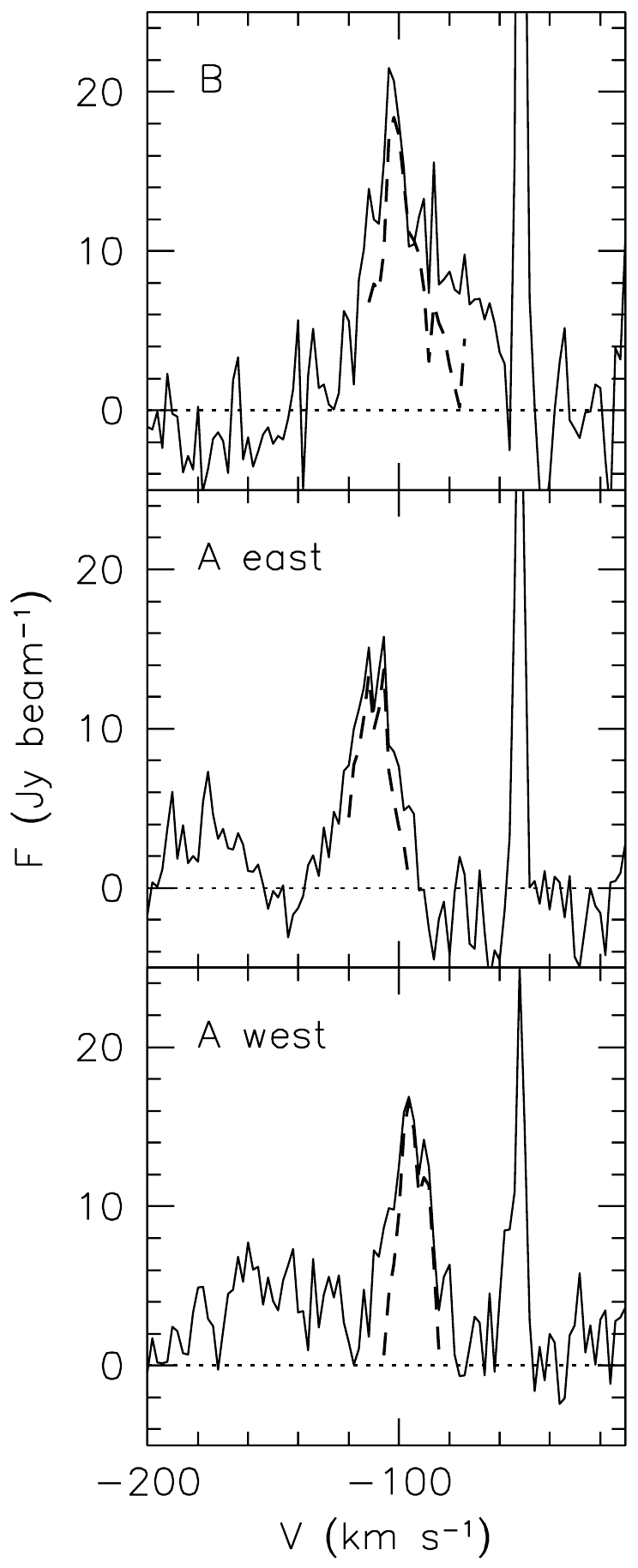}
\caption{Velocity profiles of clouds A east, A west and B, integrated
  over their area as shown in Fig.\ \ref{fig:GBT}.  Full profiles are
  derived from the unmasked D-array cube, dashed profiles from the
  masked version. The peak at $\sim -50$ \kms is due to Galactic
  foreground emission.
  \label{fig:profs}}
\end{figure}

An interesting question to explore is whether the SE clumps have
properties consistent with low-mass galaxies or UHVCs. We extract
integrated velocity profiles of clouds A and B from the D-array cube,
where we treat cloud A as two separate objects, hereafter ``A east''
and ``A west''. We use the D-array zeroth-moment map as a mask to
define the area over which to extract the profiles. We extracted
profiles from both the masked cube (used to create the moment maps)
and the unmasked cube. The three sets of profiles are shown in
Fig.~\ref{fig:profs}. The profiles are narrow and well-defined. For
the A-clouds, the masked and unmasked profiles agree well with each
other. For the B-cloud, the unmasked profile is significantly wider
than the masked profile. Inspection of the cube shows that this extra
emission is due to a nearby diffuse \HI feature, though it is not
clear whether this emission belongs to the cloud (and is extra
confirmation that these clouds do not exist in isolation).

We measure the velocity widths $W_{20}$ and $W_{50}$ of the masked and
unmasked profiles at 20 percent and 50 percent of the peak flux. These
are listed in Table \ref{tab:mdyn}. We use these widths to calculate
indicative dynamical masses $M_{\rm dyn}\sin i = (W/2)^2 R / G$ of
these clouds, where $W$ is the velocity width $W_{20}$ or $W_{50}$,
$R$ is the radius of the cloud, $G$ is the gravitational constant and
$i$ the inclination of the cloud.  For the radius we simply use half
of a cloud's extent along its major axis. These radii are also listed
in Table \ref{tab:mdyn}, along with the resulting indicative dynamical
masses.  These values assume the clouds are fully gravitationally (rotationally)
supported.  Other estimates for the dynamical mass, which assume that
the clouds are fully or partially pressure supported, tend to give
higher values for $M_{\rm dyn}$, so the values given
in Table \ref{tab:mdyn} are lower limits.

Comparing the \HI masses with the indicative dynamical masses (we find
no evidence for a stellar component in SDSS images), we find that the
dynamical to \HI mass ratios are high: for the $W_{20}$ values
we find an average ratio of $59.8 \pm 21.7$ (where we have omitted the
value of the unmasked B-cloud profile), and for the $W_{50}$ values we
find a ratio of $23.7 \pm 11.4$.

Using the GBT \HI masses instead, would decrease these ratios by a
factor $\sim 3$ (cf.\ Sect.\ \ref{sec:clouds}), but this would still
give ratios substantially larger than unity.

At first glance, these results would indicate that the clouds are
dark-matter-dominated, and good candidates for low-mass galaxies.
Indeed, the velocity widths, \HI masses and dynamical masses are fully
consistent with those of the UHVC candidates and the dwarf galaxies
Leo T and P, as described in \citet{adams16}.

\begin{deluxetable*}{llccccrr}
  \tabletypesize{\scriptsize} \tablewidth{0pt} \tablecaption{Comparison of properties of \HI clouds\label{tab:mdyn}}
  \tablehead{\colhead{Cloud}&\colhead{Type} & \colhead{$M_{\rm HI}$} & \colhead{$R_{\rm HI}$} & \colhead{$W_{20}$} & \colhead{$W_{50}$}& \colhead{$M_{\rm dyn}^{W20}\sin i$} & \colhead{$M_{\rm dyn}^{W50}\sin i$}\\[2pt]
\colhead{} & \colhead{} & \colhead{$(\times 10^6$\,\msun)} & \colhead{(kpc)}  & \colhead{(\kms)} & \colhead{(\kms)} & \colhead{($\times 10^6$\,\msun)} & \colhead{$(\times 10^6$\,\msun)}}
\startdata
A east & masked   & 2.0 & 2.4 & 25 & 15 & 87.2 & 31.4\\ 
       & unmasked & 2.0 & 2.4 & 34 & 24 & 161.3 & 80.3\\ 
A west & masked   & 2.8 & 3.1 & 22 & 14 & 87.2 & 35.3\\ 
       & unmasked & 2.8 & 3.1 & 32 & 20 & 184.5 & 72.1\\ 
B      & masked   & 3.2 & 3.3 & 36 & 17 & 248.6 & 55.4\\ 
       & unmasked & 3.2 & 3.3 &(58)\tablenotemark{a}& 24 & 645.2 & 110.5 \\
\enddata
\tablenotetext{a}{This value is probably affected by emission not belonging to the cloud.}
\end{deluxetable*} 

There is, however, one big difference. The clouds discussed here are
not isolated, but part of a larger, tidal structure. In contrast, the
UHVC candidates and Leo T and P galaxies are isolated.  Interpreting
their velocity width is therefore less ambiguous than for clouds A and
B whose velocity widths could also be explained by processes other
than rotation. For example, the tidal arm in which the clouds are
embedded shows a north-south velocity gradient of $\sim 50$ \kms, and
part of this gradient could be reflected in the clouds. Streaming
motions of the gas in the arm are a possibility as well. Finally, as
discussed earlier, it is not clear whether the clouds are separate,
isolated physical entities. More likely they are density enhancements
in the tidal arm that could possibly evolve into tidal dwarfs.

The fact that the clouds are only observed in or near tidal
structures, and not in the rest of the mosaic volume (which has after
all been observed to the same depth and sensitivity) is another
argument against the interpretation of the velocity widths as evidence
for these clouds being dark-matter-dominated objects.  We therefore
find no unambiguous evidence for dark-matter
dominated clouds in our survey volume.

\section{Star formation, Damped Lyman-$\alpha$ absorbers and velocity dispersions}

\subsection{Star formation in and around the M81 triplet\label{sec:sf}}

So far, we have discussed the \HI distribution in and around the M81
triplet focusing on the dynamics, and implications for cloud masses
and dark matter content. \HI is, however, also strongly linked to star
formation. For example, in many nearby galaxies, whenever the \HI
column density in the disk reaches a certain density threshold, star
formation is seen to occur (e.g., \citealt{skillman87}). More
recently, similar thresholds have been observed also in the outer
disks of galaxies using deep H$\alpha$ imaging (e.g., \citealt{ferguson98}) or
GALEX ultraviolet observations (e.g., \citealt{bigiel10}).

Star formation is linked to \HI through the molecular phase of
  the gas. Recent detailed studies of the molecular gas in the triplet
  galaxies show a large variety in the morphology and kinematics of
  this gas. CO observations of M82 show that the molecular gas extends
  far beyond the disk, with the well-known gas outflows dominating the
  distribution \citep{leroy15}. The CO in M81 is much less prominent
  and mainly associated with the dust in the spiral arms
  \citep{sanchez11,brouillet91}.  The CO in NGC 3077 is discussed in
  \citet{walter02b} and \citet{meier01}. As is the case with the \HI
  as well, much of the molecular gas associated with NGC 3077 is found
  outside the optically bright part of the galaxy
  \citep{walter99,walter06}.

As for the stars themselves, the M81 triplet is close enough that its luminous stellar populations
can be resolved using large ground-based telescopes (e.g.,
\citealt{mouhcine09, barker09, okamoto15}).  These studies find that
the young stars closely follow the \HI distribution as presented in
\citet{yun94}, while the older stars are found in extended, tidally
disturbed halos, well beyond the \HI distribution.

Here we re-examine this correlation using the new \HI observations
presented in this paper and deep resolved star photometry obtained
with the Subaru 8m telescope and the Suprime-Cam instrument
\citep{miyazaki02}.  This instrument consists of 10 CCDs of $2048
\times 4096$ pixels arranged in a $2 \times 5$ pattern, with a pixel
scale of $0.2''$ and a total field of view of approximately $34'
\times 27'$ (including long edge inter-chip gaps of 16-17
arcsec and short edge gaps of 5-6 arcsec).  The dataset consists of 7
distinct pointings with Suprime-Cam, covering a total area of
$\sim 1.8$ square degrees, with exposure times of $\sim
6300-7600$s in the Johnson $V$-band and $\sim 4300-4500$s in the
Sloan $i'$-band.  The seeing was in the range $0.7''$-$1.1''$.  One
field was obtained during the period 2005 January 7-8, while the
remaining six fields were obtained during 2010 January 16-18.  All
data were acquired, processed and calibrated onto Johnson-Cousins $V, I$
using the same set-up and procedure
as the 2005 data, which have previously been presented in
\citet{barker09}.  Finally, following \citet{bernard12}, PSF-fitting
photometry was performed on the individual exposures using the standalone
version of the \textsc{daophot/allstar/allframe} suite of programs
\citep{stetson94}.  Extinction corrections were applied to individual
sources using the maps of \citet{schlegel98} and the coefficients
provided by \citet{schlafly11} for an $R_V=3.1$ reddening law.

To isolate OB stars within the object catalog, we select all
high-quality stellar sources with extinction-corrected magnitudes $I_0
< 25.25$ and extinction-corrected colors $ -1.0 < (V-I)_0 < 0.2$.  As
shown by \cite{barker09}, these cuts isolate main sequence stars with
ages of 10--150 Myr.  To compensate for the worse seeing conditions
during the observations of the central field covering the northern
part of M81, leading to a lower completeness fraction in the selected
color and magnitude ranges, we applied a small correction factor of
3.9 to the weighting scheme in this region.  In order to facilitate a
direct comparison with our \HI data, we create an image of the young
stellar distribution by representing each object with a Gaussian equal
to the natural-weighted beam size of the \HI data and scale the
amplitude using the $V$-band brightness. This better illustrates the
relative density and brightnesses variations of the young star
population compared to simply plotting the individual star counts.

\begin{figure*}
  \centering
\includegraphics[width=0.9\hsize]{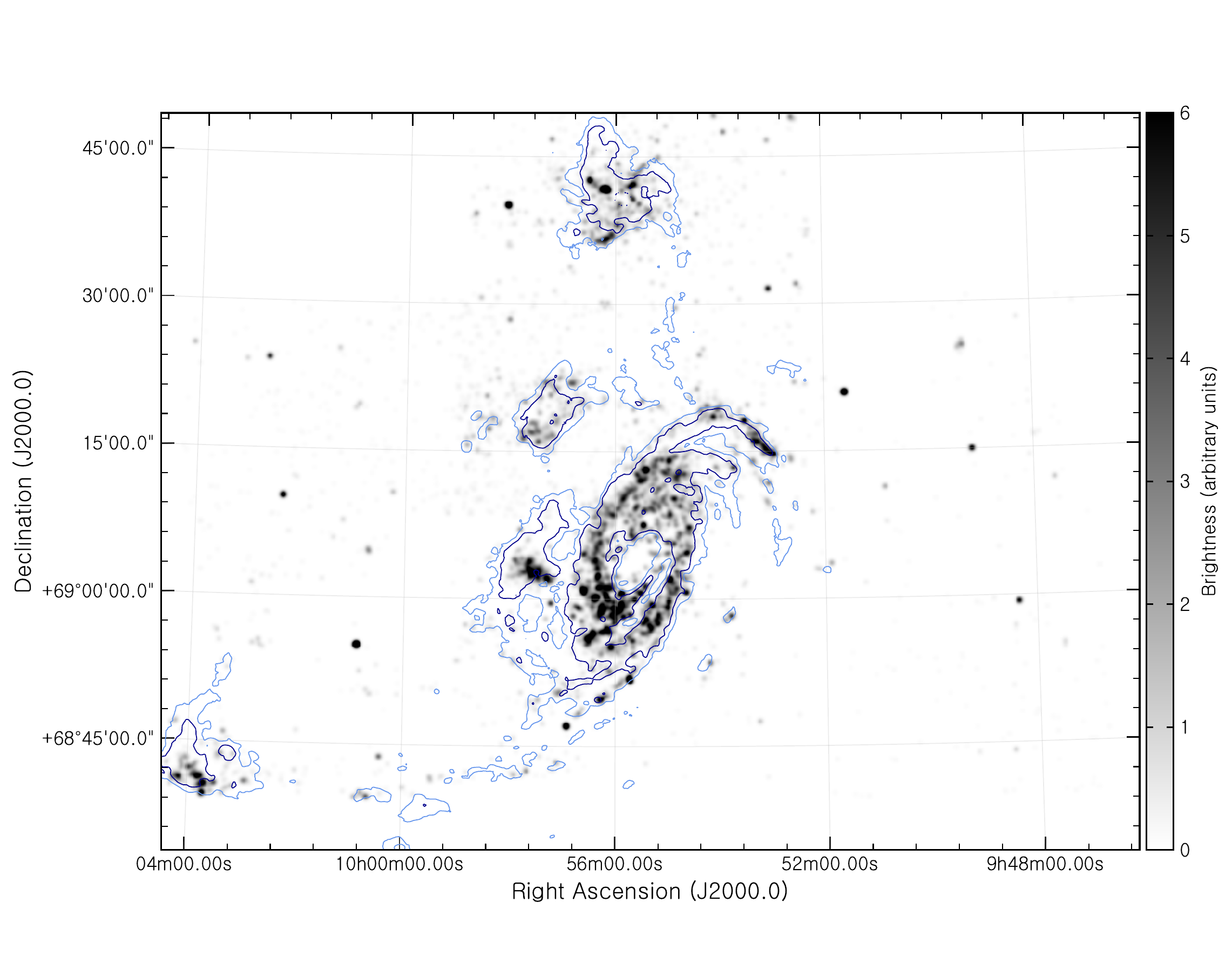}
\caption{Luminosity-weighted distribution of OB stars from the Subaru
  Suprime-Cam catalog.  Note the two stellar arcs above and below the
  M82 disk. Overplotted are the $3 \cdot 10^{20}$ cm$^{-2}$ (light
  blue) and the $6 \cdot 10^{20}$ cm$^{-2}$ (dark blue) \HI column
  density contours.
  \label{fig:obstar}}
\end{figure*}

The luminosity-weighted distribution of OB stars is shown in
Fig.\ \ref{fig:obstar}. An interesting feature is the pair of arcs
that can be seen above and below the disk of M82. These arcs were also
noted by \citet{okamoto15}, and the southern arc was also discussed in
detail in \citet{davidge08} and \citet{sun05}.  \citet{davidge08}
estimates that the stellar mass of the arc is between $3\cdot 10^5$
\msun and $2\cdot 10^6$ \msun, with an age of $\sim 5 \cdot 10^7$
years, consistent with the typical interaction timescale in the
triplet. \citet{davidge08} speculates that the arc could have formed
due to triggered star formation as a result of the M82 outflows and
notes that it coincides with a similar arc seen in GALEX UV-data
\citep{hoopes05}.  A comparison of Fig.\ \ref{fig:obstar} with our
integrated \HI map shows that the arc also coincides with a ridge in
the \HI surface density. The northern arc, while equally well defined
in the young stars distribution, does however show no such
corresponding features in the UV or \HI.

A comparison of the stellar distribution in Fig.\ \ref{fig:obstar}
with the \HI distribution in Fig.\ \ref{fig:mom0} shows a close
correspondence between the OB star concentrations and the majority of
the \HI peaks. Some peaks can still be identified that do not have
corresponding star formation, but due to the time scale needed for
star formation, one would not expect a one hundred percent match.
This correspondence is of course a well-known result, first
demonstrated in the early work by \citet{skillman87}, indicating that
a minimum column density is needed for star formation. Much work has
been done in trying to quantify and understand this star formation
threshold. A full review of this field is beyond the scope of this
paper, and we refer to the overview by \citet{kennicutt12}. Estimates
of the star formation threshold value usually arrive at values of a
few times $10^{20}$ cm$^{-2}$. A derivation on theoretical grounds is
given in \citet{schaye04}, who argue for a value of $3-10 \cdot
10^{20}$ cm$^{-2}$ over a range in gas fractions, metallicity, UV
radiation intensity and amount of turbulence.

In Fig.\ \ref{fig:obstar}, we overplot the $3 \cdot 10^{20}$ cm$^{-2}$
contour on the young stellar distribution. Projection effects due to
the geometry of the M81 triplet increase the observed surface
densities from those that would be seen when face-on.  Although it is
difficult to correct for this for the triplet as a whole, we know that
in the M81 disk, with an inclination of 57$^{\circ}$, the observed
surface densities are close to a factor of two higher than the face-on
ones. We therefore also overplot the $6 \cdot 10^{20}$ cm$^{-2}$
contour.

The latter contour tightly encloses the star formation in the main
galaxies, while the lower $3 \cdot 10^{20}$ cm$^{-2}$ contour also
encompasses some of the fainter stars in the bridge between M81 and
NGC 3077. Looking at the stellar luminosities we find that the $6
\cdot 10^{20}$ cm$^{-2}$ contour encompasses $\sim 65$ percent of the
total OB star luminosity, while the $3 \cdot 10^{20}$ cm$^{-2}$
contour contains $\sim 81$ percent of the total OB star flux.  These
results are consistent with the findings of \citet{barker09}, who
examine the evidence for a star formation threshold by comparing the
\HI and OB star distribution in M81 over a significantly smaller
area. The fact that the relation between the number and luminosities
of OB stars and the \HI column density holds for the entire field
analysed here suggests that star formation threshold is indeed a local
phenomena, independent of whether the gas and stars are in a disk
environment or not.

\subsection{Comparison with Damped Lyman-$\alpha$ absorbers}

Direct observations in emission of the cosmological evolution of \HI
in galaxies at high redshifts will probably stay outside our reach
until the advent of future radio telescopes such as the Square
Kilometre Array.  Absorption line observations towards quasars, on the
contrary, are already giving us valuable clues about high-redshift
\HI. Much of the information is gleaned from absorption lines in
Lyman-$\alpha$. These reveal the signature of many intervening \HI
clouds over a large range in column density. The detections with the
highest column densities, $N_{\rm HI} > 2 \cdot 10^{20}$ cm$^{-2}$,
are known as Damped Lyman-$\alpha$ Absorbers (DLAs, \citealt{wolfe86};
see also the review by \citealt{wolfe05}).

The DLAs are thought to probe the high-$z$ equivalents of gas disks
observed in 21-cm at $z=0$ \citep{wolfe95,wolfe05}.  Although many of
the observed characteristics of the DLAs could be reproduced by
simulations (e.g., \citealt{rahmati14}), the single largest
discrepancy was the inability to reproduce the observed distribution
of velocity width measurements, quantified by the $\Delta V_{90}$
parameter. For these absorption observations, this parameter is
defined as the difference between the velocities where the cumulative
optical depth profile crosses the 5 percent and the 95 percent levels.

Early simulations obtained distributions peaking at low $\Delta
V_{90}$ values of $\sim 50$ \kms without the high velocity-width tail
of $\Delta V_{90} > 100$ \kms found in the observations (e.g.,
\citealt{pontzen08}). This implied that more energy was needed to
increase the gas velocities. The favored method of achieving this is
through feedback (e.g., stellar winds, superwinds). Simulations which
included post-hoc feedback showed that this could substantially
increase the observed $\Delta V_{90}$ distribution (e.g.,
\citealt{tescari09}).  Later simulations have added in these feedback
prescriptions and can now much better reproduce the observed $\Delta
V_{90}$ distribution (e.g., \citealt{barnes14,bird15}). \citet{bird15}
show that in the Illustris simulation, the largest $\Delta V_{90}$
objects are found to come from sightlines intersecting multiple
individual HI structures.

If feedback does indeed play a major role in determining the velocity
widths of the DLAs, then an interesting question is whether we can
find low-redshift counterparts of the DLAs and study in detail the
processes giving rise to the observed $\Delta V_{90}$ distribution.
Using the \HI emission data from the THINGS survey, \citet{zwaan08}
derived the probability distribution function of $\Delta V_{90}$ of
galaxies in the local universe.  This can be directly compared with
the DLA $\Delta V_{90}$ distribution if the low-ion metals used to
trace the DLAs are distributed like the neutral gas. \citet{zwaan08}
argue that this is a reasonable assumption, though \citet{prochaska02}
note that small-scale spatial variations in metallicity might affect
this comparison. Like \citet{prochaska02}, we here proceed under the
assumption that these differences are not important to the kinematical
properties we are studying here.

\citet{zwaan08} find that the distribution of $\Delta V_{90}$ for the
THINGS galaxies is sharply peaked around $\sim 30$ \kms, a factor two
lower than the observed peak in the distribution for higher-redshift
DLAs as found by \citet{prochaska08}. They use the \citet{yun93a} \HI
data, along with VLA archival data, to determine the distribution of
$\Delta V_{90}$ in M82 and find values similar to those found in
DLAs. They conclude that superwinds and tidal features can to a large
extent explain the higher $\Delta V_{90}$ values in DLAs.  The
presence of tidal features and outflows in the M81 triplet means it
could be considered a local example of such high-redshift objects. Our
observations cover the full spatial as well as spectral extent of the
triplet, and are thus suitable to derive the M81 triplet velocity
width distribution function which we can directly compare with the
\citet{zwaan08} and DLA results.

We therefore revisit the $\Delta V_{90}$ analysis with our M81 triplet
data set as well as a more recent DLA analysis by
\citet{neeleman13}. The latter study is based on high-resolution
spectra of 100 \HI-selected DLA systems, and includes a careful
treatment of selection effects.

For the analysis we use the masked natural-weighted C+D mosaic, but
regridded to a the pixel size equal to $34''$, or the geometric mean
of the beam size.  This means each pixel is independent. For each
spatial position we determine $\Delta V_{90}$ by constructing the
normalized cumulative velocity profile, and measuring the velocities
where the 5 percent and 95 percent levels occur.  Note that for a
Gaussian profile $\Delta V_{90}$ is equivalent to $3.28\sigma$ where
$\sigma$ is the dispersion.

\begin{figure}
  \centering
  \includegraphics[width=\hsize]{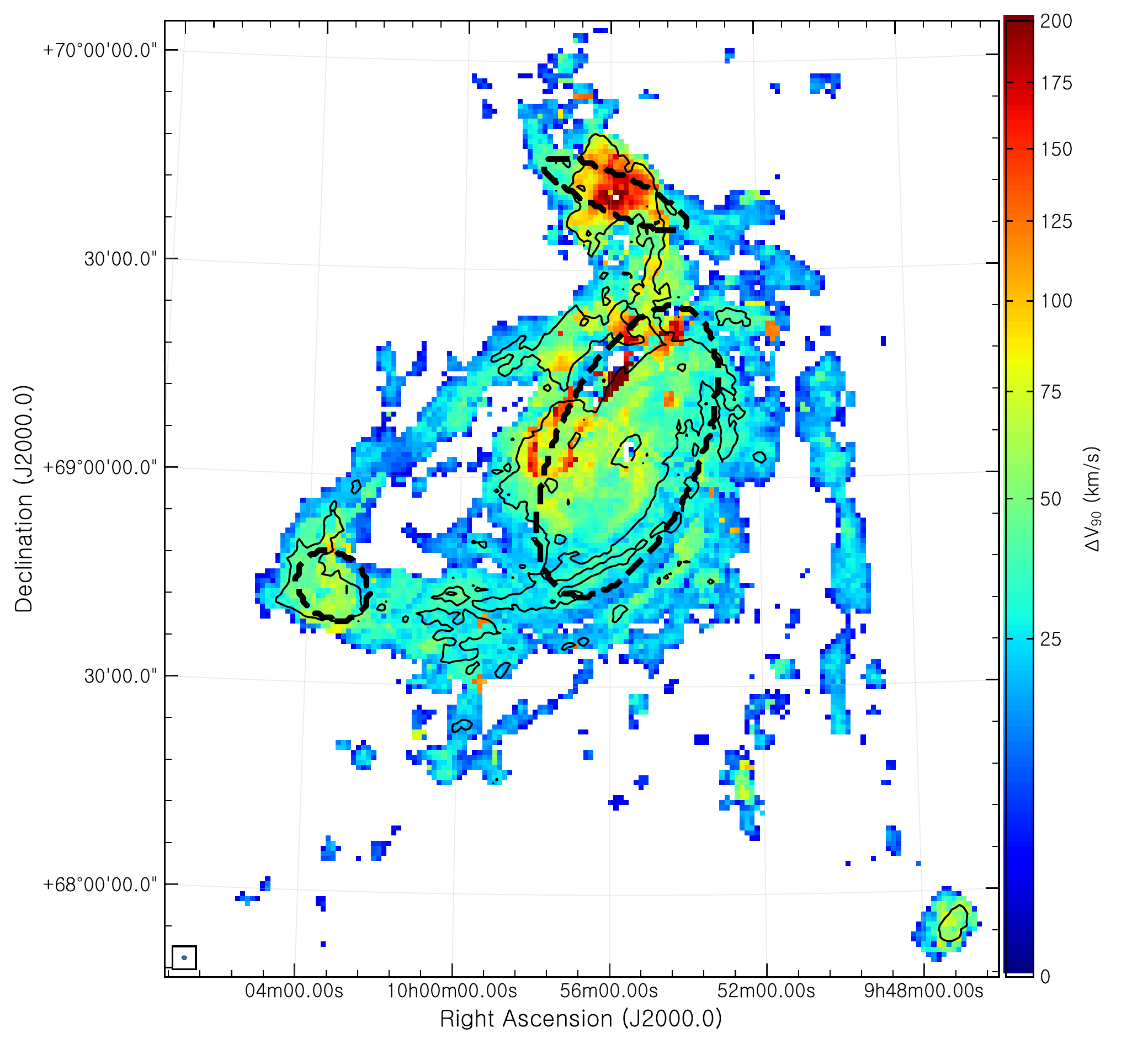}
  \caption{Map of $\Delta V_{90}$. The color map uses a square-root
    stretch. The thin black contour indicates the $2 \cdot 10^{20}$
    cm$^{-2}$ level.  The black dashed ellipses indicate the optical
    disks, defined here as having a radius of $2R_{25}$ where the
    optical positions, sizes and orientation parameters have been
    taken from HyperLEDA.
  \label{fig:v90map}}
\end{figure}

Figure \ref{fig:v90map} shows the $\Delta V_{90}$ map of the M81
triplet. It shows similar features as the second-moment map in
Fig.\ \ref{fig:mom2}, with the highest values also here occuring in
and around M82 and the interface between M81 and M82.  Typical values
in the southern part of the triplet are $\sim 20-40$ \kms, while in
the northern part these increase to $\sim 160-200$ \kms.  As discussed
in Sect.\ \ref{sec:moms}, these higher values are due to a combination
of genuinely high velocity spreads (due to outflows) and multiple
components along the line of sight.

A difference map of the $\Delta V_{90}$ and the second-moment maps
(assuming that $\Delta V_{90} = 3.28\sigma$) has a median value of 0.1
\kms. The only large differences occur in the regions with extreme
velocity dispersions ($\ga 70$ \kms). In the southern disk of M81,
where no such extreme dispersions are found, the average difference
between the maps is $\sim -0.1 \pm 0.7$ \kms, showing that this
analysis is largely independent of the way the velocity width is
defined. We will return to this in Sect.\ \ref{sec:veldispcoldens}.

Also shown in Fig.\ \ref{fig:v90map} is the $2 \cdot 10^{20}$
cm$^{-2}$ column density contour, which encloses the areas of the
system that would give rise to DLA absorption. We also indicate the
approximate areas of the optical disks (defined here as having a
maximum radius of $2R_{25}$) of the three main galaxies, using the
optical positions and orientation parameters from HyperLEDA. We use
$R_{25} = 10.9'$ for M81, $5.5'$ for M82, and $2.6'$ for NGC 3077.

Note that the majority of the DLA area (i.e., column density higher
than $2 \cdot 10^{20}$ cm$^{-2}$) occurs within the optical disks. Only
a relatively small fraction, mainly between M81 and M82, and in the
M81-NGC 3077 bridge, is found in the tidal features outside the disks.

\begin{figure}
  \centering
  \includegraphics[width=0.9\hsize]{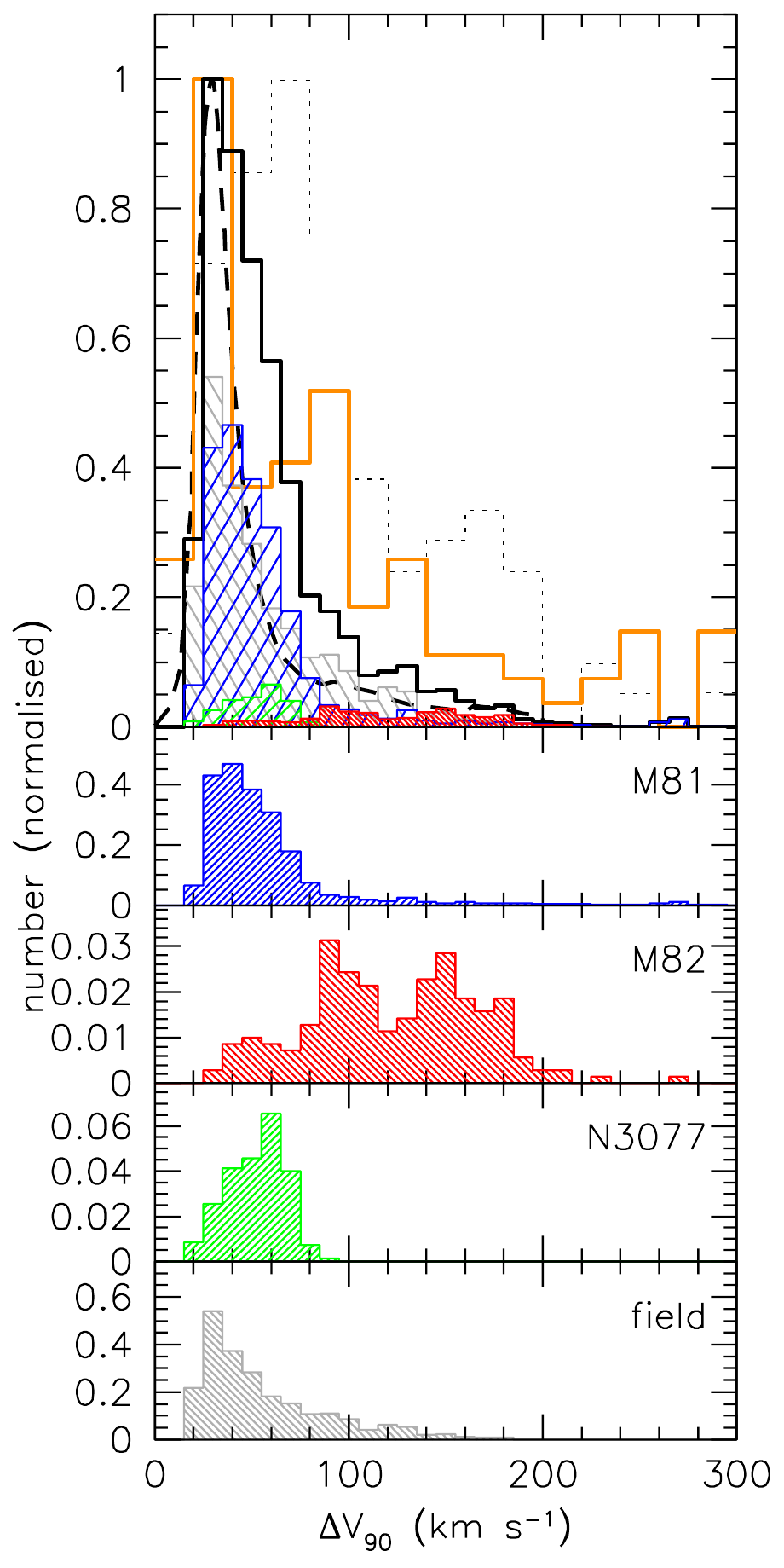}
  \caption{Histograms of $\Delta V_{90}$. The top panel shows the
    distribution for the M81 triplet selecting only column densities
    above $2 \cdot 10^{20}$ cm$^{-2}$ (thick black full lines); a
    sample of DLA absorbers from \citealt{prochaska08} (thin black
    dotted lines); a recent DLA sample from \citet{neeleman13} (orange
    full lines) and the THINGS sample (\citealt{zwaan08}, thick black
    dashed lines). The hatched histograms indicate the four
    environments found in the triplet. Blue: optical disk of M81
    within $2R_{25}$, red: M82 optical disk within $2R_{25}$, green:
    optical disk of NGC 3077 within $2R_{25}$, gray: field, i.e., any
    part of the triplet not included in the optical disks. The four
    bottom panels show zoom-ins on each of the four environment
    histograms. Histograms are created using maps where the pixel size
    is equal to the beam size, so that pixels are independent.
  \label{fig:v90histo}}
\end{figure}

Figure \ref{fig:v90histo} shows the histogram of $\Delta V_{90}$
values as found within the $2 \cdot 10^{20}$ cm$^{-2}$ contour in
Fig.\ \ref{fig:v90map}. Also shown are the $\Delta V_{90}$
distribution as found by \citet{zwaan08} for the THINGS galaxies, and
the measured values for DLA absorbers as presented in
\citet{prochaska08}, as well as the more recent measurements from
\citet{neeleman13}.  The M81 triplet is characterised by a sharp peak
and a long tail towards higher velocities. The peak occurs at $\sim 30$
\kms, identical to what was found for the THINGS galaxies. This
reflects the fact that in terms of area, the triplet is dominated by
the M81 disk, a rotationally supported disk just like the THINGS
galaxies. The \citet{neeleman13} DLA values also strongly peak at
$\sim 30$ \kms. They suggest that this peak corresponds to the
``low-cool'' part of the DLA population. These DLAs are thought to
reside in smaller dark matter halos, where infalling gas is not
shock-heated (``cold accretion'') resulting in the formation of
neutral gas disks \citep{wolfe08}. This is consistent with the current
properties of M81 and the other THINGS galaxies.

The tail towards higher velocities is mainly caused by the
high $\Delta V_{90}$ values found near M82. These values cover a
similar range in high $\Delta V_{90}$ values as found for the DLAs.
We can illustrate this in more detail by looking at the $\Delta
V_{90}$ distribution as a function of environment. We define three
disk environments using the definitions of the optical disks of the
three major galaxies as shown in Fig.\ \ref{fig:v90map}, and select
the high-column density areas within these disks.  The other
high-column density material outside these disks we define as being in
a field or intra-disk environment. The $\Delta V_{90}$ histograms of
each of these four environments are shown in Fig.\ \ref{fig:v90histo}.

The M81 and the field distributions are similar at low velocities.
The high-velocity tail of the field distribution is due to gas outside
the optical disk along the minor axis of M82 with intrinsically higher
dispersions, and gas in the M81-M82 interface with to multiple
components along the line of sight (cf.\ Sect.\ \ref{sec:moms}). It is
interesting that the same origin for the high $\Delta V_{90}$ was
recently also found in results from numerical simulations
\citep{bird15}.

The range in values found for NGC 3077 is similar to that for M81,
though the peak of the NGC 3077 distribution occurs at $\sim 60$ \kms.
This is mostly due the fact that the NGC 3077 ``disk'' as indicated in
Fig.\ \ref{fig:v90map} encompasses part of the NGC 3077-M81 bridge, where
dispersions are higher due to the presence of multiple components 

M82 exhibits a much larger range in $\Delta V_{90}$ values which
mostly reflect intrinsically high dispersions. It covers the same
velocity range as the DLAs.

Note that as M82 can be regarded as an edge-on disk, \emph{some} of
the high $\Delta V_{90}$ values found there can be due to rotational
broadening of the emission line. In a rotating disk, this effect will,
however, only be important for lines of sight close to the (projected)
center of the galaxy. Figure~\ref{fig:v90map} shows that the area with
high-$\Delta V_{90}$ values is much larger, and also extends further
vertically than radially. Close inspection of Fig.\ \ref{fig:v90histo}
shows that for $140$ \kms $ < \Delta V_{90} < 200$ \kms, only about
half of the high-$\Delta V_{90}$ values are found in the M82
environment as defined above. The other half occur outside this
environment, but still in the general vicinity of M82. We conclude
that the majority of high-$\Delta V_{90}$ values found near M82 are
associated with outflows, with rotational broadening playing a less
important role.


Figure \ref{fig:v90histo} thus shows that the M81 triplet can
reproduce most of the velocity widths found in DLAs. The only
exception are velocity widths $\ga 200$ \kms, which are very rare in
the triplet, but which do occur in DLAs. Nevertheless, we can ask the
question in how far we can reproduce the observed DLA $\Delta V_{90}$
distribution using only M81-like and M82-like galaxies. Restricting
ourselves to $\Delta V_{90} < 200$ \kms, we find that an increase of a
factor between 2 and 5 of the number of high-$\Delta V_{90}$ values as
found near M82, gives a distribution close to that of the
\citet{neeleman13} DLA distribution.

\subsection{Velocity dispersion and column density\label{sec:veldispcoldens}}

\begin{figure}
  \includegraphics[width=0.9\hsize]{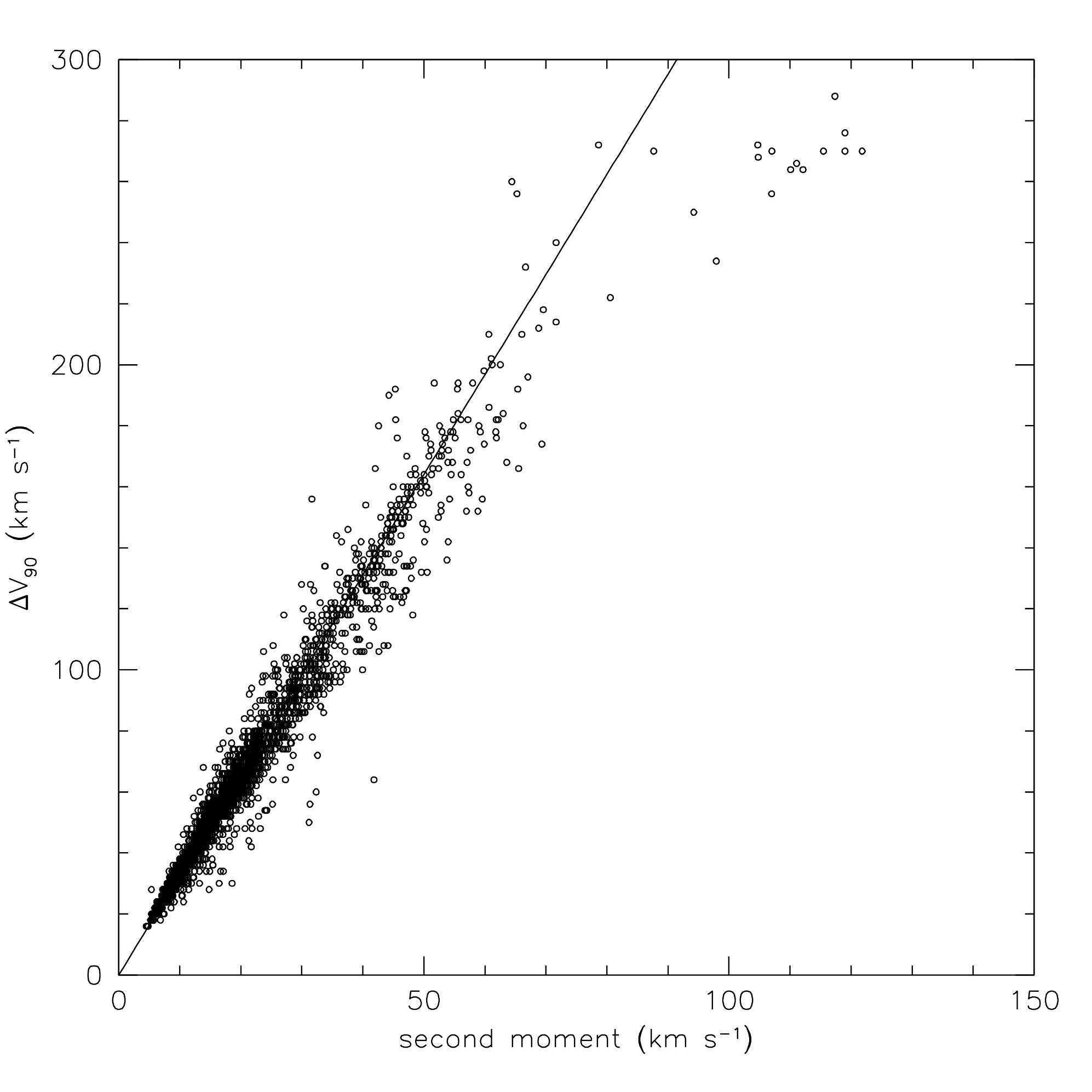}
  \caption{Plot of $\Delta V_{90}$ against the corresponding
    second-moment value. The line indicates the relation $\Delta
    V_{90} = 3.28 \sigma$ expected for a Gaussian profile, where
    $\sigma$ is the dispersion (second moment) of the profile.
    \label{fig:v90disp}}
\end{figure}

Focussing more on the properties of the M81 triplet itself, we can ask
what the distribution of velocity dispersions and column densities is
in the four disk/field environments defined above. We use velocity
dispersions here, rather than $\Delta V_{90}$ values, as the former
are easier to determine and can be directly related to literature
values. Also, as noted above, for a Gaussian profile $\Delta V_{90} =
3.28\sigma$. In our data set, the majority of the profiles are (close
to) Gaussian and the two types of velocity width can be directly
intercompared. This is illustrated in Fig.\ \ref{fig:v90disp}. The
dozen points with second-moment values $>100$ \kms correspond to the
high-dispersion ridge already discussed in Sect.\ \ref{sec:mom}, where
the high values are caused by distinct, multiple and widely-separated
components. These non-gaussian profiles obviously have a different
impact on the second-moment and $\Delta V_{90}$ values.

We now also include column densities below $2 \cdot 10^{20}$
cm$^{-2}$, allowing us to more fully address questions such as whether
the \HI velocity dispersion depends on column density, and whether
tidal material can be identified based purely on the column density
and/or velocity dispersion.

We use the zeroth- and second-moment natural-weighted C+D array maps
presented in Figs.\ \ref{fig:mom0} and \ref{fig:mom2}, but, as above,
regridded to a pixel size of $34''$ so that one pixel is approximately
equal to the beam size and yields an independent measurement.  We
furthermore subdivide the disks in an inner disk ($0<R<R_{25}$) and an
outer disk ($R_{25}<R<2R_{25}$).


\begin{figure*}
  \includegraphics[width=0.49\hsize]{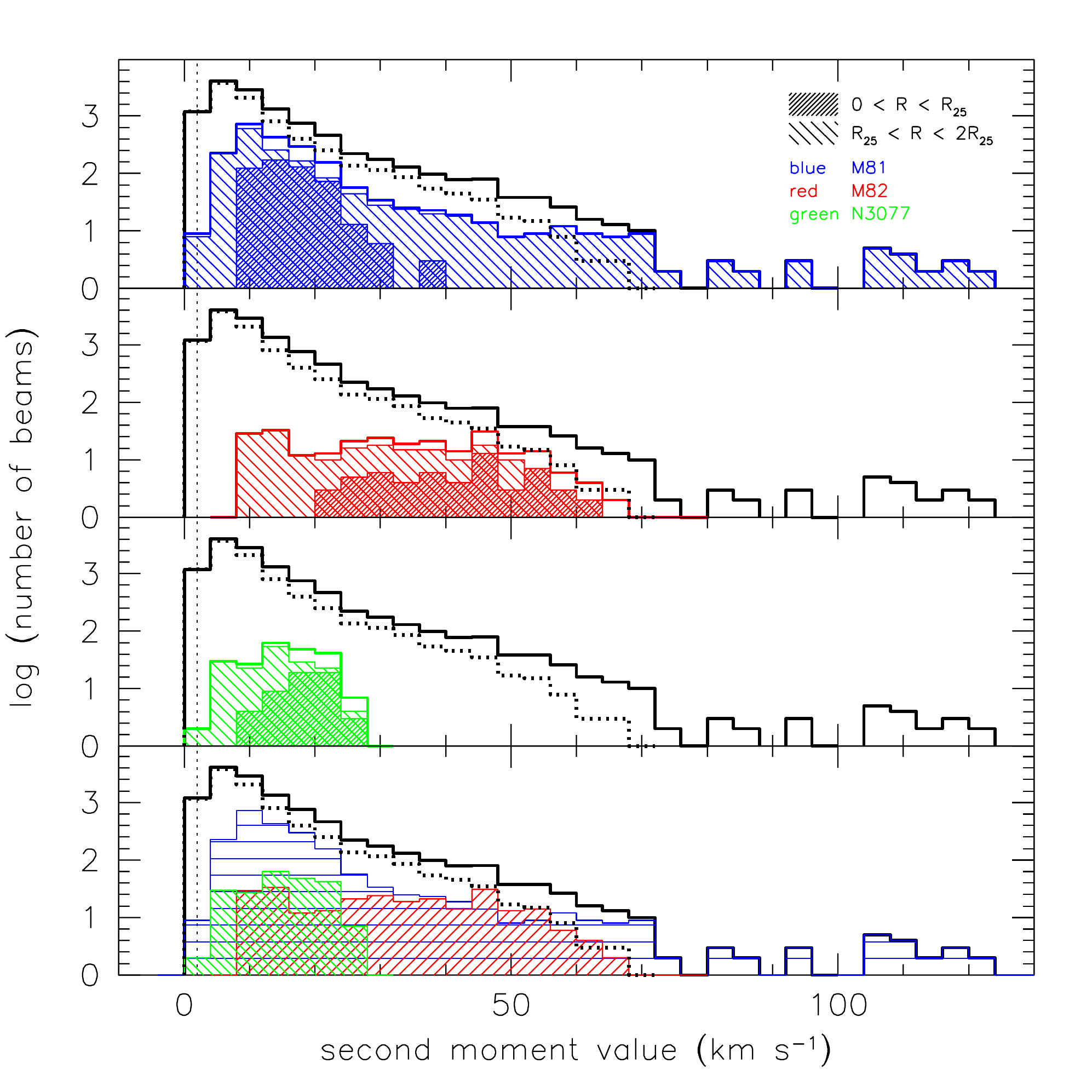}
  \includegraphics[width=0.49\hsize]{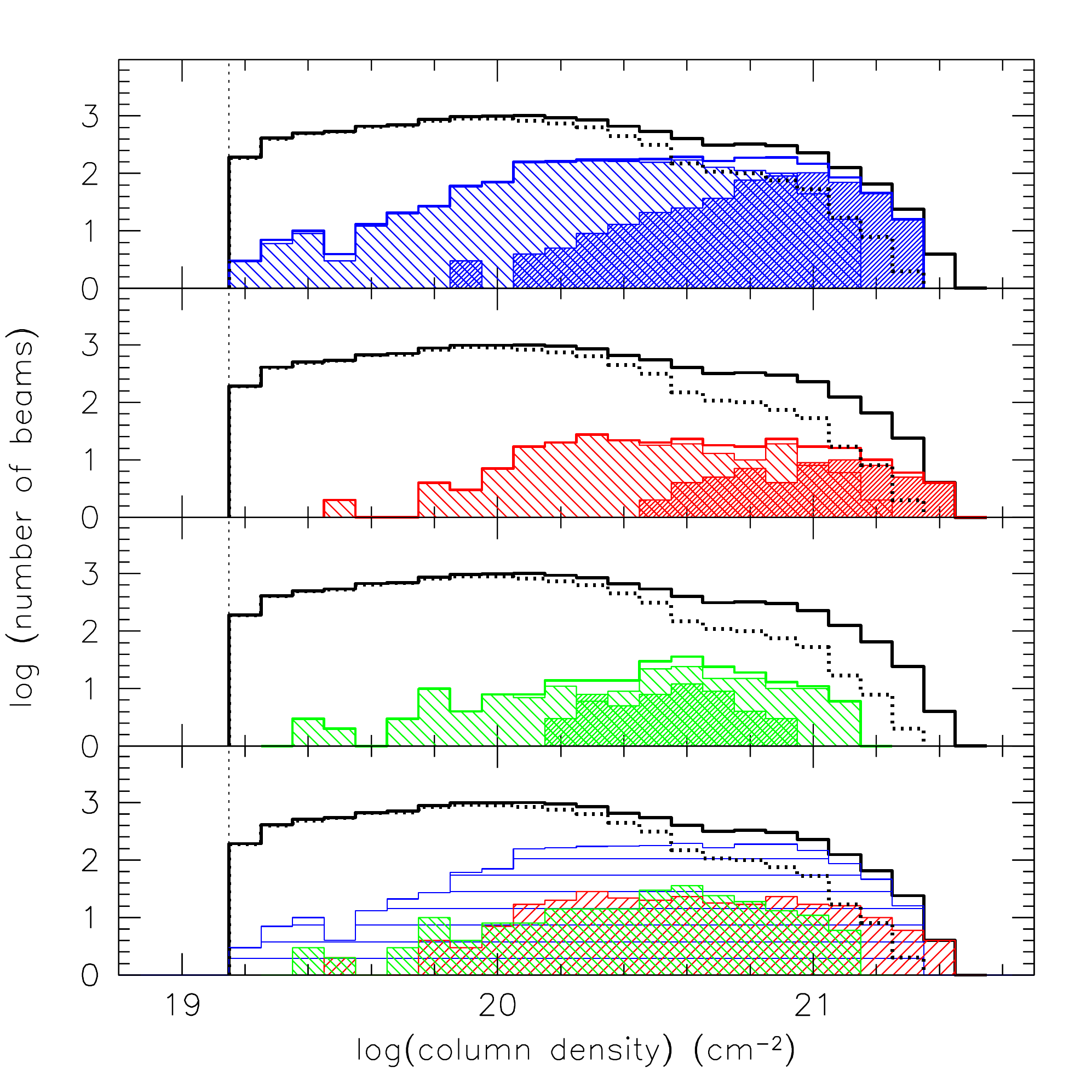}
  \caption{Histograms of the velocity dispersions (left panel) and
    column densities (right). In all sub-panels the full, thick
    histogram represents the entire observed area; the dotted
    histogram the field environment; the blue histogram in the top
    sub-panel shows the M81 disk environment; the red histogram in the
    second sub-panel from the top the M82 environment and the green
    histogram in the third subpanel from the top the NGC 3077
    environment. In these three sub-panels, single hatched histograms
    indicate the outer disk, cross-hatched histograms the inner disk.
    The bottom sub-panels compare the three environments. The blue
    horizontally-hatched histogram represents the M81 environment, the
    red $45^{\circ}$-hatched histogram the M82 environment, and the
    green $-45^{\circ}$-hatched histogram the NGC 3077 environment.
    The vertical dotted lines in all sub-panels indicate the
    respective velocity dispersion and column density limits of the data.
    \label{fig:coldispindiv}}
\end{figure*}

Figure \ref{fig:coldispindiv} shows histograms of the column densities
and velocity dispersions in the entire $3^{\circ} \times 3^{\circ}$
field. We see that the velocity dispersion histogram peaks close to
the spectral resolution limit of the datacube. The median value of the
velocity dispersion is 8.6 \kms.   M82 lacks the low velocity
dispersion peak, while NGC 3077 only shows relatively low dispersions.

\begin{figure*}
  \centering
  \includegraphics[width=0.7\hsize]{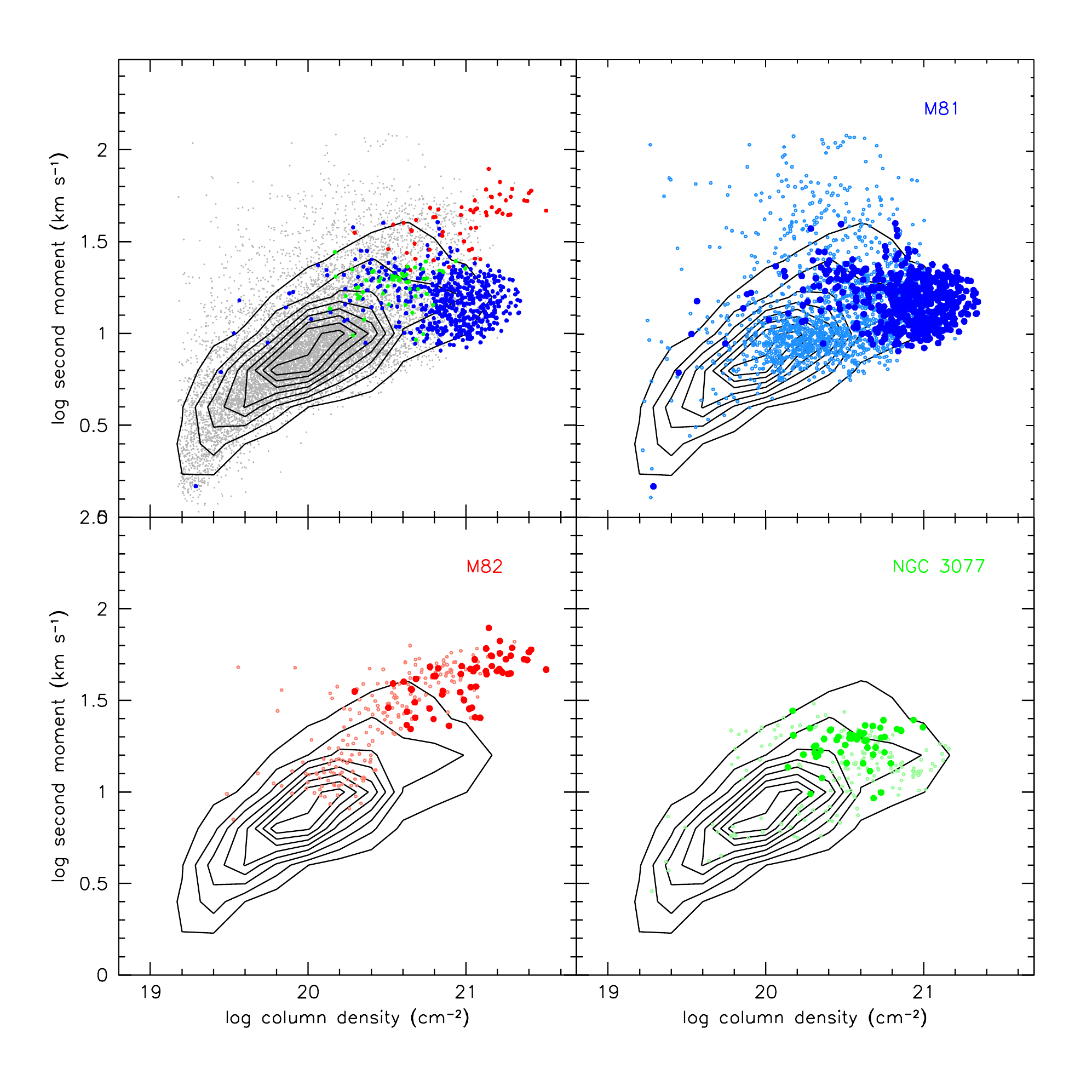}
  \caption{Pixel-pixel comparisons of column densities and velocity
    dispersions. The top-left panel compares the distribution of the
    entire mosaic (gray points, with black density contours
    overplotted), and the inner disks of M81 (blue), M82 (red) and NGC
    3077 (green). The other three panels compare the inner (large
    filled symbols) and outer disks (small open symbols) of the three
    main galaxies with the total distribution (contours). Top-right:
    M81 inner disk (blue) and outer disk (light-blue). Bottom-left:
    M82 inner disk (red) and outer disk (light-red). Bottom right: NGC
    3077 inner disk (green) and outer disk (light green).
    \label{fig:coldisp}}
\end{figure*}

In terms of column densities, we see that the field on average shows
lower column densities than the galaxies, which of course is not
unexpected from inspection of the moment maps. The column density
distributions of M81, M82 and NGC 3077 are similar in shape.

We can split these distributions up further and investigate
differences between inner and outer disk.  For M81 there are a large
number of high dispersion values in the outer disk, compared to the
inner disk. These are not physical, but indicate the presence of
multiple components, as already discussed.  In M82 the velocity
dispersions of the inner disk are higher than in the outer disk, just
as in NGC 3077. Note that the spread and average values for M82 are
significantly higher than those of NGC 3077.  The column density
histograms show that for M81 and M82 the highest column densities are
found in the inner disk. This is not the case for NGC 3077, which is
due to the offset of the main \HI component from the optical center of
the galaxy.

Finally we compare the column densities and velocity dispersions, see
Fig.\ \ref{fig:coldisp}. The top-left panel compares the column density-velocity
dispersion distribution of the entire field with those of the inner
disks of the three main galaxies. The bulk of the \HI has a column
density of $\sim 10^{20}$ and a velocity dispersion of $\sim 7$ \kms,
but there is a clear trend of velocity dispersion with column density.
The three galaxies all sit at the upper tip of this distribution, but
each occupy a different average position. The bulk of the \HI in the
inner disk of M81 has a column density of $\sim 10^{21}$ cm$^{-2}$ and
a velocity dispersion of $\sim 14$ \kms. The inner disk of M82 sits at
a similar column density level, but has a higher velocity dispersion
of $\sim 45$ \kms, though with a large spread. NGC 3077 sits at an
intermediate position with an average column density of $\sim 3 \cdot
10^{20}$ cm$^{-2}$ and a velocity dispersion of $\sim 20$ \kms.

The other three panels in Fig.\ \ref{fig:coldisp} compare the inner
and outer disks of the three galaxies with the total distribution. The
outer disk of M81 has, not unexpectedly, a lower velocity dispersion
and column density than the inner disk, but note the significant
number of points that have very high velocity dispersions.  Comparing
this with the typical column densities of the M82 high-dispersion
locations, we can conclude that high velocity dispersions at column
densities below a few times $10^{20}$ cm$^{-2}$ indicate with a high
probability that these are due to multiple gas components.

M82 also shows a clear column density-velocity dispersion relation,
though we find a significant number of outer disk points with high
velocity dispersions. This is, to a large extent, due the gas present
along the minor axis (i.e., above and below the disk), which makes
distinction between inner and outer disk somewhat artifical.

NGC 3077 shows no clear trend, and no real distinction between inner
and outer disk. In this case, interpretation is complicated further by
the offset between the \HI and optical components.

In summary, the three inner disks of the triplet galaxies occupy
distinct positions in column density-velocity dispersion space. High
velocity dispersions at high column densities most likely reflect
intrinsically high dispersions, while similar dispersions at low
column densities are due to the presence of multiple components at
different velocities. 

\section{Summary}

We have presented a new, high-resolution, 105-pointing \HI mosaic of
$3^{\circ} \times 3^{\circ}$ centered on the M81 triplet, M81, M82 and
NGC 3077 obtained with the VLA in its C- and D-configurations. This is
the first radio synthesis data set that maps the entire volume of the
triplet at high spatial and spectral resolution. These data can serve as
input for further sophisticated modeling of the interaction and
evolution of the triplet (e.g., \citealt{oehm17}).

Our main results are summarized as follows:

\begin{itemize}

\item We do not find a large population of free-floating \HI clouds
  down to an \HI mass limit of $\sim 10^4$ \msun. While small clouds
  and \HI complexes are detected they only occur close (spatially and
  spectrally) to the main \HI tidal features of the triplet,
  suggesting they are all debris of the interaction that shaped the
  triplet. A detailed investigation of the \HI masses of these clouds
  show that they are likely embedded in extended low-column density
  tidal features.

\item Comparison with a sensitive GBT \HI mosaic of the same area by
  \citet{chynoweth08} shows that the VLA mosaic has detected most of
  the \HI in the southern part of the mosaic (i.e., the southern part
  of M81 and NGC 3077). In the northern part of the mosaic (M82 and
  the the M81-M82 transition region) the GBT has detected a
  significant excess of flux most likely associated with M82. This
  probably indicates the presence of a low-column density \HI
  component associated with the M82 outflows.

\item Using additional data we show that low-column density features
  detected by the GBT beyond the south-eastern edge of our VLA mosaic
  are resolved into clouds. In turn, we detect a small \HI cloud
  beyond the extent of the GBT mosaic, suggesting that a low-column
  density \HI tail resulting from the interaction may extend
  further south-east beyond the areas mapped by the VLA and GBT.

\item A comparison of the velocity widths and \HI masses of these
  clouds seems consistent with them being dominated by dark
  matter. Their properties are, in that regard, very similar to those
  of UHVCs or the smallest gas-rich dwarf galaxies. However, given
  their association with tidal features it is more likely that the
  velocity widths should not be interpreted in terms of gravitational
  support. It is possible that these clouds will eventually evolve in
  tidal dwarf galaxies.

\item We compare the observed \HI column densities with a Subaru
  Suprime-Cam map of the resolved young stellar population of the
  triplet. The majority of the OB star distribution is found within
  the $6 \cdot 10^{20}$ cm$^{-2}$ contour. After taking projection
  effects into account, this is consistent with theoretical
  predictions for the star formation threshold surface density value.

\item We derive the distribution of $\Delta V_{90}$ of the \HI
  profiles and compare these with that observed for DLAs to
  investigate whether the triplet can be regarded as a local version
  of the high-$z$ objects that cause the DLA absorption. We find that
  the peaks of the distributions coincide at low $\Delta V_{90}$
  values, consistent with the interpretation that the low $\Delta
  V_{90}$ values occur in objects that will evolve in neutral gas
  disks. High $\Delta V_{90}$ values are found around M82, and these
  cover the entire range in $\Delta V_{90}$ found in DLAs up to 200
  \kms. This is consistent with high $\Delta V_{90}$ values being
  caused by feedback, outflows or multiple components along the line
  of sight. For the triplet to also reproduce the relative fraction of
  high- versus low- $\Delta V_{90}$ values found in DLAs, the
  frequency of the values found near M82 needs to be increased by a
  factor 2-5, presumably indicating that in DLAs the relative
  importance of feedback and outflow effects is somewhat more
  important than in the triplet.
  
\end{itemize}

\acknowledgements

We thank the referee for constructive comments that have made a
significant difference to this paper.

The National Radio Astronomy Observatory is a facility of the National
Science Foundation operated under cooperative agreement by Associated
Universities, Inc.

Funding for the Sloan Digital Sky Survey IV has been provided by the
Alfred P. Sloan Foundation, the U.S. Department of Energy Office of
Science, and the Participating Institutions. SDSS-IV acknowledges
support and resources from the Center for High-Performance Computing
at the University of Utah. The SDSS web site is www.sdss.org.

SDSS-IV is managed by the Astrophysical Research Consortium for the 
Participating Institutions of the SDSS Collaboration including the 
Brazilian Participation Group, the Carnegie Institution for Science, 
Carnegie Mellon University, the Chilean Participation Group, the French Participation Group, Harvard-Smithsonian Center for Astrophysics, 
Instituto de Astrof\'isica de Canarias, The Johns Hopkins University, 
Kavli Institute for the Physics and Mathematics of the Universe (IPMU) / 
University of Tokyo, Lawrence Berkeley National Laboratory, 
Leibniz Institut f\"ur Astrophysik Potsdam (AIP),  
Max-Planck-Institut f\"ur Astronomie (MPIA Heidelberg), 
Max-Planck-Institut f\"ur Astrophysik (MPA Garching), 
Max-Planck-Institut f\"ur Extraterrestrische Physik (MPE), 
National Astronomical Observatories of China, New Mexico State University, 
New York University, University of Notre Dame, 
Observat\'ario Nacional / MCTI, The Ohio State University, 
Pennsylvania State University, Shanghai Astronomical Observatory, 
United Kingdom Participation Group,
Universidad Nacional Aut\'onoma de M\'exico, University of Arizona, 
University of Colorado Boulder, University of Oxford, University of Portsmouth, 
University of Utah, University of Virginia, University of Washington, University of Wisconsin, 
Vanderbilt University, and Yale University.

\clearpage





  

\end{document}